\documentclass[10pt]{article}

\usepackage{latexsym}
\usepackage[dvips]{epsfig} 
\usepackage{amssymb,amsmath,amsthm,mathrsfs} 
\usepackage{graphicx}
\usepackage{lineno}

\usepackage[authoryear,sort&compress]{natbib}
\usepackage{url}
\usepackage{enumerate}
\usepackage{colordvi,color,pspicture}
\usepackage{psfrag}
\usepackage{rotating}
\usepackage{algorithmic}

\newcommand{\X}{\mathcal{X}}

\newcommand{\I}{\mathbf{I}}

\newcommand{\V}{\mathcal{V}}

\newcommand{\E}{\mathbf{E}}
\newcommand{\Var}{\mathbf{Var}}

\newcommand{\al}{\alpha}

\DeclareGraphicsExtensions{.jpg, .eps}
\theoremstyle{plain}

\theoremstyle{definition}

\theoremstyle{remark}

\begin{document}

\title{Technical Report \# KU-EC-08-2:\\
The Use of Labeled Cortical Distance Maps for Quantization and Analysis of
Anatomical Morphometry of Brain Tissues}
\author{
E. Ceyhan$^{1,2^\ast}$, M. Hosakere$^{2}$, T. Nishino$^{3,4}$, J.
Alexopoulos$^{3}$, R.D. Todd$^{5}$, K.N. Botteron$^{3,4}$,\\
M.I. Miller$^{2,6,7}$, J.T. Ratnanather$^{2,6,7}$
}
\date{\today}

\maketitle
\begin{center}
$^{1}$\textit{Dept. of Mathematics, Ko\c{c} University, 34450, Sar{\i}yer, Istanbul, Turkey.}\\
$^{2}$\textit{Center for Imaging Science, The Johns Hopkins University, Baltimore, MD 21218.}\\
$^{3}$\textit{Dept. of Psychiatry, Washington University School of Medicine, St. Louis, MO 63110.}\\
$^{4}$\textit{Dept. of Radiology, Washington University School of Medicine, St. Louis, MO 63110.}\\
$^{5}$\textit{Dept. of Genetics, Washington University School of Medicine, St. Louis, MO 63110.}\\
$^{6}$\textit{Institute for Computational Medicine, The Johns Hopkins University, Baltimore, MD 21218.}\\
$^{7}$\textit{Dept. of Biomedical Engineering, The Johns Hopkins University, Baltimore, MD 21218.}\\
\end{center}

\noindent
*\textbf{corresponding author:}\\
Elvan Ceyhan,\\
Dept. of Mathematics, Ko\c{c} University,\\
Rumelifeneri Yolu, 34450 Sar{\i}yer,\\
Istanbul, Turkey\\
e-mail: elceyhan@ku.edu.tr\\
phone: +90 (212) 338-1845\\
fax: +90 (212) 338-1559\\

\noindent
\textbf{short title:} Using Labeled Cortical Distance Maps for Morphometric Quantization

\noindent
\textbf{keywords:} computational anatomy, depression, labeled cortical
distance map (LCDM), morphometry, pooled distances

\newpage

\begin{abstract}
\noindent
Anatomical shape differences in cortical structures in the brain can be
associated with various neuropsychiatric and neuro-developmental diseases or disorders.
Labeled Cortical Distance Map (LCDM), a recently devised tool,
can be a powerful tool to quantize such morphometric differences.
In this article, we investigate
various issues regarding the analysis of LCDM distances in relation to morphometry.
The length of the LCDM distance vector provides the number of voxels
(approximately a multiple of volume (in $mm^3$));
median, mode, range, and variance of LCDM distances are all suggestive of size, thickness,
and shape differences.
Various statistical tests are employed to
detect left-right morphometric asymmetry,
group differences, and stochastic ordering (i.e., cdf differences) of these quantities.
However these measures provide a crude summary based on LCDM distances
which may convey much more information about the tissue in question.
To utilize more of this information, we pool (merge) the LCDM distances
from subjects in the same group or condition.
We check for the similarity of the distributions of LCDM distances for subjects
in the same group using the kernel density plots, and also investigate the
influence of the outliers (i.e., subjects with extremely different LCDM
distance distributions).
The statistical methodology we employ require normality and within and between sample independence.
We demonstrate that the violation of these assumptions have mild influence on the tests.
We specify the types of alternatives the parametric
and nonparametric tests are more sensitive for.
We also show that the pooled
LCDM distances provide powerful results for group differences in
distribution, left-right morphometric asymmetry of the tissues, and
variation of LCDM distances.
As an illustrative example, we use gray matter
(GM) tissue of ventral medial prefrontal cortices (VMPFCs) from subjects
with major depressive disorder, subjects at high risk, and control subjects.
We find significant evidence that VMPFCs of subjects with depressive
disorders are different in shape compared to those of normal subjects.
Although the methodology used here is applied on the LCDM distances of GM of VMPFC,
it is also valid for morphometric measures of other organs
or tissues and distances similar to LCDM distances.
\end{abstract}

\newpage
\section{Introduction}
\label{sec:intro}
Quantification of morphometric properties of neocortical tissues is a major
component of Computational Anatomy [1-21].
Our group recently developed the Labeled Cortical Distance Mapping (LCDM) techniques
[22] which was shown to be useful in identifying cortical thinning in the cingulate cortex in subjects with
Alzheimer's Disease [23] and in subjects with schizophrenia [24] in comparison to control subjects.

Cortical thinning has been observed in other regions in a variety of
neuro-developmental and neuro-degenerative disorders (see above references for examples).
In particular, functional imaging studies implicate the
ventral medial prefrontal cortex (VMPFC) in major depressive disorders (MDD)
[25, 26] which have been correlated with
shape changes observed in structural imaging studies [27, 28].
The prefrontal cortex, together with amygdala and hippocampus,
plays an important role in modulating emotions and mood.
Structural imaging studies
in MDD have largely focused on adult onset with only few focused on early
onset MDD which has been associated with structural deficits in the
subgenual prefrontal cortex, a subregion of the VMPFC [28].
Furthermore, the whole VMPFC has been examined in a twin study of early onset MDD [29].

Several studies of the VMPFC and related structures have been obtained from
analysis of the cortex as a whole [4,17, 30, 31] whereas others have pursued a more localized analysis attempts
to deal with the highly folded gray/white matter cortex [32].
In this way the laminar shape of the
brain tissue can be quantified in great detail. Two aspects of the laminar
shape are structural formation (like surface and form of the tissue) and
scale or size (like volume and surface area).
Throughout the article, we
call all aspects of laminar shape as the \textit{morphometry} of the tissue (including shape and
size), the surface structure and form will be referred as ``shape'' and
scale will be referred as ``size''.

\begin{figure}[htbp]
\centering
\rotatebox{90}{ \resizebox{2.5 in}{!}{\includegraphics{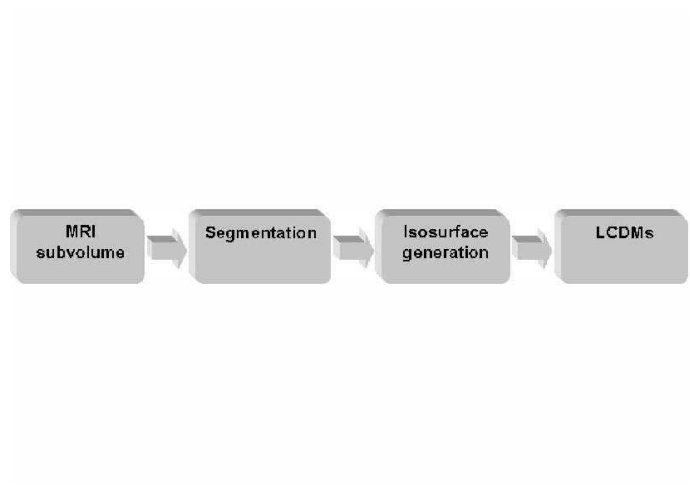} } }
\caption{
\label{fig:PFC-meth1}
A schematic view of flowchart of LCDM measurement procedure. }
\end{figure}

\begin{figure}[htbp]
\centering
\rotatebox{90}{ \resizebox{2.5 in}{!}{\includegraphics{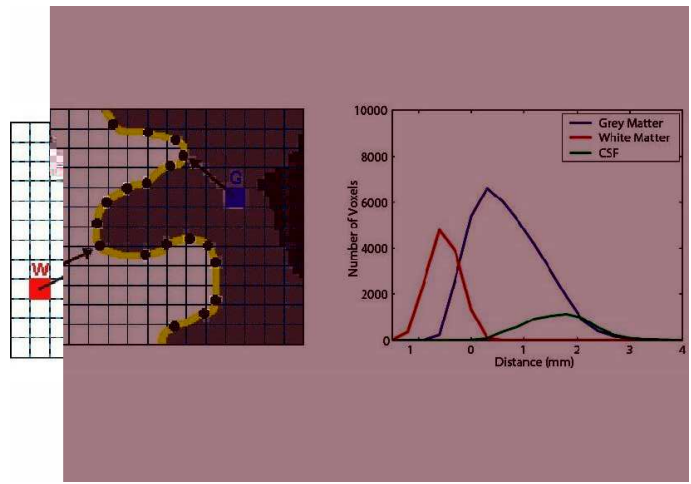} } }
\caption{
\label{fig:PFC-meth2}
A two-dimensional illustration of normal
distances from a GM and a WM voxel to the GM/WM interface (left) and
non-normalized histograms of LCDM distances of GM, WM, and CSF tissues. }
\end{figure}

The first step in creating LCDM metrics involves segmenting MRI subvolumes
of the tissue in question.
Then every voxel is labeled by tissue type as
gray matter (GM), white matter (WM), and cerebrospinal fluid (CSF).
For every voxel in the image volume, its (normal) distance from the center of
the voxel to the closest point on GM/WM surface is computed.
A signed distance is used to indicate the location of each voxel with respect to the
GM/WM surface; distances are positive for GM and CSF voxels, and negative for WM voxels.
See Figure \ref{fig:PFC-meth1} for a schematic flowchart of the LCDM procedure
and Figure \ref{fig:PFC-meth2} for a two-dimensional illustration of LCDM distance calculation
and non-normalized histograms of the (signed) distances for GM, WM, and CSF.

As an illustrative example, we investigate GM tissue in VMPFCs in a study of
early onset depression in twins. Previously, we analyzed various
morphometric measures (i.e., volume, descriptive statistics based on LCDM
distances such as median, mode, range, and variance) and demonstrated that
except for left-right asymmetry and correlation between left and right
measures, these variables usually \textit{failed} to discriminate the
healthy subjects from depressed ones [33].
One reason for this is the fact that the
subjects in our data set are age-matched female twins, who potentially have
VMPFCs similar in size. Moreover, this might be partly due to the size of
groups; i.e., if we had more participants in the study, these measures would
have been more likely to yield significant differences between the groups.
On the other hand, by only using a descriptive summary statistic of
numerous of distances for each person, we essentially lose most of the
information provided by LCDM measures.
In this article we provide a strategy
to avoid this information loss by pooling the LCDM distances.
We use the pooled (by condition or group) distances to detect morphometric differences
such as differences in shape, size, thickness variation, and left-right asymmetry.
However there is a downside to pooling,
the pooled distances do not have within sample independence,
as the distances of neighboring voxels
for each voxel at a particular hemisphere of a person are dependent.
Moreover, there is also dependence between distances from left and right
VMPFCs of each subject.
We demonstrate that within sample dependence does
not affect the tests in terms of empirical significance levels (or Type I errors) or power;
and left-right dependence only makes the asymmetry tests
less powerful than they could have been.

We describe the acquisition of LCDM distances for VMPFCs in Section \ref{sec:data-acquisition},
the methods we employ in Sections \ref{sec:stat-tests},
describe the example data set in Section \ref{sec:example-data},
provide analysis of volumes and descriptive measures based on LCDMs in Section \ref{sec:vol-desc},
outlier detection in Section \ref{sec:outlier-detection},
analyze the pooled distances in Section \ref{sec:results},
and present the influence of assumption of violations in Section \ref{sec:assumption-violations}.
In [33] we computed simple descriptive
measures for each left and right VMPFCs of the subjects
and analyzed these measures for group differences.
These LCDM-based descriptive statistics are also analyzed in more detail in this article.
For more technical detail on LCDM see, e.g., [36],
where accuracy of LCDMs and variability of cortical mantle distance profiles are also discussed.
In [36], LCDM is used for detecting differences in cingulate due to dementia of
Alzheimer's type (DAT).
The usual Welch's $t$-test is applied for volume comparisons,
Wilcoxon rank sum test is applied for group comparisons on randomly selected subsamples
from LCDM distances.
In this article, rather than subsampling or summarizing,
we use the entire LCDM distance set by pooling the
distances for each group and investigate the validity of the underlying
assumptions for the tests used.
Since in this article we focus on the use of
LCDM distances, rather than the clinical implications of the genetic
influence (due to twinness), we ignore the twin influence for most of the
current analysis.

\section{Methods}
\label{sec:methods}
\subsection{Data Acquisition}
\label{sec:data-acquisition}
A cohort of 34 right-handed young female twin pairs between the ages of 15
and 24 years old were obtained from the Missouri Twin Registry and used to
study cortical changes in the VMPFC associated with MDD.
Both monozygotic and dizygotic twin pairs were included,
of which 14 pairs were controls (Ctrl) and 20 pairs had one twin affected with MDD,
their co-twins were designated as the High Risk (HR) group.

Three high resolution T1-weighted MPRAGE magnetic resonance scans of this
population were acquired using a Siemens scanner with $1\,mm^{3}$ isotropic resolution.
Images were then averaged, corrected for intensity inhomogeneity
and interpolated to $0.5 \times 0.5 \times 0.5 \,^{3}$ isotropic voxels.

Following Ratnanather et al. [32], a region of interest (ROI)
comprising the prefrontal cortex stripped of the basal ganglia,
eyes, sinus, cavity, and temporal lobe was defined manually and segmented
into gray matter (GM), white matter (WM), and cerebrospinal fluid (CSF) by
Bayesian segmentation using the expectation maximization algorithm [34].
A triangulated representation of the cortex at the
GM/WM boundary was generated using isosurface algorithms [35].

See Figure \ref{fig:PFC-meth1} for the schematic flowchart of LCDM measurement procedure
and Figure \ref{fig:PFC-meth2} for an illustration of normal distances from GM and WM voxels to the
interface in a two-dimensional setting and non-normalized histograms of LCDM distances
for GM, WM, and CSF tissues of a cingulate tissue.
The GM tissue comprises most of the cortex; and by construction, most of GM
distances are positive, most of WM distances are negative, and all of CSF
distances are positive.
The mismatch of the signs for some GM and WM voxels
close to the GM/WM boundary are due to the way the surface is constructed in
relation to how the pixels are labeled, such that a surface is always
intersecting pixels, and has to maintain a somewhat smooth shape.
So some appropriately labeled GM and WM pixels may fall on a side of surface that
they should not belong to; however, these mislabeled voxels constitute a
small number of voxels and do not affect our overall analysis.
Let $V$ be the regular lattice of voxels defining the region of interest,
$S\left(\Delta \right)$ be the triangulated graph representing the smooth boundary
at the GM/WM surface.
Then the distance computation algorithm is specified as [36-38].

\begin{algorithmic}
\FORALL{$v_i \in \V$}
\STATE $s_{closest} \leftarrow $ a point in $S(\Delta)$ such that

\FORALL{$s_j \in S(\Delta)$}
\STATE $d(s_{closest},v_i) \le d(s_j,v_i)$
\ENDFOR
\STATE $D_i \leftarrow d(s_{closest},v_i)$

\ENDFOR
\end{algorithmic}
where $d\left( {\cdot ,\cdot} \right)$ stands for the usual Euclidean
distance, $v_i$ is the $i^{th}$ voxel, $s_j$ is the $j^{th}$ point
in $S\left( \Delta \right)$ and $D_i$ is the $i^{th}$ distance
(i.e., distance for $i^{th}$ voxel). That is, an LCDM distance is a set distance
function $d\;\mbox{:}\;\mbox{}v_i \in \;V\to \;d\left( {\mbox{centroid}(v_i ),S\left( \Delta \right)} \right)$,
which is the distance between the
centroid (or center of mass) of $v_i$ and the set $S\left( \Delta \right)$.
More precisely,
$$
D_i \;:=\;\;d\left( {\mbox{centroid}(v_i ),S\left( \Delta \right)}
\right)\;=\;\mathop {\min}\limits_{s\in S\left( \Delta \right)} \;\left\|
{\mbox{centroid}(v_i )-\;s} \right\|_2
$$
To distinguish the distances for voxels from different tissue types, we
denote the distance for $i^{th}$ voxel in tissue type label as
$D_i \left( {label} \right)$ for $label\in {\kern 1pt}\left\{ {\mbox{WM,GM,CSF}} \right\}$.

Volume and GM distributions as a function of position from the GM/WM interface can be derived from LCDMs.
The GM volume is simply the total number of GM voxels (times the volume of a single voxel).
As variability of total GM volume
can be a confounding factor in studying cortical thickness, normalizing LCDMs of each
individual by its total GM volume generates Cortical Mantle Distance (CMD) profiles.
Integrating the density function results in a distribution function that represents
the percentage of GM as a function of distance from the cortical surface.

\subsection{Statistical Tests}
\label{sec:stat-tests}
We use various morphometric measures of left (and right) VMPFCs in our analysis.
LCDM distance measures are sufficient to determine the volumes (in $mm^3$) of VMPFCs.
For each person, we also record the median, mode, range, and variance of LCDM distances.
For left (and right) volume, median, mode, range, and variance comparisons between groups,
we can not apply Kruskal-Wallis (K-W) test for equality of the distributions of these measures for all groups,
because there is an inherent (genetic) dependence between MDD and HR groups,
since cotwin of each MDD subject is by definition a HR subject.
So we use Wilcoxon rank sum test to compare MDD and Ctrl VMPFC,
and also HR and Ctrl VMPFC.
On the other hand, we use Wilcoxon signed rank test for MDD and HR VMPFC,
due to dependence of the samples.
Then we adjust these $p$-values for simultaneous pairwise comparisons by
Holm's correction method.
See [45] for the tests and [39] for Holm's correction.
We resort to the non-parametric tests only,
when the assumptions of normality and homogeneity of
the variances (HOV) are not met.
When the assumptions are met
and only the parametric tests yield significant results,
we use the parametric counter-parts
(Welch's $t$-test for independent samples and paired $t$-test).

For morphometric asymmetry of left and right VMPFCs,
we compare these measures between left and right VMPFCs
(overall and for each group) by
Wilcoxon signed rank test which is a paired (for differences of measures) test
(see, e.g., [40]).

We perform correlation analysis between left and right morphometric measures
using Spearman's rank correlation coefficient and the corresponding test
against the correlation coefficient being nonzero.
We also compare cdfs (cumulative distribution functions)
between groups for left (and right) VMPFCs
using Kolmogorov-Smirnov (K-S) test (see, e.g., [40]).

When we pool (i.e., merge) the LCDM distances by group, there is an inherent dependence
between LCDM distances due to the spatial correlation between neighboring
voxels of a left or right VMPFC.
For the parametric tests (ANOVA $F$-test and $t$-test),
the assumptions of normality and within sample independence are
violated, while for nonparametric tests (K-W test and Wilcoxon tests) only
within sample independence is violated.
Since more assumptions are violated for the parametric tests,
the nonparametric tests are expected to be more appropriate in our analysis.
However, since the correlation structure is similar for each person (hence for each group),
its influence on both parametric and nonparametric tests is negligible.
See Section \ref{sec:assumption-violations} for a
Monte Carlo simulation study to justify the use of these tests for such data structures.
We use Kruskal-Wallis (K-W) test for equality of the
distributions of the pooled distances for all left groups and ANOVA
$F$-tests (with and without HOV) for equality of the means for all left groups;
if K-W test yields a significant $p$-value, then we use pairwise Wilcoxon rank
sum test to compare the pairs of the groups;
similarly, if one of the ANOVA $F$-tests is significant,
then we use pairwise $t$-test to compare the pairs of the groups.
Then we adjust these $p$-values for simultaneous pairwise
comparisons by Holm's correction method [39].
We perform similar analysis
for right groups. For morphometric asymmetry of left and right VMPFCs, we
compare these measures between left and right VMPFCs (overall and for each
group) by Wilcoxon rank sum test (see, e.g., [40]) and Welch's $t$-test.
Although there is an inherent dependence on the MDD and HR VMPFC or left and right
distances, we do not use Wilcoxon signed rank test (for dependent samples)
or matched pair $t$-test, because the distances for the left and right VMPFCs
can not be matched (paired).
For the same reason, we cannot perform correlation analysis on these groups.
We also compare cdfs between groups
for left (and right) VMPFCs using Kolmogorov-Smirnov (K-S) test (see, e.g., [40]).

The test of equality or homogeneity of the variances (HOV) of pooled
distances is also important.
Because, variance differences
between groups might be indicative of differences between the variations of
the morphometry of VMPFCs.
Therefore, we perform HOV test by using Levene's
test with absolute dispersion around the median, which is also known as
Brown-Forsythe's (B-F) HOV test (see, e.g., [41]).

\subsection{Example Data Set}
\label{sec:example-data}
\begin{figure}[htbp]
\centering
\rotatebox{90}{ \resizebox{2.5 in}{!}{ \includegraphics{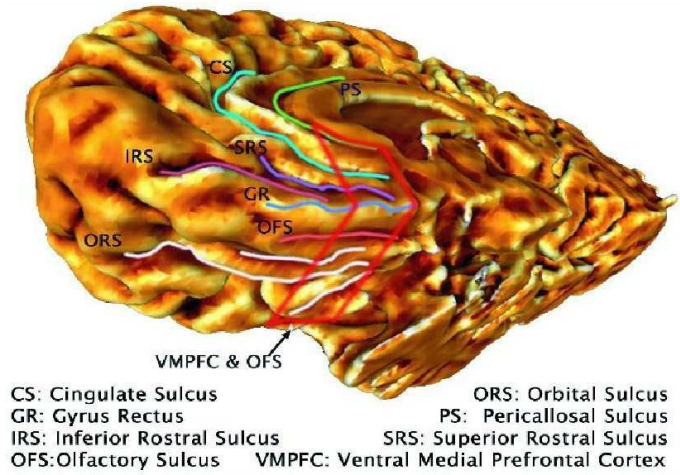} } }
\caption{
\label{fig:VMPFC-loc}
The location of VMPFC in the brain.
}
\end{figure}

As an illustrative example,
we use GM of left and right VMPFCs.
The prefrontal cortex, together with the amygdala and hippocampus
plays an important role in modulating emotions and mood.
For the location of VMPFC in brain, see Figure \ref{fig:VMPFC-loc}.
Abnormalities have been demonstrated in structure and function of the prefrontal cortex
due to depression [25,26].
Previous structural imaging studies on Major Depressive Disorder (MDD)
have largely focused on adult onset, while only few have focused on early onset MDD.
Botteron et al. [44] have conducted structural imaging studies on
early onset of MDD in the ventral medial prefrontal cortex (VMPFC) region of twins.
Structural deficits in the subgenual prefrontal cortex have been shown to
implicate early onset of MDD [29].

For convenience in notation, we suppress the label argument in the
distances, as we only consider the GM tissue.
Let $D^L$ be the set of LCDM
distances, $D_{ijk}^L$, which is the distance associated with $k^{th}$
voxel in GM of left VMPFC of subject $j$ in group $i$ for $j = 1,{\ldots},n_{i}$ and
$i = 1,2,3$ (group 1 is for MDD, group 2 for HR, and group 3 for Ctrl).
Right VMPFC distances $D^R$ are denoted similarly as$D_{ijk}^R$.
The LCDM distances for GM in left and right VMPFCs, $D^L$ and $D^R$, are plotted by
subject in Figure \ref{fig:raw-dist-LR}.
The automated segmentation is more reliable for the GM
close to the GM/WM surface due to the high level of contrast.
However, there are still voxels which, although labeled appropriately, have the incorrect
sign, that is, some of $D_{ijk}^L$ and $D_{ijk}^R$ are negative for each
subject. Large distances are potentially less reliable, due to the
diminishing contrast around the boundary of GM and CSF compartments. Thus,
based on Figure \ref{fig:raw-dist-LR} and prior anatomical knowledge on VMPFCs (e.g., [42]),
we only keep distances larger than $-0.5 \textit{ mm}$ so
that mislabeled WM is excluded from the data, and the upper limit is set to
$5.5\, mm$, so that the error due to less reliable large distances is reduced.
Observe that in the left VMPFC distances,
MDD subjects 8 and 11, HR subjects 3 and 18, and Ctrl subjects 6, 10, 11, 12, and 22 seem to be more
different with Ctrl subjects 11 and 12 being ``thinner" while the rest are ``thicker" than other left VMPFCs.
On the other hand, in the right VMPFC distances,
VMPFC of MDD subject 4, HR subject 5, and Ctrl subjects 12 and 21 seem to be more different,
with VMPFCs of HR subject 5 and Ctrl subject 12 being thinner and the others being thicker than
the rest of the right VMPFCs.
Note that Figure \ref{fig:raw-dist-LR} provides a preliminary assessment of reliability of LCDM distances,
since it does not provide the distributional behavior of the distances, but
only problems with small (negative) and large distances.
As a technical aside, we note that only 0.16 {\%} of left distances and 0.14
{\%} of right distances are below -0.5 \textit{mm}; on the other hand, only 0.22 {\%}
of left distances and 0.07 {\%} of right distances are above $5.5 \textit{mm}$.

\section{Analysis of Volumes and Descriptive Measures of LCDM Distances of VMPFCs}
\label{sec:vol-desc}
Volume is a measure of size of VMPFCs.
Let $V^L_{ij}$ be the volume of left VMPFC of subject $j$ in group $i$,
for $j=1,\ldots,n_i$ and $i=1,2,3$.
Right VMPFC volumes are denoted similarly as $V^R_{ij}$.
For each person, number of LCDM distances recorded
yield the number of voxels, which in turn yields a multiple of the volumes in $mm^3$,
since each voxel is a cube of size $0.5\times 0.5 \times 0.5$ $mm^3$.
That is,
$$V^L_{ij}=0.125 \times \sum_{k}\I\left(D^L_{ijk} \in [-.5,5.5]\right)$$
where $\I(\cdot)$ stands for the indicator function.
Right VMPFC volumes can be obtained similarly.

\subsection{Analysis of Volumes of VMPFCs}
\label{sec:vol}
\begin{figure}[htbp]
\centering
\rotatebox{-90}{ \resizebox{3.5 in}{!}{ \includegraphics{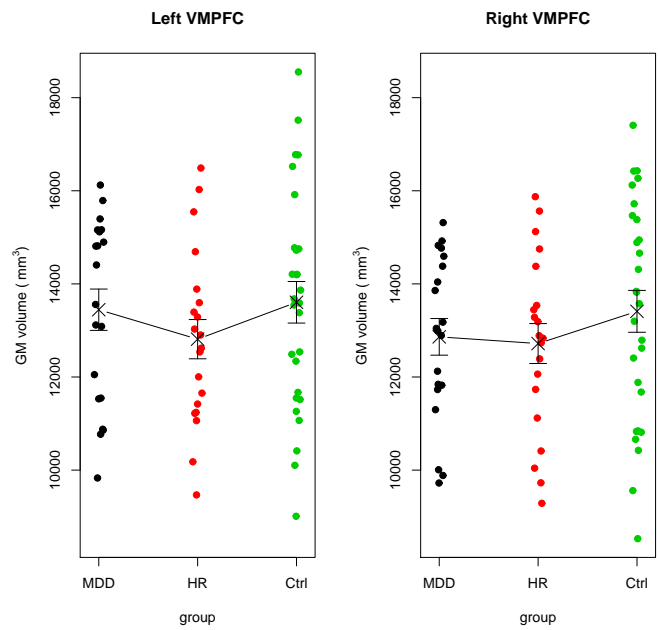} } }
\caption{
\label{fig:Scatter-factor-vol}
Depicted are the (slightly jittered) scatter plots of the volumes of the left and right VMPFCs.
The crosses, $\times$, are located at the mean volume value for each group.
}
\end{figure}

We analyze volumes of VMPFCs for various purposes:
(i) to provide an outline of the methodology we will employ for other quantities in this article,
(ii) to compare the volume results with other comparisons,
and (iii) to check the volume (size or scale) differences due to groups.
We note the group of each volume value;
for example, if a volume is of a VMPFC of a person in group MDD,
then the corresponding group is MDD.
See Figure \ref{fig:Scatter-factor-vol}
for the (jittered) scatter plot of the volumes by group
for left and right VMPFCs.
See also Table \ref{tab:describe-vol}
for the sample sizes, means, and standard deviations
of the volumes, overall and for each group,
for left and right VMPFCs.
Observe that left VMPFC volumes are larger than the right VMPFC volumes.
Moreover the mean volume measures
for the left VMPFC seem to be more different between groups compared to right VMPFCs.

To find which pairs, if any, manifest significant differences,
we perform pairwise comparisons by Wilcoxon rank sum
tests for MDD,Ctrl and HR,Ctrl pairs and Wilcoxon signed rank test
for MDD,HR pair for left VMPFC volumes using Holm's correction.
The $p$-values for the pairwise tests for left and right VMPFC volumes
are presented in Table \ref{tab:pairwise-pval-vol},
where $p_W$ stands for $p$-value for Wilcoxon rank sum test,
$p_t$ stands for $p$-value for Welch's $t$-test,
significant results are marked with an asterisk (*) and
$(\ell)$ stands for the alternative that first group volumes tend to be less than second group volumes
and $(g)$ stands for the alternative that first group volumes tend to be greater than second group volumes.
More precisely, given two groups of random variables $X$ and $Y$,
$(\ell)$ alternative implies
$F_X(x)>F_Y(x)$ where $F_X$ and $F_Y$ are the distribution functions for $X$ and $Y$, respectively,
or $\E[X]<\E[Y]$ where $\E[\cdot]$ stands for expectation or $P(X>Y)<P(X<Y)$.
Notice that all of these various forms of alternatives
convey the idea that ``X tends to be smaller than Y" in some way [40].
Observe that none of the pairs
manifest significant differences in distribution of volumes.
For the $t$-test, $(\ell)$ stands for the alternative
that the first group mean volume is less than the second group mean,
and $(g)$ stands for the alternative that the first group mean is greater than the second group mean.
None of the pairs indicate significant differences in mean volumes.

Since MDD,HR volumes are dependent,
we only test for the homogeneity or equality of the variances of volumes of
MDD,Ctrl and HR,Ctrl pairs.
We observe that HOV of MDD and Ctrl left VMPFC volumes is not rejected ($p=.3610$),
and likewise for HR and Ctrl left VMPFC volumes ($p=.3202$).
Similarly, HOV of MDD and Ctrl right VMPFC volumes is not rejected ($p=.1038$),
and likewise for HR and Ctrl right VMPFC volumes ($p=.1038$).
This suggests that the spread or variation in volumes
of left (and right) VMPFC within
groups is not significantly different from each other for the groups considered.

We also test for the differences between left and right VMPFC volumes, i.e.,
left-right volumetric asymmetry.
For the associated $p$-values, see Table \ref{tab:paired-pval-vol},
where $p_W$ stands for the $p$-value for Wilcoxon signed rank test,
and $p_t$ stands for paired $t$-test.
For testing overall left-right asymmetry,
we pool all the left volumes into one set,
and all the right volumes into another.
Observe that left VMPFC volumes are significantly larger than right VMPFC volumes
($p_W=.0064$ and $p_t=.0087$).
Among the groups,
only MDD VMPFC shows significant volumetric left-right asymmetry with $p_W=.0360$ and $p_t=.0233$.
That is,
the depressed subjects tend to have more left-right volumetric asymmetry
compared to HR and Ctrl subjects, in such a way that left volumes tend to be significantly larger
than the right volumes in MDD subjects.

Spearman's rank correlation coefficients,
denoted $\rho_S$, and the associated $p$-values
between the left and right scales are
given in Table \ref{tab:correlation-vol},
where MDDL refers to volumes of left VMPFC of MDD patients, MDDR, HRL, and HRR
are defined similarly.
Observe that there is significant (positive) correlation between the
left and right volumes for overall, MDD, HR, and Ctrl VMPFCs.
Correlation implies that, for example,
when left volumes increase or decrease,
so do the right volumes.
On the other hand,
MDD,HR left volumes are mildly correlated,
while MDD,HR right volumes are not significantly correlated.

We also compare the cumulative distribution function of the volumes by group,
which may also provide a stochastic ordering,
for MDD,Ctrl and HR,Ctrl pairs only, since the MDD,HR pairs are dependent.
See Table \ref{tab:cdf-pval-vol} for the associated $p$-values.
Observe that none of the $p$-values is significant.

Notice that except for left-right asymmetry and correlation between left and right volumes
(for each group),
none of the comparisons is significant at .05 level.
But this does not necessarily imply lack of VMPFC shape differences between groups,
as volume only measures an aspect of size.
Next, we analyze various descriptive measures (summary statistics)
based on GM LCDM distances.
The lack of significant group differences in volumes might be due to the fact that
the data consists of age-matched female subjects,
whose VMPFCs might be very similar in size.
Furthermore, if the number of subjects per group is increased,
then it is more likely to see significant group differences,
if they exist.

\subsection{Analysis of Descriptive Measures of LCDM Distances of GM of VMPFCs}
\label{sec:desc-lcdm}
In this section we analyze some other measures which are more directly associated
with LCDM distances.
We find the descriptive measures (summary statistics) of the LCDM distances for each person.
Among the descriptive statistics we analyze are
the median, mode, range, and variance of the LCDM distances.
We conduct the tests that we used for volumes in Section \ref{sec:vol}
on these descriptive measures.

Note that each of these descriptive measures conveys information about
some aspect of morphometry (shape and size) of VMPFCs.
We provide these analysis to demonstrate how LCDM distances
can be used as a simple comparative tool.

The {\bf median} of LCDM distances for VMPFCs yields a central distance measure,
or distance from ``center" of VMPFC GM to the GM/WM interface.
The median distance for left and right VMPFC gray matter for subject $j$ in group $i$
are denoted as $med\left(D^L_{ij}\right)$ and $med\left(D^R_{ij}\right)$, respectively.
We use the median distance rather than the mean distance here,
because LCDM distances are skewed right, so median is a better
measure of centrality as it is more robust to extreme values compared to the mean.
See Figure \ref{fig:raw-dist-LR} for right skewness of distances in the scatterplot
and Figure \ref{fig:kernel-dens-LR} for the kernel density plots for LCDM
distances for left and right VMPFC by subject.
The tests indicate that there is no group differences
in the distributions of the median distances for both left and right VMPFCs,
no significant left-right asymmetry,
and no significant difference between the cdfs of median distances of
groups for both left and right VMPFCs.
HOV is rejected for HR and Ctrl left median distances with $p=.0261$ only.
Furthermore, MDD,HR, and Ctrl-left,Ctrl-right median distances are
significantly positively correlated,
while MDD,HR left (and right) median distances are not.

The {\bf mode} of a data set as a descriptive statistic is the most frequent observation in the data set.
To make it more meaningful for our data,
we rounded the distances to 1 decimal place.
Hence mode corresponds to the tenth of a millimeter that
contains the most number of GM voxel distances.
For instance, if mode of a subject is 2.2 mm
the most number of distances are in the interval $[2.2\,mm,2.3\,mm]$ compared to other intervals.
More precisely, the mode of LCDM distances for VMPFC
yields the distance from the ``widest" strip parallel to the GM-WM interface.
The mode of distances for left and right VMPFC gray matter for subject $j$ in group $i$
are denoted as $mode\left(D^L_{ij}\right)$ and $mode\left(D^R_{ij}\right)$, respectively.
The tests indicate that there is no group differences
in mode of the distances for both left and right VMPFC,
no significant left-right asymmetry,
no significant variance difference of mode of distances between groups,
and no significant difference between the cdfs of the modes of the distances
between groups for both left and right VMPFCs.
There is mild positive correlation between Ctrl-left,Ctrl-right modes,
and MDD,HR right modes.

The {\bf range} of LCDM distances for VMPFCs yields a rough measure of ``thickness" of GM of VMPFC.
We use the range (maximum LCDM distance minus minimum LCDM distance)
rather than the maximum distance here,
although conceivably the latter is also a reasonable choice.
The range of distances for left and right VMPFC gray matter for subject $j$ in group $i$
are denoted as $range\left(D^L_{ij}\right)$ and $range\left(D^R_{ij}\right)$, respectively.
The tests indicate that there is no group differences
in range of distances (thickness) for both left and right VMPFCs,
no significant differences between the cdfs of groups for both left and right VMPFCs,
and HOV of ranges of groups is not rejected.
Left distance ranges are significantly larger than right distance ranges,
overall and for each group.
Furthermore,
there is mild positive correlation between Ctrl-left,Ctrl-right ranges,
and MDD,HR right ranges.

The {\bf variance} of LCDM distances for VMPFC yields a measure of ``variation" of size of GM in VMPFC.
We use the variances, rather than standard deviations here,
since both yield the same results under rank based non-parametric tests.
The variance of distances for left and right VMPFC gray matter for subject $j$ in group $i$
are denoted as $\Var\left(D^L_{ij}\right)$ and $\Var\left(D^R_{ij}\right)$, respectively.
The tests indicate that there is no group difference
in variance of distances (size variation) for both left and right VMPFC,
and no significant difference between the cdfs of variances of distances
between groups for both left and right VMPFCs,
and HOV of variances of groups is not rejected.
There is significant left-right asymmetry in variance of distances,
with left variances significantly larger than right variances,
overall and for each group.
Furthermore,
there is mild correlation between MDD left and right variances.

Although descriptive statistics of LCDM distances measure some morphometric aspect of VMPFCs,
they usually fail to discriminate the healthy VMPFCs from depressed VMPFCs.
One reason for this is the fact that
the subjects in our data set are age-matched female twins,
who potentially have similar size VMPFCs.
Moreover, this might be partly due to the size of groups;
i.e., if we had more participants in the study,
these measures are more likely to yield significant differences between the groups.
On the other hand,
by only using a descriptive summary statistic of thousands of distances for each person,
we essentially lose most of the information LCDM measures convey.
To avoid this over-summarization,
we will use all the distances in our analysis in the next section.

One could also use other descriptive measures such as inter-quartile range (IQR),
skewness, and kurtosis of the LCDM distances.

\section{Pooling LCDM Distances}
\label{sec:pooling-LCDM-distances}

Since the descriptive measures of LCDM distances are summary statistics,
they tend to oversimplify the data[33], and hence we lose most
of the information conveyed by the LCDM distances.
To avoid this information
loss, we pool LCDM distances of subjects from the same group or condition;
that is, we pool the LCDM distances of all left MDD VMPFCs in one group,
those of all left HR VMPFCs in one group, and those of all left Ctrl VMPFCs in one group.
That is,
\[
D_{i\ell}^L \;=\;
\bigcup_{j=1}^{n_i} D_{ijk}^L
\]
where $D_{i\ell}^L$ is the $\ell ^{th}$ distance value for distances from
subjects in group $i$.
We pool the right VMPFC LCDM distances in a similar fashion
and denote the right pooled distances as $D_{i\ell}^R$.

One of the underlying assumptions is that the distances from VMPFC of
subjects with MDD have the same distribution, those of HR have the same
distribution, and so do those of Ctrl group.
In other words, we assume that
$D_{ijk}^L$ are identically distributed for all $j$= 1,{\ldots},$n_{i}$ and
$i= 1,2,3$.
So, $D_{ijk}^L \;\sim \;F_i^L$ for all $j,k$, and likewise $D_{ijk}^R\;\sim \;F_i^R$ for all $j,k$.
Hence the pooled distances are distributed as
$D_{i\ell}^L \;\sim \;F_i^L$ and $D_{i\ell}^R \; \sim \;F_i^R$ for $i=1,2,3$.
We take this action under the presumption that the morphometry of
VMPFCs are affected by the condition in a similar way and hence age and
gender matched subjects with the same condition should have VMPFCs similar in morphometry.
As a precautionary step, we find the \textit{extreme (outlier)} subjects;
i.e., the subjects whose VMPFCs have much different distributions than the rest in
Section \ref{sec:outlier-detection} by (subjectively) comparing
the kernel density estimates.

\subsection{Outlier Detection by Using Kernel Density Plots}
\label{sec:outlier-detection}
\begin{figure}[htbp]
\centering
\rotatebox{-90}{ \resizebox{2.3 in}{!}{ \includegraphics{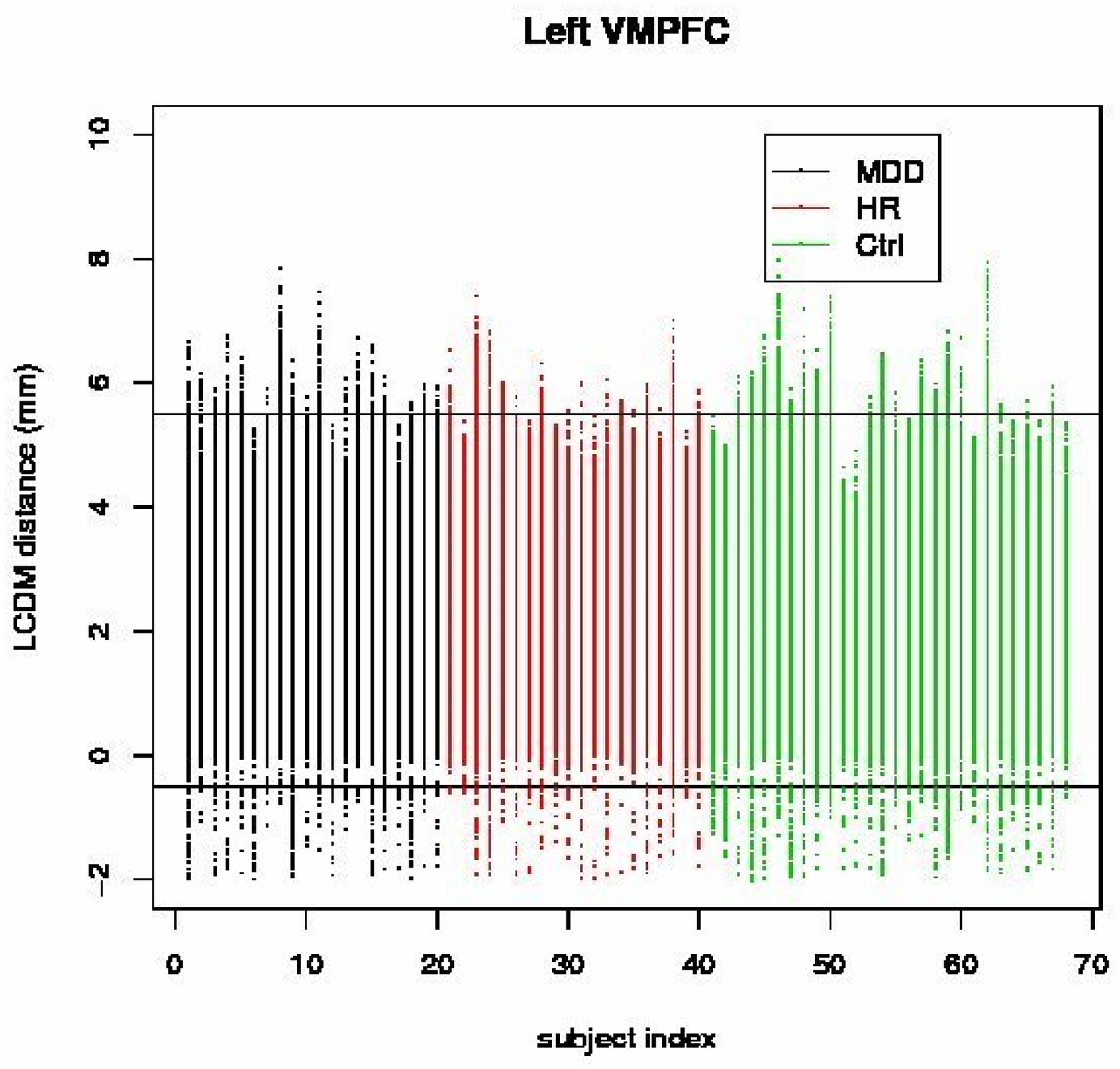} } }
\rotatebox{-90}{ \resizebox{2.3 in}{!}{ \includegraphics{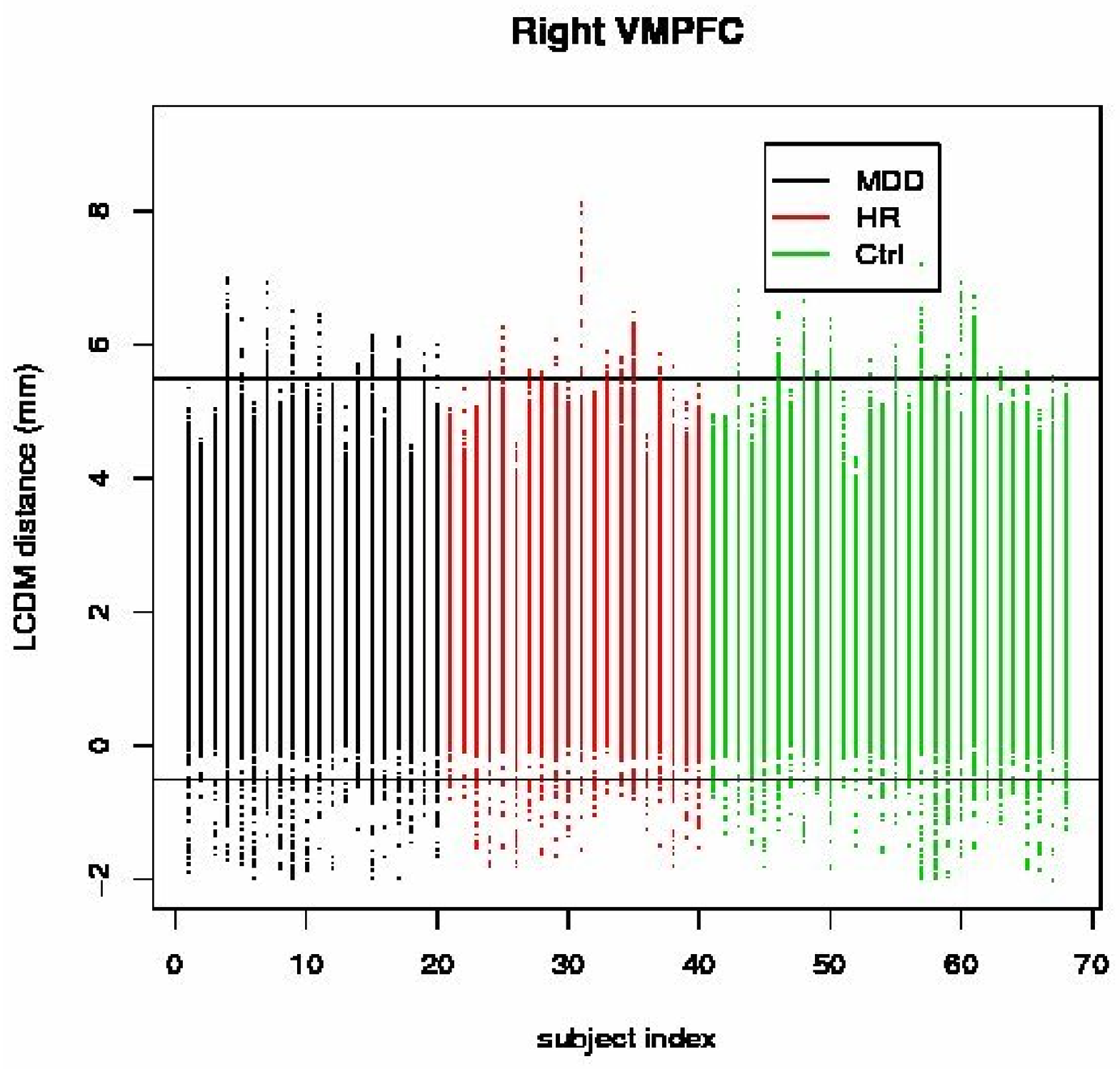} } }
\caption{
\label{fig:raw-dist-LR}
Depicted are the scatter plots of the LCDM distances for the left and right VMPFCs
by subject and color-coded for group.
The horizontal lines are located at -0.5 and 5.5 $mm$.
}
\end{figure}

\begin{figure}[htbp]
\centering
\rotatebox{-90}{ \resizebox{2.1 in}{!}{ \includegraphics{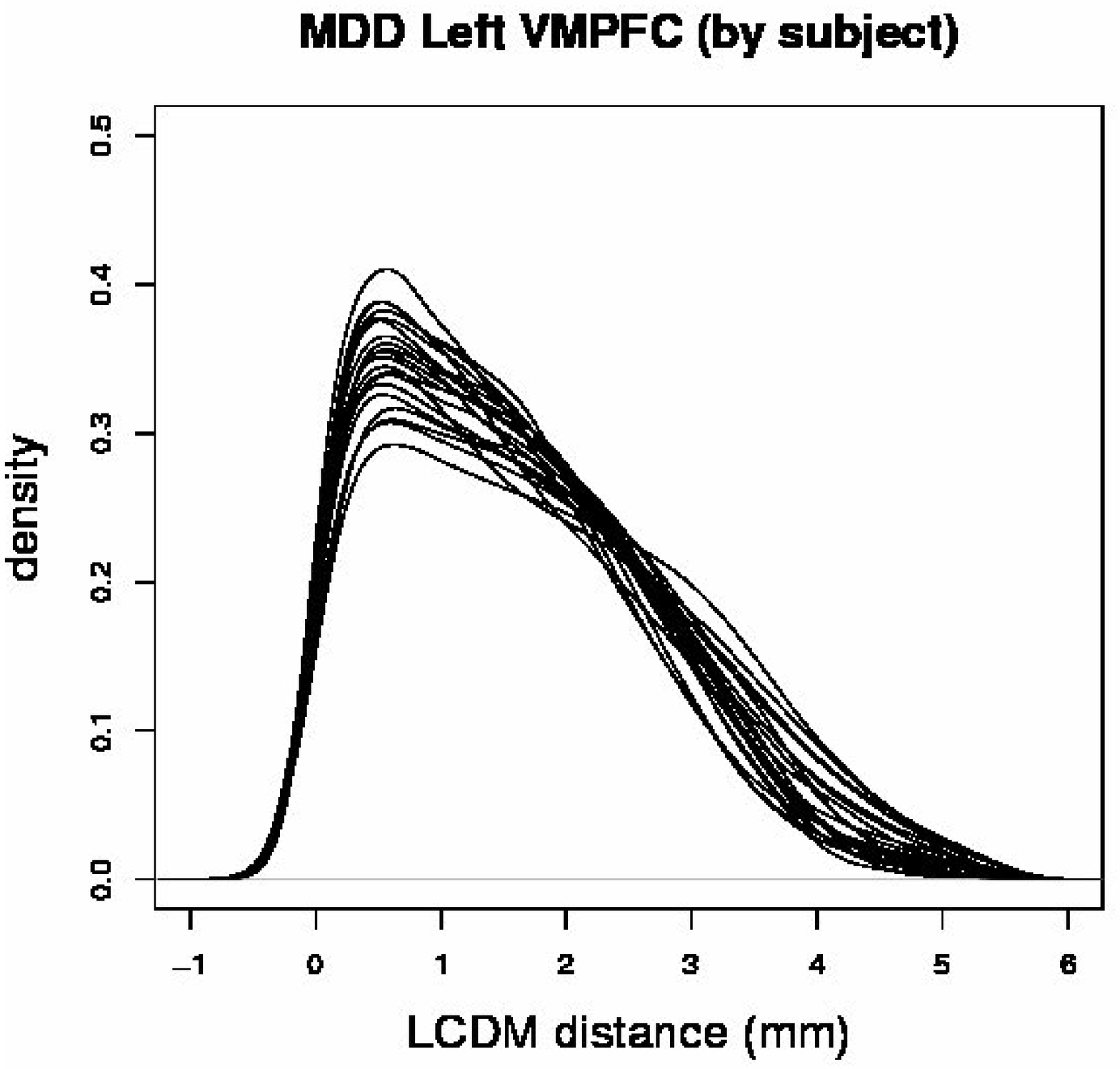} } }
\rotatebox{-90}{ \resizebox{2.1 in}{!}{ \includegraphics{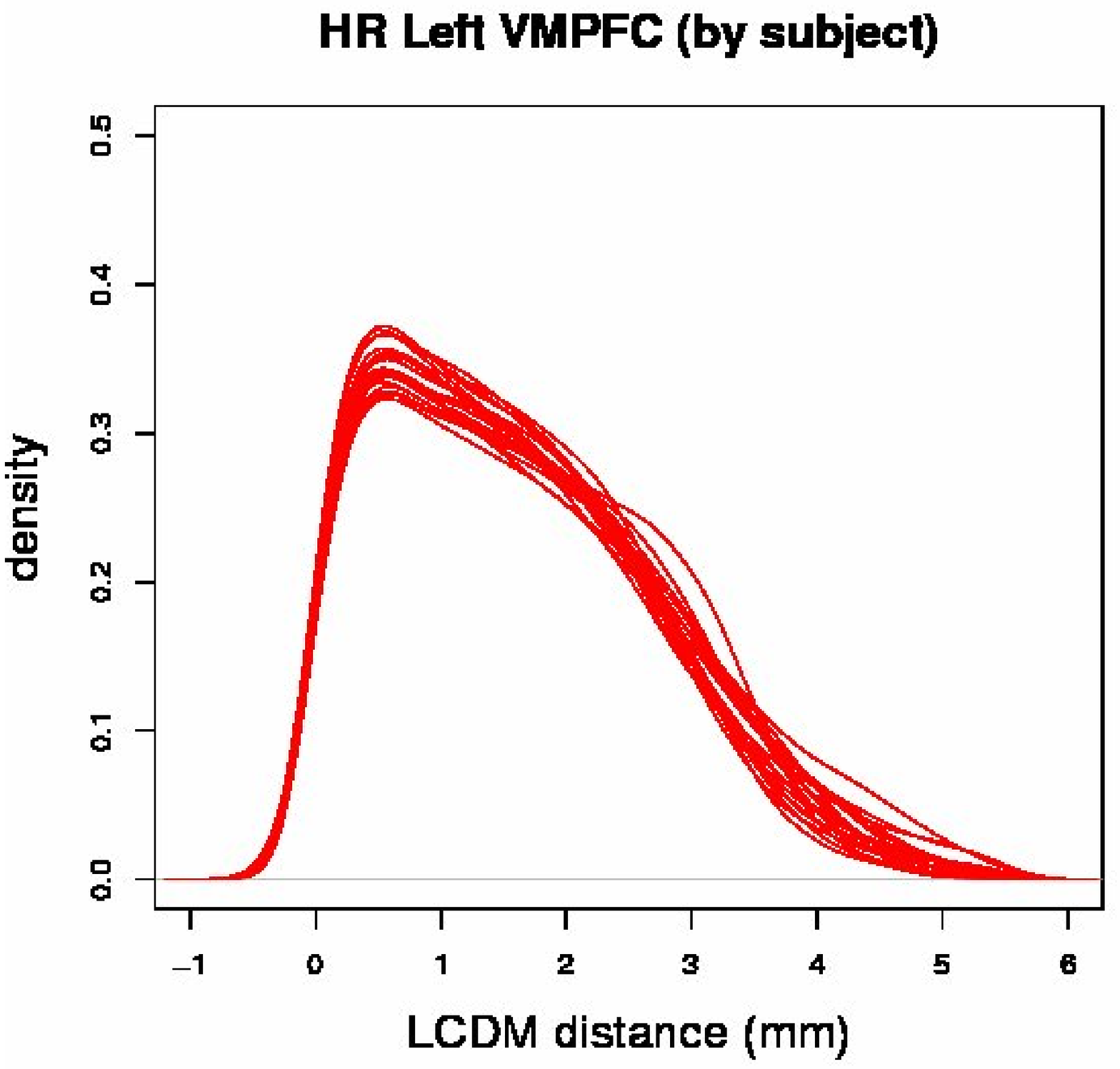} } }
\rotatebox{-90}{ \resizebox{2.1 in}{!}{ \includegraphics{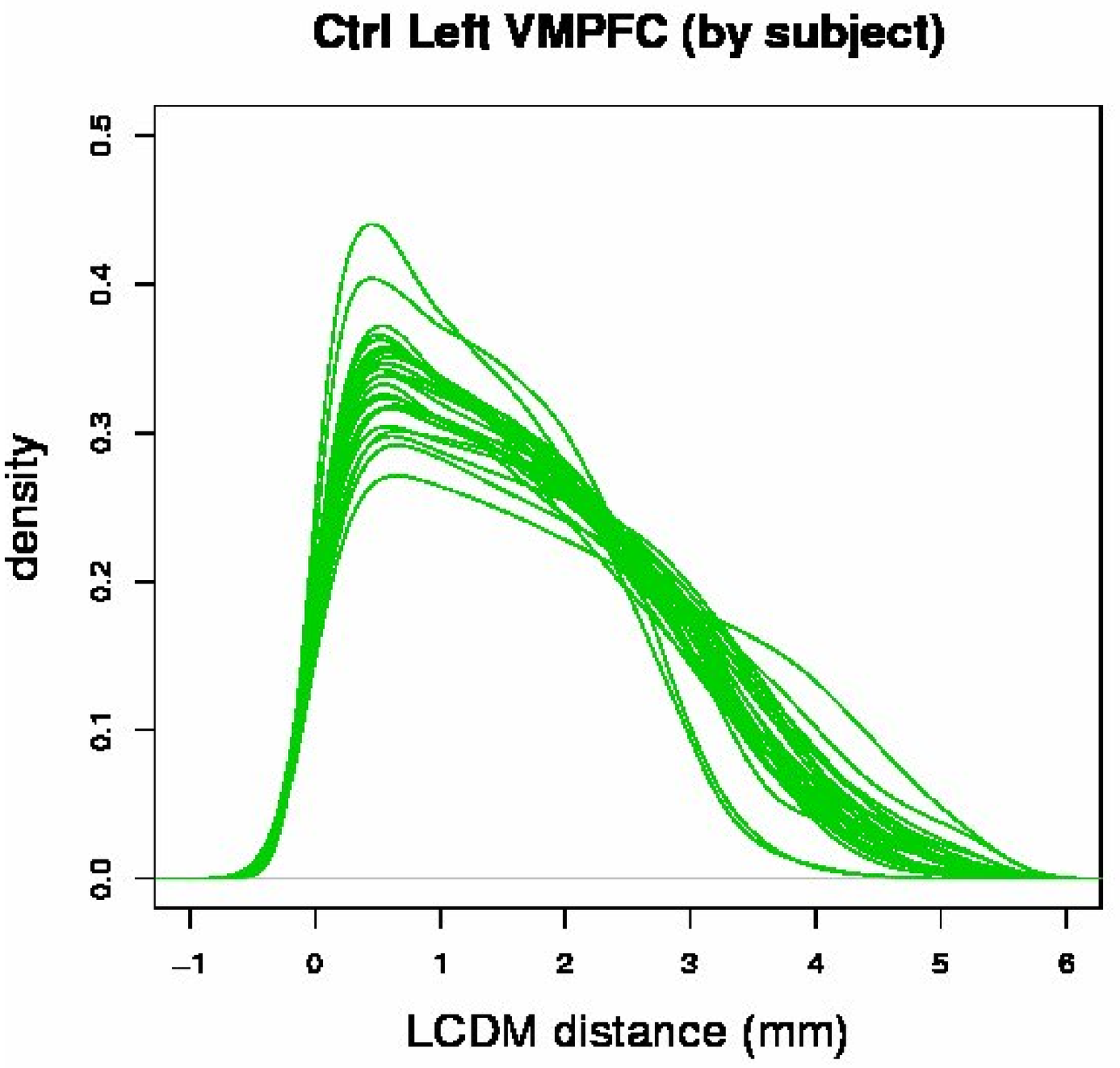} } }
\rotatebox{-90}{ \resizebox{2.1 in}{!}{ \includegraphics{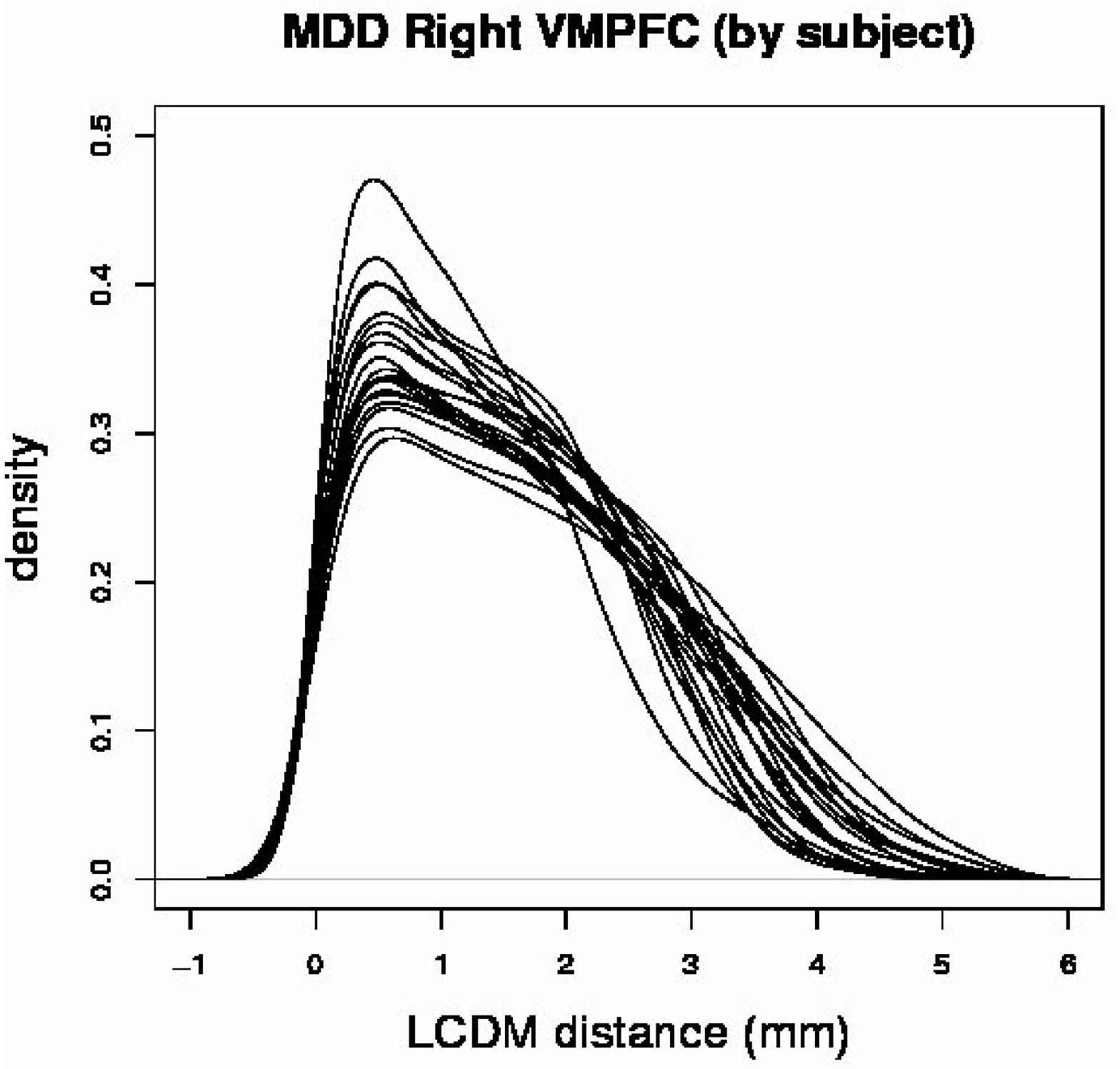} } }
\rotatebox{-90}{ \resizebox{2.1 in}{!}{ \includegraphics{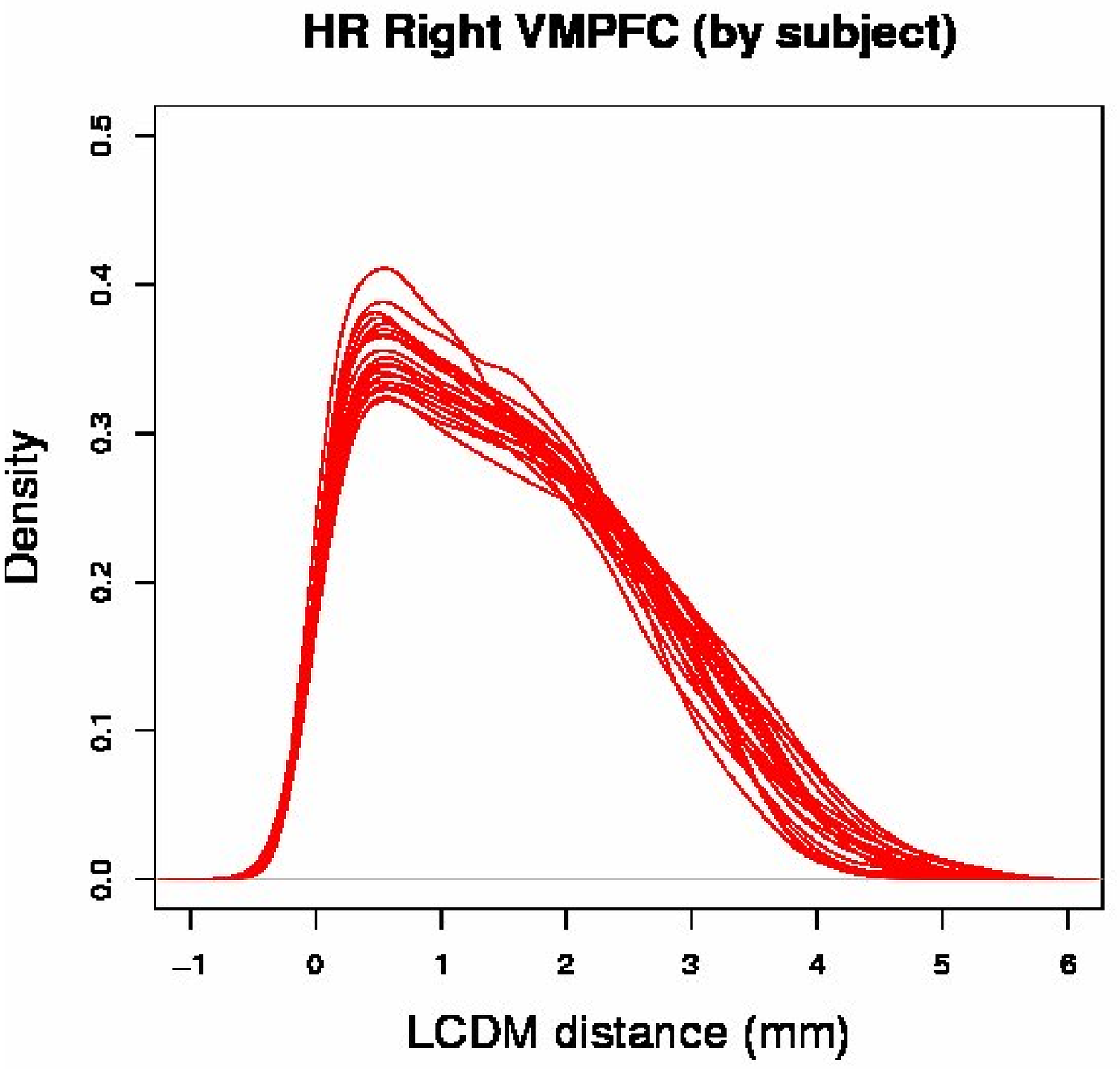} } }
\rotatebox{-90}{ \resizebox{2.1 in}{!}{ \includegraphics{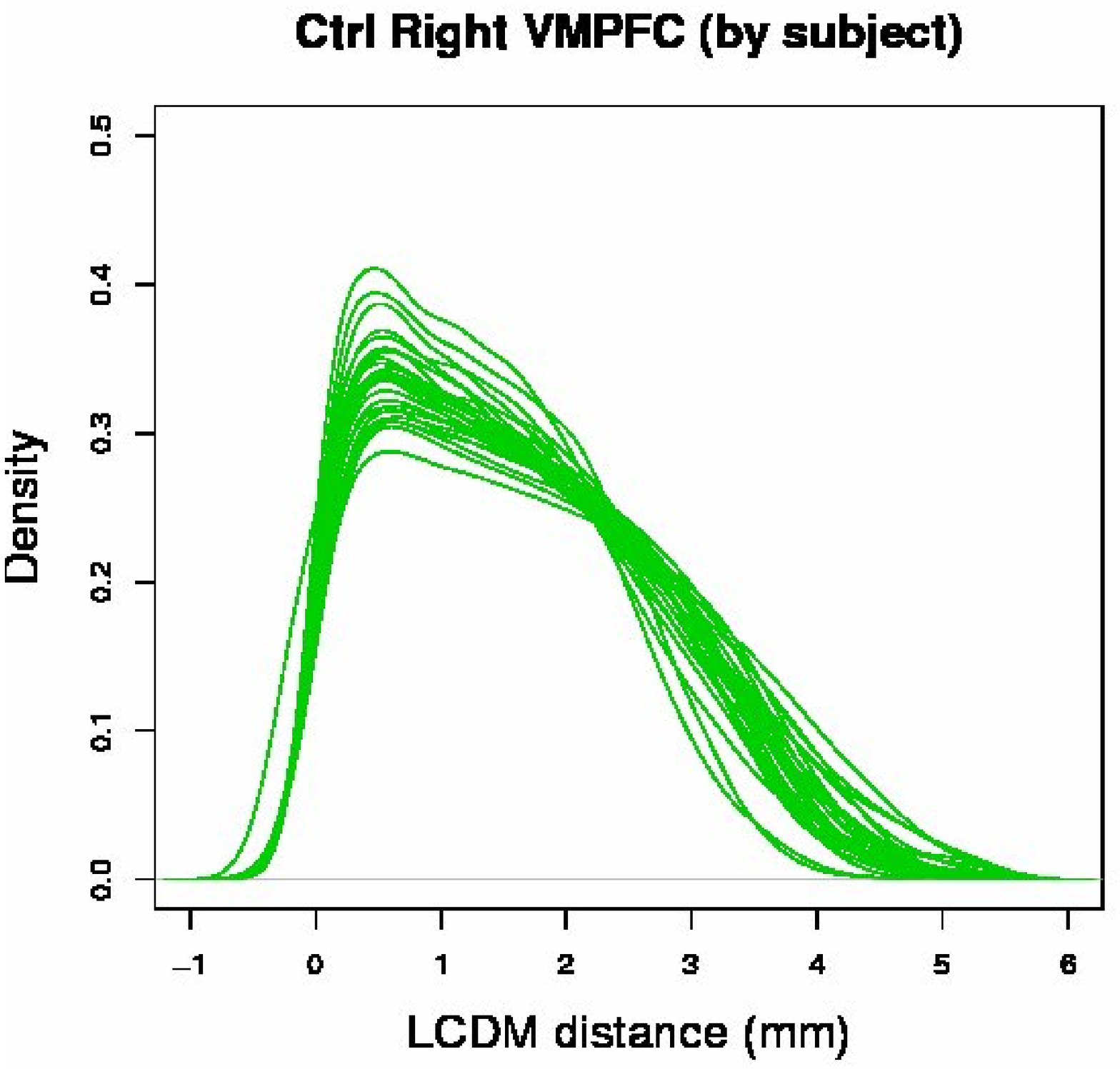} } }
\caption{
\label{fig:kernel-dens-LR}
Depicted are the plots of the kernel density estimates of the LCDM distances for the left and right VMPFCs
by subject.}
\end{figure}

When pooling the distances for subjects at a particular group, we implicitly
assume that the distances for subjects in the same group have identical
distributions. As a precautionary step, we find the \textit{extreme} (\textit{outlier}) subjects;
i.e., the subjects with VMPFCs having much different distributions than the rest.
In Figure \ref{fig:raw-dist-LR}, observe that in the left VMPFC distances, MDD subjects 8 and 11,
HR subjects 3 and 18, and Ctrl subjects 6, 10, 11, 12, and 22 seem to be
more different with Ctrl subjects 11 and 12 being ``thinner'' while the rest
are ``thicker'' than other left VMPFCs.
On the other hand, in the right
VMPFC distances, VMPFC of MDD subject 4, HR subject 5, and Ctrl subjects 12
and 21 seem to be more different, with VMPFCs of HR subject 5 and Ctrl
subject 12 being thinner and the others being thicker than the rest of the
right VMPFCs.
Note that Figure \ref{fig:raw-dist-LR} provides only a preliminary assessment of
reliability of LCDM distances, since it does not provide the distributional
behavior of the distances, but only points the problems with small
(negative) and large (positive) distances.
Hence the kernel density
estimates (or normalized histograms) can serve as a better exploratory tool
to detect outliers.
See Figure \ref{fig:kernel-dens-LR} for the kernel density estimates of LCDM
distances plotted by subject.
Notice that these kernel density estimates are
normalized for volume, as each density curve has the same unit area under it.
Recall also that Figure \ref{fig:raw-dist-LR} provides some insight on kurtosis (the
thickness of left and right tails) of the distributions of LCDM distances.
Using both Figures \ref{fig:raw-dist-LR} and \ref{fig:kernel-dens-LR}, we observe that LCDM distances for some subjects
have very different distributions than the others; i.e., they are outliers.
However, although Figure \ref{fig:raw-dist-LR} provides information about the tails
(i.e., small or large distances), Figure \ref{fig:kernel-dens-LR} is more reliable to detect the outliers as it
is normalized for volume and provides information for all distance values.
The VMPFC of outlier subjects are extremely different in shape from the
remaining subjects in each group.
Hence an outlier VMPFC in a group does not
represent an average VMPFC in that group and this discrepancy (extremeness)
might be due to some other factor affecting that subject only.
A careful investigation shows that, among the left VMPFCs, MDD subjects 1 and 9, HR
subjects 3 and 12, and Ctrl subjects 6, 10, 11, 12, and 19 are outliers,
while among the right VMPFCs, MDD subjects 7, 11, 13, 15, 18, 19, and 20, HR
subjects 4, 8, 14, and Ctrl subjects 4, 10, 11, 12, 25, and 26 seem to be the outliers.
Therefore, we remove these subjects in our pooled distance
data sets, but perform analysis on both the pooled distances with all
subjects included and the pooled distances without the outliers and remark
on how the outliers influence the comparisons. Notice that HR left subject
3, Ctrl left subjects 6, 10, 11, 12, and Ctrl right subject 12 are the subjects
deemed as outliers by both of Figures \ref{fig:raw-dist-LR} and \ref{fig:kernel-dens-LR}.
Observe also that LCDM
distances (more precisely the normalized histograms or kernel density
estimates) can be used as an exploratory tool to detect the outliers.

Here it may seem a little excessive to use the term ``outlier'' as the
number of subjects that are defined as outliers seem to be numerous.
Approximately 15{\%} of left-VMPFCs are defined as outliers, while over
20{\%} of right-VMPFCs are classified as outliers. This might seem too high
to treat these subjects as outliers, thereby suggesting that some form of
mixed distribution modeling may be more appropriate.
Although keeping the
outliers changes the results considerably compared to results from deleting
the outliers, for the methods based on pooling the distances, we recommend
removing the outliers, perhaps in a more conservative manner than ours,
because the basic premise of pooling is based on the similarity of the
distance distributions for VMPFCs of the same group.
The issue of modeling
the distances with mixed distributions is a topic of ongoing research.

\begin{figure}[htbp]
\centering
\rotatebox{-90}{ \resizebox{3. in}{!}{ \includegraphics{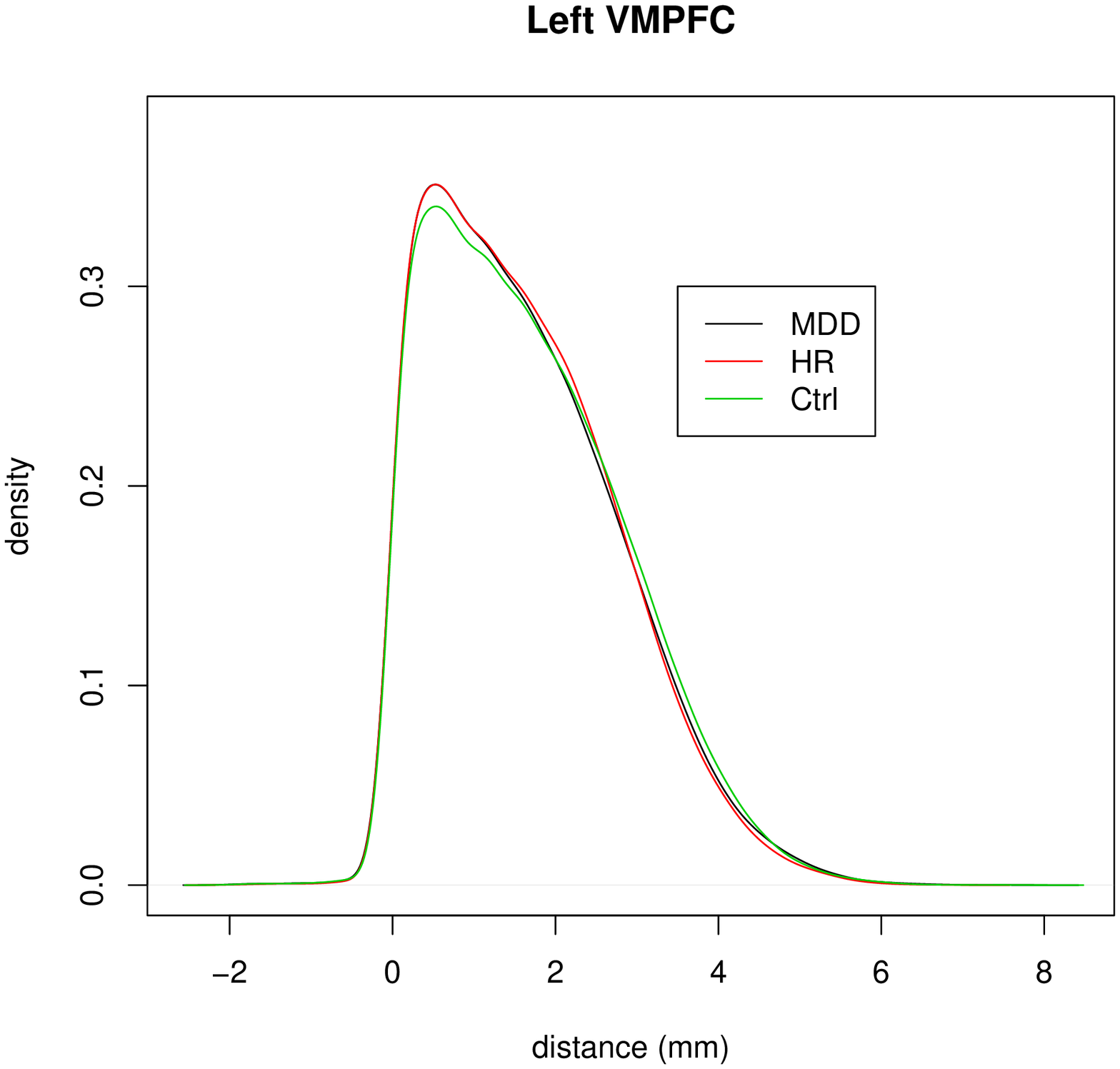} } }
\rotatebox{-90}{ \resizebox{3. in}{!}{ \includegraphics{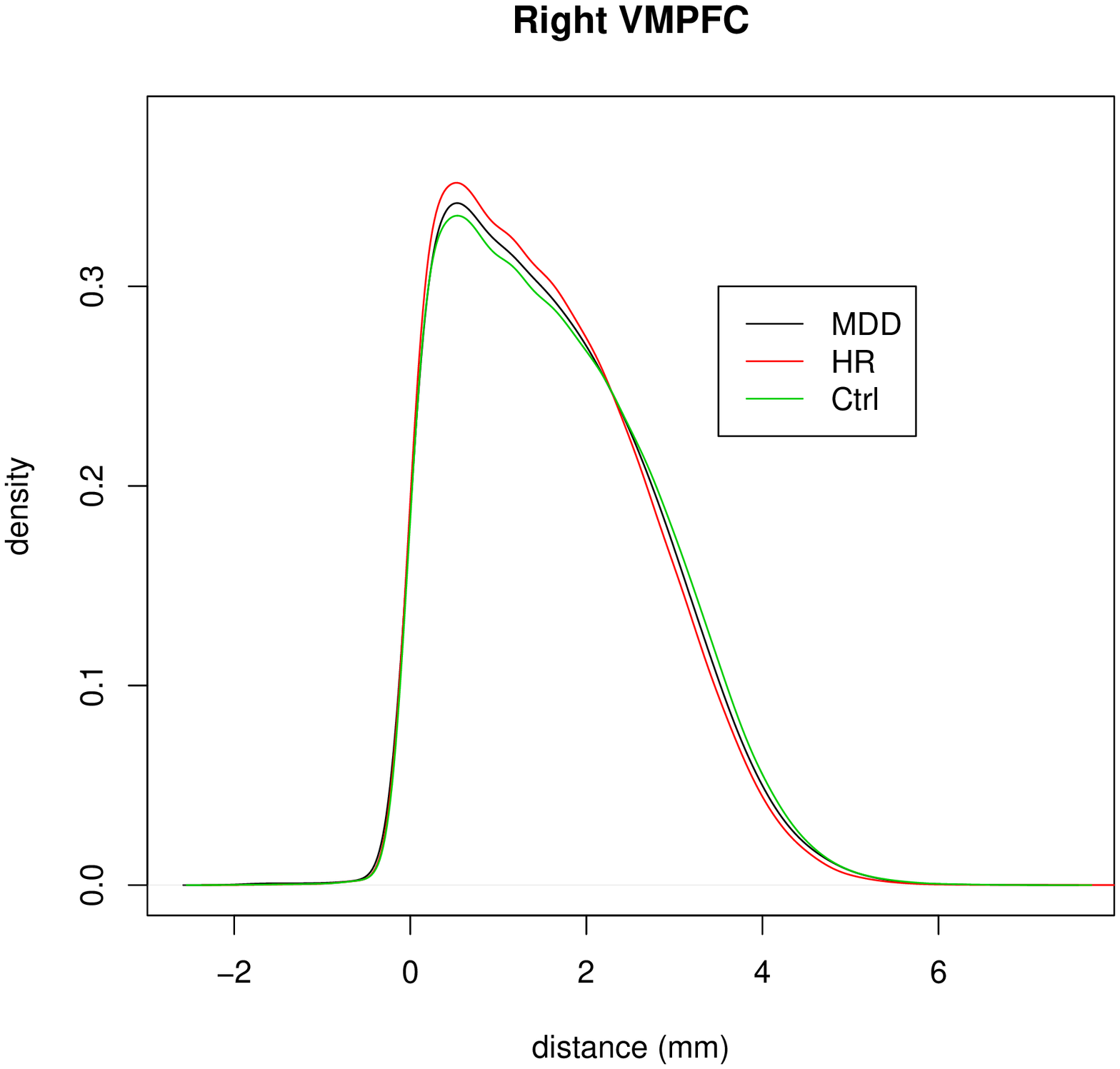} } }
\caption{
\label{fig:kernel-dens-pool}
Depicted are the plots of the kernel density estimates of the pooled LCDM distances by group
when extreme subjects are removed for the left and right VMPFC.}
\end{figure}

See Figure \ref{fig:kernel-dens-pool} for the kernel density estimates of pooled LCDM distances when
the outliers are removed.
See also Table \ref{tab:stat-pooled-LCDM} for the corresponding sample
sizes, means, and standard deviations of the pooled LCDM distances, overall
and for each group.
Observe that the density profiles of LCDM distances for
the left VMPFC of MDD and HR subjects seem to be very similar while both are
different from that of Ctrl subjects. On the other hand, there seems to be
more separation between the density profiles in right VMPFCs.
After removing
the extreme subjects, the sample sizes of LCDM distances have decreased
while the medians and standard deviations (hence the variances) for all left
and right groups have increased.
Furthermore, mean LCDM distances for left
VMPFC got smaller, while for right VMPFC the mean distances have increased.
Furthermore, the order of mean distances for left and right VMPFCs do not
change with all the subjects included and when outliers removed.
For left VMPFCs, the order of mean and standard deviations are HR $<$ MDD $<$ Ctrl
(more accurate notation would be \textit{mean}$\left( {D_2^L} \right)<$\textit{mean}$\left( {D_1^L
} \right)<$\textit{mean}$\left( {D_3^L} \right)$, which we shorten for convenience),
while the order of medians is MDD $<$ HR $<$ Ctrl. The change in the order
of means and medians is due to the levels of right skewness of the distributions.
For right VMPFCs, the order of means, medians, and standard
deviations are HR $<$ MDD $<$ Ctrl.
Thus, we observe that the outliers in
the right VMPFC, although influence the means, medians, and variances, do
not change their order.

\subsection{Results}
\label{sec:results}

\subsubsection{The Equality of the Distributions of Pooled LCDM Distances}
\label{sec:equality-distribution-LCDM}
First we address the differences in the distributions in location but not in spread.
The differences in the distributions in the location (e.g., means or
medians) of LCDM distances imply shape differences.
Hence, we test the equality of the distributions of the left (pooled) distances between groups;
i.e.,
\[
H_o :F_1^L=F_2^L=F_3^L
\]
where $F_i^L$ is the distribution function of the pooled distances for group
$i = 1,2,3$. Likewise for right pooled distances.

The left and right pooled distances for each group are significantly
non-normal with $p_L<.0001$ for each test where $p_L$ is the $p$-value for
Lilliefor's test of normality (see, e.g., [43]), possibly due to heavy
right skewness of the densities.
Moreover, HOV is rejected with $p_{BF}<.0001$ for both left and right pooled distances where $p_{BF}$ is the
$p$-value for B-F test.
Hence non-parametric tests of group comparisons are more
appropriate for this data.
Note that the above hypothesis of equality of the
distributions of the pooled distances can be attributable to the similarity
in the VMPFC shapes for all groups, but not vice versa (i.e., the equality
of the distributions does not necessarily imply morphometric similarity, but
similarity in the distance structure of GM tissue with respect to the GM/WM surface.)
Notice that LCDM distances analyzed in this fashion provide
morphometric information, on cortical mantle thickness and shape, but the
width (the length of VMPFC parallel to the GM/WM surface) is less relevant.
Because the comparison is done on the ranking of distances with respect to
the GM/WM surface.
For example, suppose two VMPFC tissues are composed of
100 and 1000 voxels of similar distances and then the test will detect no
difference, although the morphometry is obviously different.
Hence, as long as the voxels are at a similar distance from the GM/WM surface, their
abundance will not be influencing the test results.

The equality of the distributions of the distances of left VMPFCs is
rejected with $p_{KW}<.0001$ where $p_{KW}$is the $p$-value for K-W test, and
likewise for right VMPFC distances ($p_{KW}<.0001)$.
Without removing the extreme subjects (i.e., when all subjects are included), we have the same
conclusions for right and left VMPFCs with $p_{KW}<.0001$ for both.
Hence, we perform pairwise comparisons by Wilcoxon rank sum test for left (and
right) VMPFC distances, using Holm's correction for multiple comparisons.
The $p$-values adjusted by Holm's correction method for the simultaneous
pairwise comparisons for left and right VMPFC distances are presented in
Table \ref{tab:pairwise-pooled-LCDM}.
Observe that, with all subjects included and when the outliers are
removed, MDD-left and HR-left distances are not significantly different
($p_W =.3022$ for former, $p_W =.0776$ for latter where $p_W$ is the
$p$-value for Wilcoxon rank sum test), while both tend to be significantly less
than Ctrl-left distances ($p_W<.0001$ for all).
Hence, the VMPFC left distances tend to decrease due to the depressive disorders,
possibly due to a thinning in left VMPFCs.
In right VMPFC distances, with all the subjects included,
we observe that MDD and HR-right distances are not significantly
different from each other, while both tend to be significantly smaller than
Ctrl-right distances ($p_W<.0001$ for both).
When outliers are removed,
we observe that MDD-right distances tend to be significantly smaller than
HR-right distances ($p_W =.0084$) which tend to be significantly smaller
than Ctrl-right distances ($p_W<.0001$ for both).
Observe that outliers
(although do not change the order of mean pooled distances) do influence the
results, in particular for MDD and HR groups.
Looking at the kernel density
estimates in Figure \ref{fig:kernel-dens-LR}, we see that the outliers in the HR group are more
similar to MDD group, hence making the MDD and HR distance distributions
more similar than chance.
Recall that we were not able to detect these
differences by using volume, or simple descriptive measures based on LCDM [33].
Thus, the pooled LCDM distances provide comparisons that
are more powerful to detect group differences.

Since the densities of the distances are skewed right, these differences do
not reflect the order in the mean distances, but rather the order in the median distances.
Furthermore, in these analysis we ignore the influence of
possible dependence between twin pairs due to genetic similarity.

\subsubsection{Homogeneity of the Variances (HOV) of Pooled LCDM Distances}
\label{sec:HOV-LCDM}
Observe that K-W and Wilcoxon tests suggest shape differences when rejected,
in particular the direction of the alternatives might indicate cortical thinning.
Similarity of the morphometry of VMPFCs will cause similarity of
LCDM distances, which in turn implies similarity of the variances of LCDM distances.
Variance of distances is suggestive of morphometric variation in VMPFCs.
So similar shapes and sizes imply similar variances, but not vice versa.
For example, cortical thinning will reduce the variation in LCDM distances;
and the larger the spread in the boundary (surface) of VMPFC,
the larger the variance of LCDM distances.
Hence, we test the equality of the
variances of the left (pooled) distances between groups; i.e.,
\[
H_o :\Var\left( {D_{1\ell}^L} \right)=\Var\left( {D_{2\ell}^L} \right)=\Var\left( {D_{3\ell}^L} \right)
\]
where $D_{il}^L$ is the variance of the pooled distances for group $i = 1,2,3$.
Likewise for right distances.
With all the subjects included and when
extreme subjects are removed, HOV is rejected with $p_{BF}<.0001$ for both
left and right VMPFC.
See Table \ref{tab:pairwise-HOV-LCDM} for the corresponding $p$-values for
simultaneous pairwise comparisons adjusted by Holm's correction method.
With all the subjects included and when the extreme subjects are removed, the
order of the variances is HR $<$ MDD $<$ Ctrl for both left and right VMPFCs
with $p_{BF}<.0001$ for all six possible comparisons.
This implies that the morphometric variation reduces in left and right VMPFCs due to suffering
from or being at high risk for depressive disorders compared to Ctrl subjects
and is smallest for the HR subjects for both left and right VMPFCs.
Observe that both Wilcoxon rank sum tests (for location) and B-F tests
(for variances) yield significant results with the same ordering between groups
(HR $<$ MDD $<$ Ctrl), which might be due to cortical thinning among other
factors.

\subsubsection{Morphometric Left-Right Asymmetry with Pooled Distances}
\label{sec:LR-asymmetry-LCDM}
LCDM distances can also be used to detect left-right morphometric asymmetry,
which might be due to shape or size asymmetry between left and right VMPFCs.
If the left and right VMPFCs are similar, then the distributions of the left
and right VMPFCs will be similar, but not vice versa.
But, if the distributions of the left and right VMPFCs are different,
then there is evidence for morphometric left-right asymmetry, which can also be detected
by the use of LCDM distances (with Wilcoxon rank sum test).
In terms of size asymmetry, LCDM emphasizes mantle thickness asymmetry,
rather than the mantle width asymmetry.

Hence we test
\[
H_o :F_i^L=F_i^R
\]
for each $i = 1,2,3$.
See Table \ref{tab:pairwise-LR-asymmetry-LCDM}
for the associated $p$-values, which are adjusted
by Holm's correction method.
Observe that when all the subjects are
included, left distances are significantly larger than right distances,
overall, and by group with $p_W<.0001$ for each test.
When extreme subjects (outliers) are removed, Ctrl and MDD VMPFC distances yield
significant left-right asymmetry, with left distances being significantly
smaller than right distances ($p_W<.0001$ for both); and HR-left
distances are significantly larger than HR-right distances ($p_W =.0015$).
Hence, we conclude that there is significant left-right asymmetry in LCDM distances.
However, the direction of left-right asymmetry is different for
MDD and HR subjects, while it is same for MDD and Ctrl subjects.
This suggests that cortical mantle in left VMPFC is thinner for MDD and Ctrl subjects and
thicker for HR subjects compared to their right counterparts.
Notice that the inclusion of outliers (i.e., when all subjects are included)
influences MDD and Ctrl groups to the extent that the direction of the asymmetry is reversed.

\subsubsection{Stochastic Ordering of Pooled LCDM Distances}
\label{sec:stochastic-order-LCDM}
\begin{figure}[htbp]
\centering
\rotatebox{-90}{ \resizebox{2.3 in}{!}{ \includegraphics{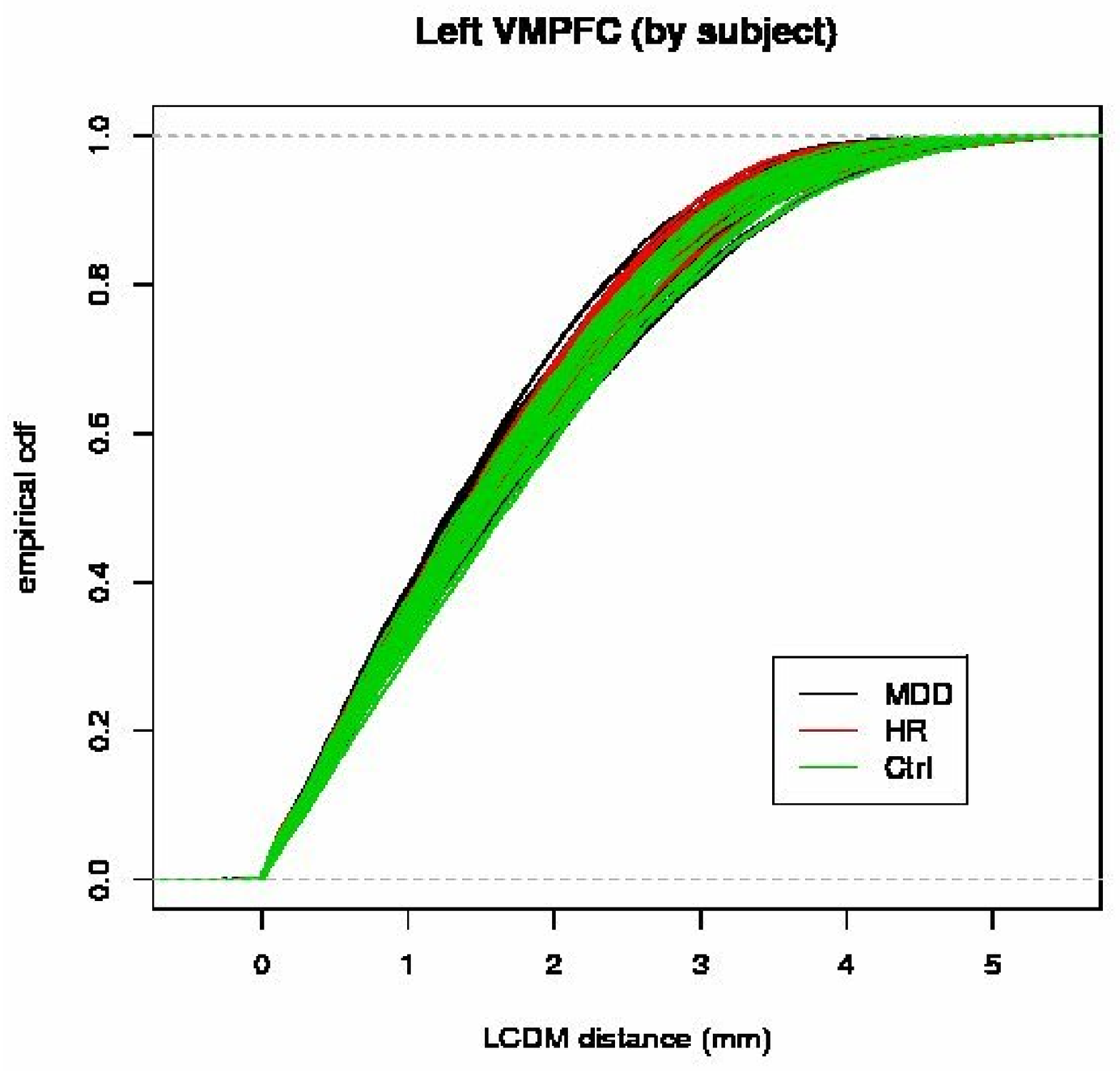} } }
\rotatebox{-90}{ \resizebox{2.3 in}{!}{ \includegraphics{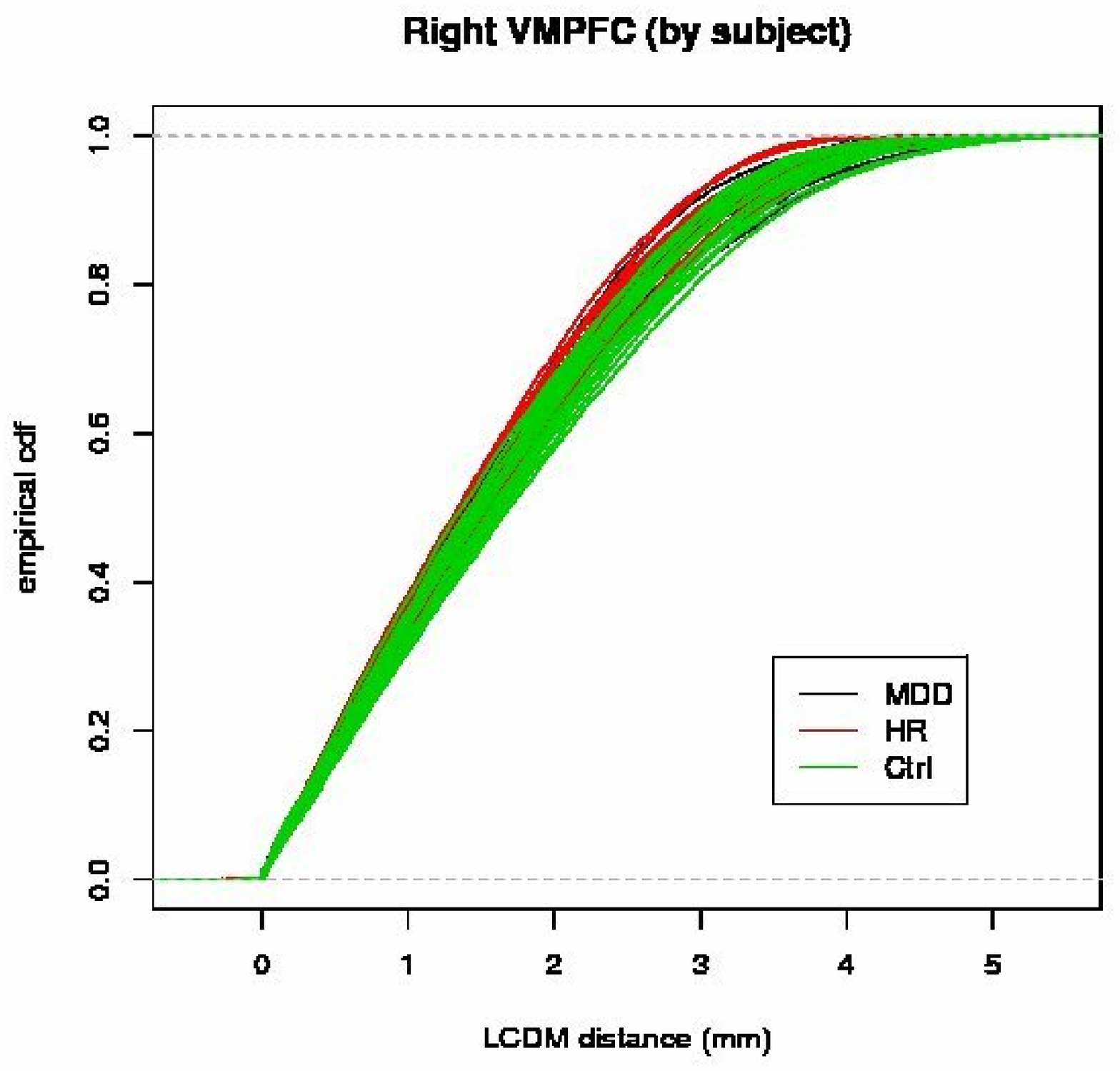} } }
\caption{
\label{fig:emp-cdf-by-subject}
Depicted are the plots of the
empirical cdfs of the LCDM distances of left and right VMPFCs by subject when extreme
subjects are removed for the left and right VMPFC (color-coded for group) .}
\end{figure}

\begin{figure}[htbp]
\centering
\rotatebox{-90}{ \resizebox{2.3 in}{!}{ \includegraphics{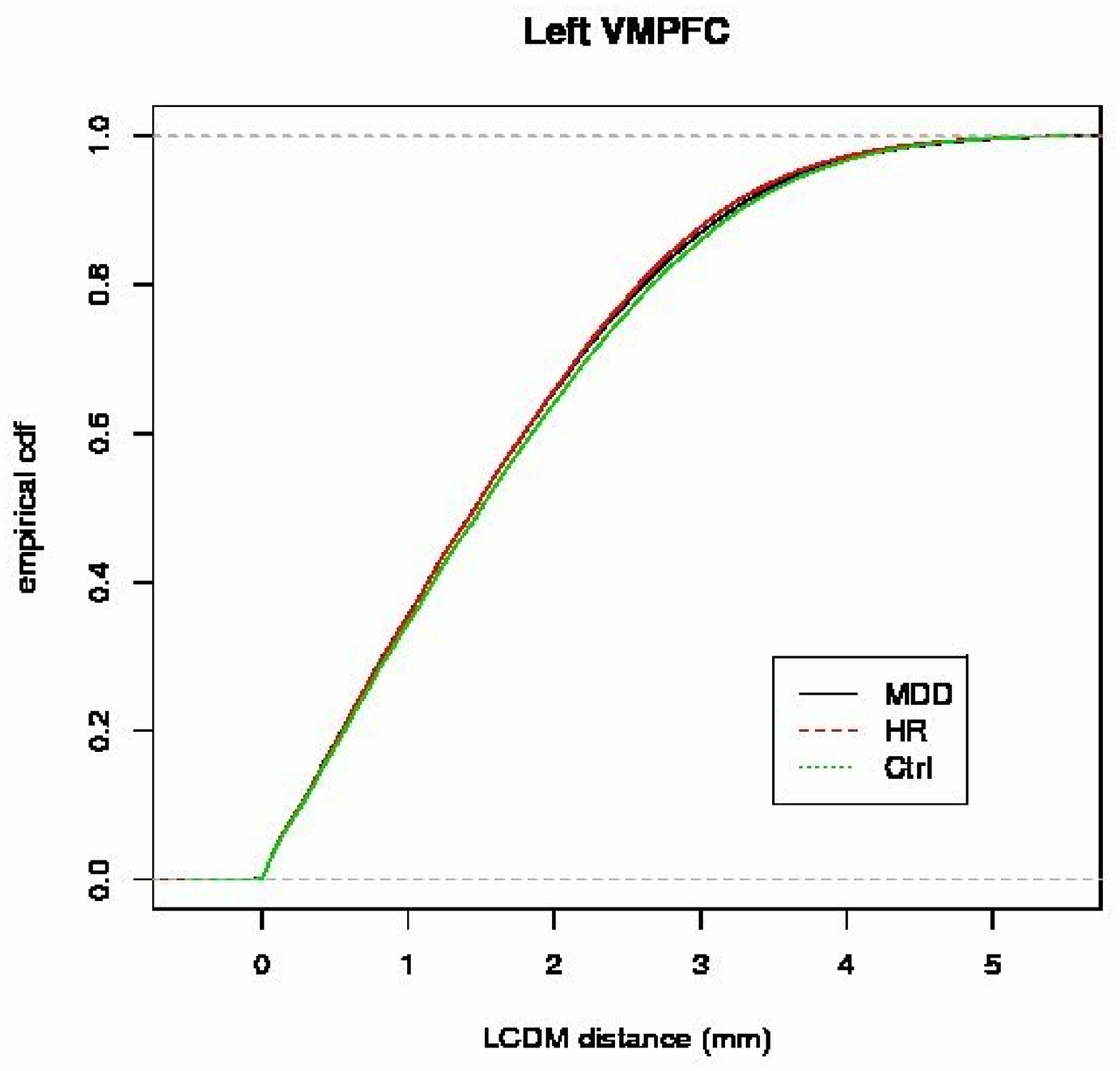} } }
\rotatebox{-90}{ \resizebox{2.3 in}{!}{ \includegraphics{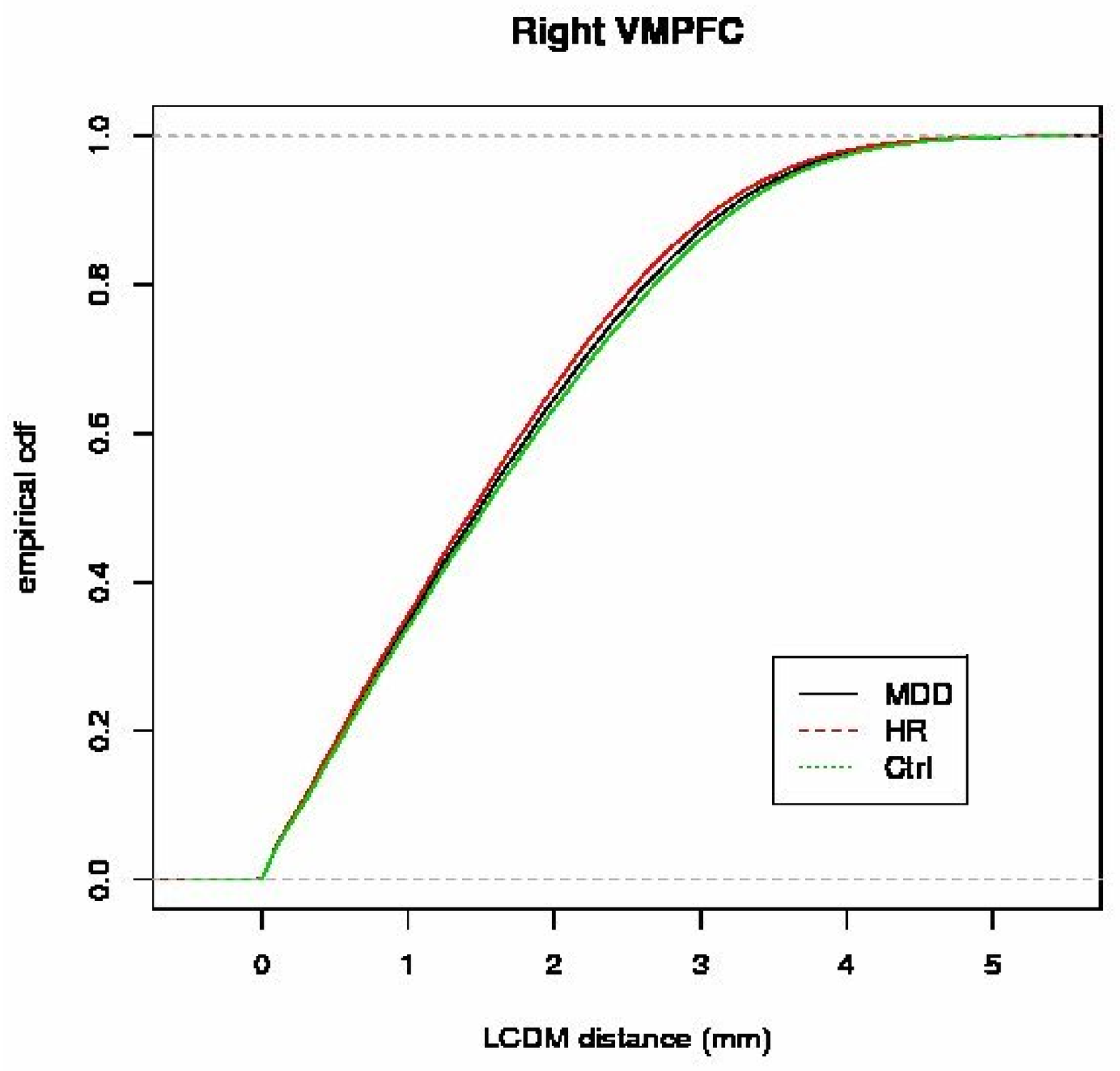} } }
\caption{
\label{fig:emp-cdf-pool}
Depicted are the plots of the empirical cdfs of the pooled LCDM distances
when extreme subjects are removed for the left and right VMPFCs.}
\end{figure}

Recall that we used Wilcoxon tests to test the null hypothesis of equality
of the distributions, i.e., $H_o :F_X=F_Y$ where $F_X $ and $F_Y$ are the
distributions of variables $X$ and $Y$, respectively.
For one-sided alternatives,
the $p$-values based on Wilcoxon test are complementary (i.e., the $p$-value for
``$<$'' and ``$>$'' alternatives add up to 1).
Hence $p$-value will be
significant for only one type of directional alternative.
Furthermore, when rejected,
Wilcoxon test implies an ordering in location parameter such as mean or median.
Stochastic ordering, if present, can be deduced from the
direction of the alternative, together with cdf plots.
See Figure \ref{fig:emp-cdf-by-subject} for the cdf plots of the
LCDM distances for each subject and
Figure \ref{fig:emp-cdf-pool} for the cdf plots of the pooled distances.
Observe that the cdf
plots for the pooled distances are not suggestive of the stochastic
ordering with the current resolution.
We can also use Kolmogorov-Smirnov (K-S) tests for $H_o :F_X=F_Y$.
Unlike Wilcoxon tests, K-S test yields
$p$-values that are not complementary for the one-sided alternatives
(i.e., they don't add up to 1).
Hence, $p$-value can be significant for both or none of the
directional alternatives.
This results from the fact that the order of the
cdfs $F_X $ and $F_Y$ can be different at different distance values on the
horizontal axis.
Moreover, if $p$-value based on K-S test is significant for
only one-sided alternative, then we can deduce stochastic ordering.
The $p$-values being insignificant or significant for both one-sided alternatives
imply lack of stochastic ordering.
But, first case implies that equality of
the distributions is retained, while the latter implies significant differences in the distributions.
So, we also apply K-S test on our data set to compare the
cumulative distribution functions of the distances by group, which might
also provide a stochastic ordering.
The null hypotheses are
\[
H_o :F_i^L=F_j^L
\]
for each ($i,j) \, \in {\{}(1,2),(1,3),(2,3){\}}$.
See Table \ref{tab:pairwise-cdf-LCDM} for the associated
$p$-values where tests for each alternative are adjusted by Holm's correction
method. Observe that, with all the subjects included, the cdf of MDD-left
distances is significantly smaller than those of Ctrl and HR-left distances.
Furthermore, the cdfs of MDD and HR-left distances are significantly
different from each other, with both sides being significant, which suggests
that the order of cdf comparisons change at different distance values.
When the extreme subjects are removed, the cdf of Ctrl-left VMPFC distances is
significantly larger than those of MDD-left and HR-left distances, and the
cdf of MDD-left distances is significantly smaller than that of HR-left distances.
Thus, we conclude that the stochastic ordering of left distances
is HR $<^{ST}$MDD $<^{ST}$Ctrl; i.e., it is more likely for HR-left
distances to be smaller compared to MDD-left and Ctrl-left distances, and
more likely for MDD-left distances to be smaller than Ctrl-left distances.
In other words, it is more likely for left VMPFCs of HR subjects to be
thinner than those of MDD subjects, which are more likely to be thinner than
those of Ctrl subjects.

With all subjects included, the cdf of MDD-right distances is significantly
smaller than HR-right distances.
But K-S test yields significant result for
both types of one-sided alternative for MDD,Ctrl and HR,Ctrl pairs.
This implies, for example, the cdfs of MDD and Ctrl-right distances are
different, hence no stochastic ordering between them.
Furthermore, the differences between the cdfs of the groups change over the distance values;
that is, for small distance values, the order is Ctrl$<$MDD$<$HR, while for
large distance values the order is HR$<$MDD$<$Ctrl.
When extreme subjects
are removed, the cdfs have the following order: Ctrl$<$MDD$<$HR.
This implies the stochastic ordering as HR $<^{ST}$MDD $<^{ST}$ Ctrl; i.e.,
it is more likely for HR-right distances to be smaller compared to MDD and
Ctrl-right distances, and more likely for MDD-right distances to be smaller
than Ctrl-right distances.
That is, it is more likely for right VMPFCs of HR
subjects to be thinner than those of MDD subjects, which are more likely to
be thinner than those of Ctrl subjects.
Thus, applying K-S test on LCDM
distances may provide the stochastic ordering of or LCDM distances or lack of it.

\section{The Influence of Assumption Violations on the Tests: A Monte Carlo Analysis}
\label{sec:assumption-violations}

\subsection{The Underlying Assumptions for the Tests}
\label{sec:assumptions}
In our analysis, we have used various (parametric and nonparametric) tests,
without addressing the validity of underlying assumptions.
Let $\X_i=\{X_{i1},\ldots,X_{i,n_i}\}$ be $m$ samples each of size $n_i$ from their respective populations.
Then, the assumptions for the \emph{K-W test} for distributional equality of
several independent samples are as follows [40]:
\begin{itemize}
\item[1.]
All samples, $\X_i$, are random sets from their respective populations; i.e.,
there is independence within each sample.
\item[2.]
There is mutual independence among various samples.
For example $\X_i$ and $\X_j$ are independent for all possible
combinations of $(i,j)$.
\item[3.]
The measurement scale is at least ordinal.
\item[4.]
Under the null hypothesis, the population distributions are identical.
That is, $X_{ij} \sim F$ for all $i=1,\ldots,m$.
\end{itemize}
The assumptions for \emph{Wilcoxon rank sum test} are same; only we have two samples.
For \emph{K-S test} for cdf comparisons,
the first three assumptions are same, but assumption 4 is:
\begin{itemize}
\item[4.]
For K-S test to be exact, the random variables are assumed to be continuous.
\end{itemize}
For discrete random variables, K-S test is still valid,
but becomes a little conservative [40].

\emph{B-F test} for HOV is the regular one-way ANOVA test
on the absolute deviations from sample medians.
That is, the usual ANOVA test is applied to samples
$\X_i^{med}=\{|X_i-med(\X_i)|\}$ for $i=1,\ldots,m$.
Hence the assumptions for B-F test are the assumptions for ANOVA F-test
on the absolute deviations from medians.
Therefore the assumptions for B-F test are:
\begin{itemize}
\item[1.]
All samples of absolute deviations from medians, $\X_i^{med}$, are random from their respective populations; i.e.,
there is independence within each sample of deviations.
\item[2.]
There is mutual independence among various samples of deviations.
For example $\X_i^{med}$ and $\X_j^{med}$ are independent for all possible
combinations of $(i,j)$.
\item[3.]
The measurement scale is at least interval and the deviations are normally distributed.
\item[4.]
The homogeneity of the variances of the deviations;
i.e., the variances of the deviations for each sample are identical.
\end{itemize}
[46] have shown that B-F test gives quite accurate error rates even when assumption 3 is violated.
However, the robustness of B-F test against assumption 4 is not clear,
since this test is for HOV and it depends on the HOV of the absolute deviations [47].

The assumptions for \emph{Wilcoxon signed rank test} are as follows [40]:
\begin{itemize}
\item[1.]
The distribution of each paired difference is symmetric.
\item[2.]
The paired differences are mutually independent.
\item[3.]
The measurement scale is at least interval.
\item[4.]
The paired differences have the same mean, which is usually zero.
\end{itemize}

Note that these assumptions are reasonably valid for the
morphometric measures like volume, surface area,
median, mode, range, and variance of the LCDM distances.
Hence, we can safely use the above tests for these measures,
except for the possible dependence between MDD and HR twins.
However, pooled LCDM distances have spatial dependence (or correlation)
hence independence within each sample does not hold,
although the other assumptions for K-W, Wilcoxon rank sum, B-F,
and K-S tests are reasonably valid.
In the next section,
we investigate the influence of assumption violations on the results by
Monte Carlo simulations.

\subsection{Simulation of Data that Resemble LCDM Distances}
\label{sec:simulation-LCDM}
\begin{figure}[htbp]
\centering
\rotatebox{-90}{ \resizebox{2.3 in}{!}{ \includegraphics{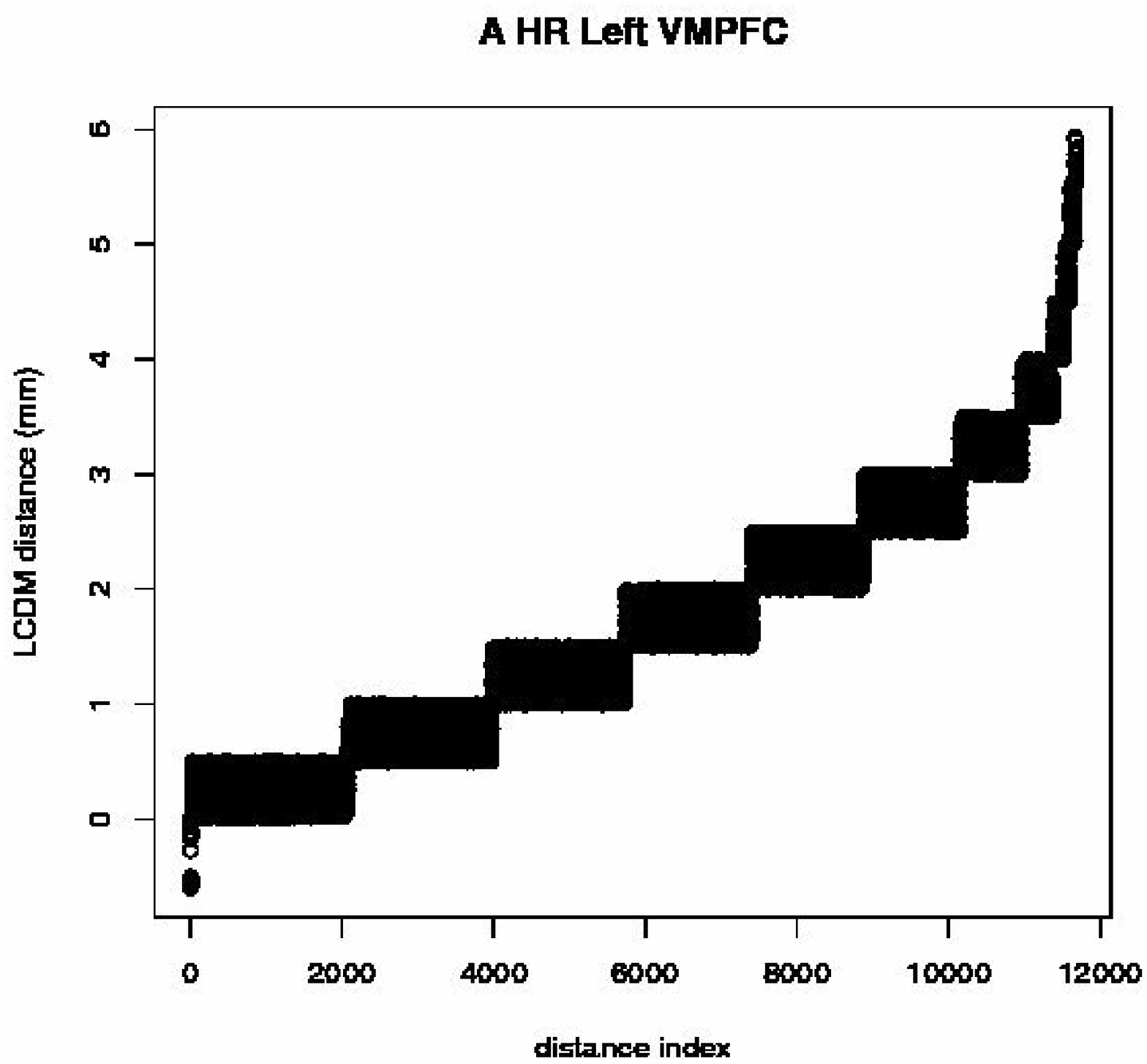} } }
\rotatebox{-90}{ \resizebox{2.3 in}{!}{ \includegraphics{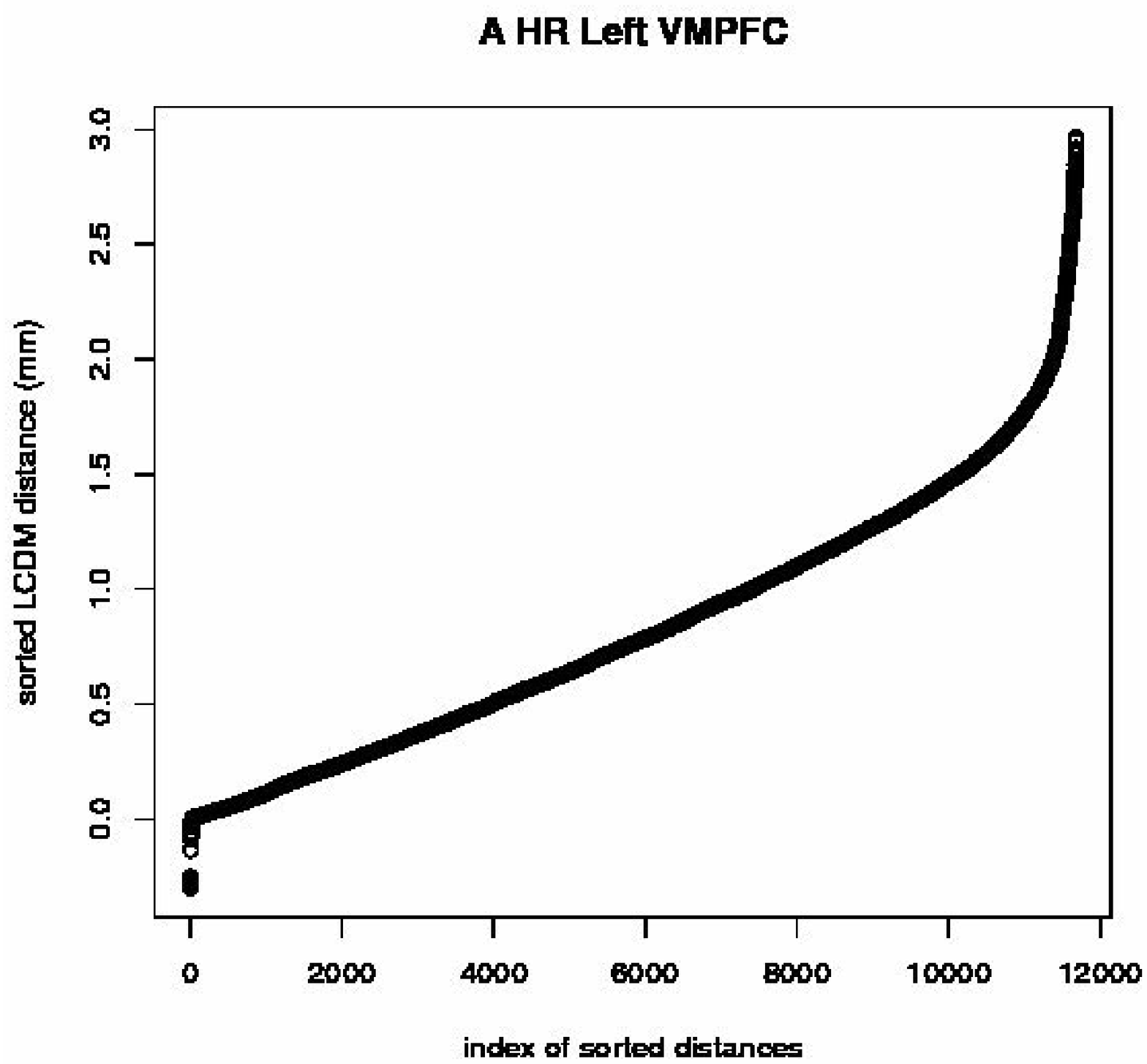} } }
\caption{
\label{fig:HR1-dist-L}
Depicted are the plots of the LCDM distances for the left VMPFC of HR subject 1.
The left plot is the distances stacked for intervals of size $0.5\, mm$
and the right plot is for the sorted distances.
}
\end{figure}

In this section, we investigate the influence of the assumption violations
due to the spatial correlation and non-normality inherent in the LCDM
distance measures on the tests.
The most crucial step in a Monte Carlo
simulation is being able to generate distances resembling those of LCDM
distances of GM tissue in VMPFCs; i.e., capturing the true randomness in
LCDM distances. For demonstrative purposes, we pick the left VMPFC of HR subject 1.
Recall that the LCDM distances for left VMPFC of HR subject 1
were denoted by $D_{21}^L$.
We rearrange the distances, $D_{21}^L$, so
that first stack of distances are in $I_0 :=\left[ {-1,0.5} \right]\, mm$, the
second stack of distances are in $I_1 :=(0.5,1.0]\, mm$, the third stack of
distances are in $I_2 :=(1.0,1.5]\, mm$, and so on (until the last stack of
distances are in $I_{11} :=(5.5,6.0]\, mm$). Let $\nu_i$ be the number of
distances that fall in $I_i$, i.e.,
$\nu_i=\left| {D_{21}^L \cap I_i} \right|$, for $i= 0,1,\ldots,11$.
Hence $\vec {\nu}=\left( {\nu_0 ,\nu_1 ,\ldots ,\nu_{11}} \right)
= (2059, 1898, 1764, 1670, 1492, 1268, 814, 417, 142, 81, 61, 16)$.
Then we merge these stacks into one group, (by appending
$D_{21}^L \cap I_{i+1}$ to $D_{21}^L \cap I_i$ for $i=1,2,3,\ldots,10)$.
See Figure \ref{fig:HR1-dist-L} where the top graph is for the merged
distances and bottom graph is for distances sorted in ascending order.

\begin{figure}[htbp]
\centering
\rotatebox{-90}{ \resizebox{2.3 in}{!}{ \includegraphics{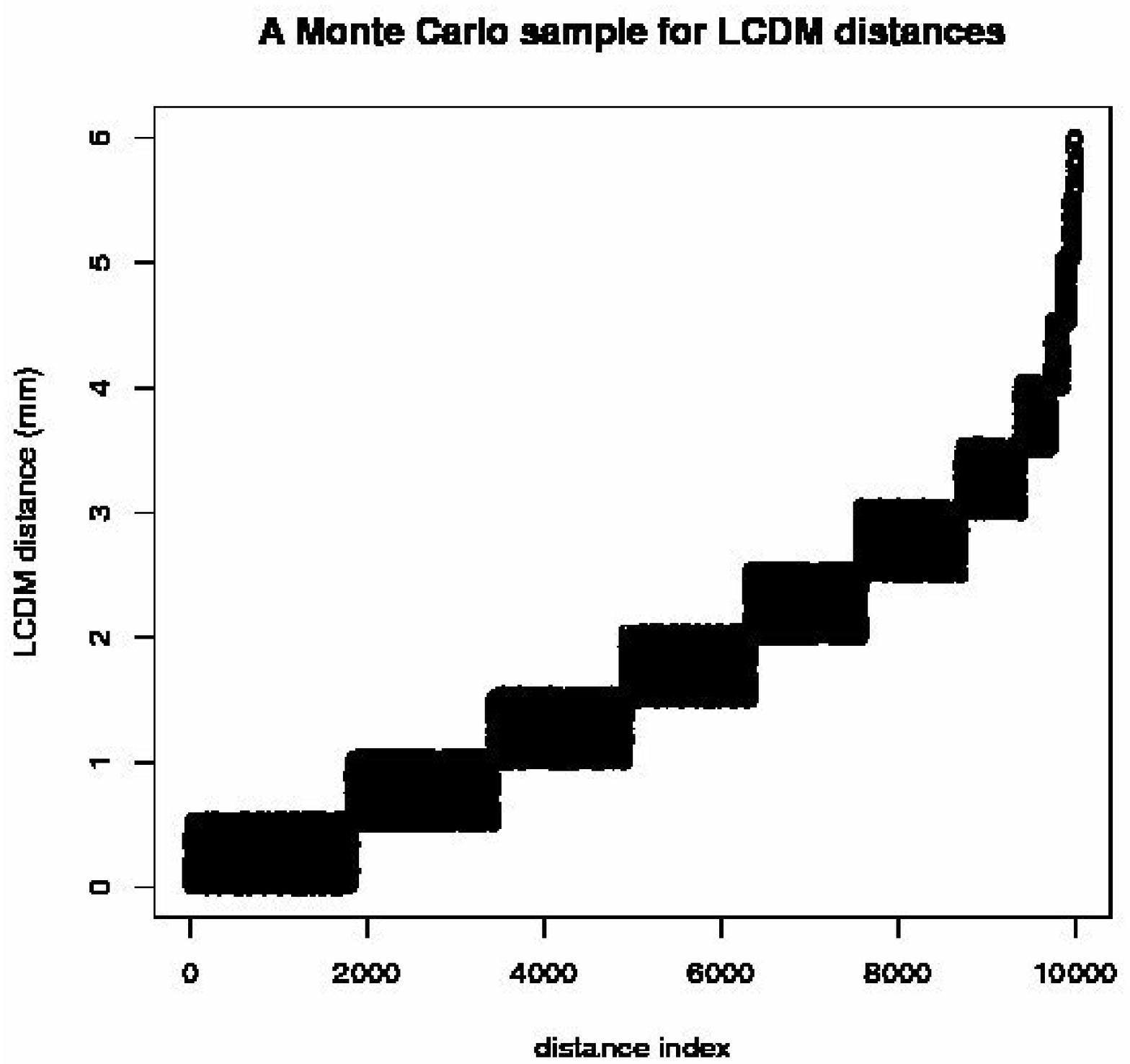} } }
\rotatebox{-90}{ \resizebox{2.3 in}{!}{ \includegraphics{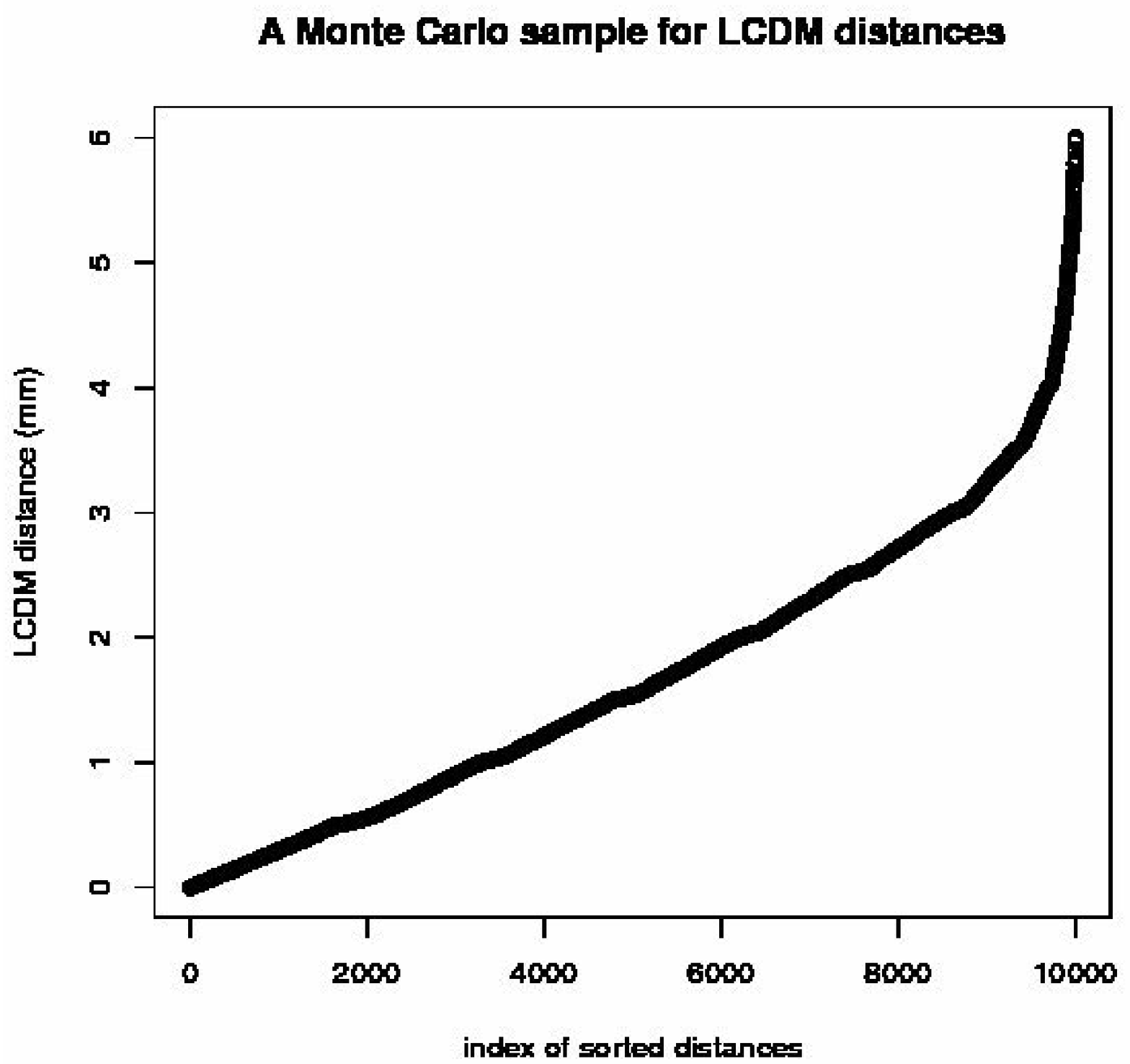} } }
\caption{
\label{fig:MC-HR1-dist-L}
Depicted are the plots of the data values generated by Monte Carlo simulation to resemble LCDM distances.
The left plot is the distances stacked for intervals of size $0.5$
and the right plot is for the sorted distances.
}
\end{figure}

A possible Monte Carlo simulation for these distances can be performed as follows.
We generate $n$ numbers in $\{0,1,2,\ldots,11\}$ proportional
to the above frequencies, $\nu_i$, with replacement.
The number of distances for left VMPFC of HR subject 1 is 11659, so we generate
$n=10000$ such numbers.
Then we generate as many ${\cal U}(0,1)$ numbers for
each $i \in \left\{ {0,1,2,\ldots,11} \right\}$ as $i$ occurs in the generated
sample of 10000 numbers, and add these uniform numbers to $i$.
Then we divide each distance by 2 to match the range of generated distances with $[0,6.0]\textit{ mm}$
which is the range of $D_{21}^L$.
More specifically, we independently
generate $n$ numbers from $\left\{ {0,1,2,\ldots,11} \right\}$ with the discrete
probability mass function $P_N \left( {N_j=i} \right)={\nu_i} \mathord{\left/ {\vphantom {{\nu_i} {11659}}} \right.
\kern-\nulldelimiterspace} {11659}$ for $i= 1,2,{\ldots},11$ and $j=1,2,{\ldots},n$.
Let $n_i$ be the frequency of $i$ among the $n$ generated
numbers from $\left\{0,1,2,\ldots,11\right\}$ with distribution $P_N$,
for $i=0,1,2,{\ldots},11$.
Hence $n=\sum\nolimits_{i=0}^{11} {n_i}$.
Then we generate $U_{ik} \sim U\left( {0,1} \right)$ for $k=1,2,\ldots,n_i$ for each $i$,
and the desired distance values are $d_{ik}=\left( {i+U_{ik}}\right)/2$.
Hence the set of simulated distances is
\[
D_{mc}=\left\{ {d_{ik}=\left( {i+U_{ik}} \right)/2:U_{ik} \sim U\left(
{0,1} \right)\;\text{for   } k=1,,N_i \;\text{   and   } \;N_i \,\sim P_N \;\;\text{for   }
\;i=0,1,2,\ldots,11} \right\}.
\]
A sample of the distances generated in this fashion is plotted in Figure \ref{fig:MC-HR1-dist-L},
where the top plot is for the distances as they are generated at each bin
(stack) of size $0.5\textit{ mm}$, the bottom plot is for the distances sorted in ascending order.
Comparing Figures \ref{fig:HR1-dist-L} and \ref{fig:MC-HR1-dist-L}, we observe that the Monte Carlo
scheme described above generates distances that resemble LCDM distances for
left VMPFC of HR subject 1.
Therefore, the distances generated in this
fashion together with modification of some parameters such as $\nu_i$
resemble the distances of VMPFCs from real subjects.

\subsubsection{Empirical Size Estimates for the Multi-Sample Case}
\label{sec:emp-size-multi-sample}
For the null hypothesis of multi-sample case which states the equality of
the distribution of LCDM distances, we generate three samples ${\cal X}$,
${\cal Y}$, and ${\cal Z}$ each of size $n_x$, $n_y$, and $n_z$, respectively.
Each sample is generated as described above with the sample
sizes for bins (stacks) have been selected to be proportional to the
frequencies $\vec {\nu}=\left( {\nu_0 ,\nu_1 ,\ldots ,\nu_{11}}
\right)= (2059, 1898, 1764, 1670, 1469, 1268, 814, 417, 142, 81, 61, 16)$,
i.e., the left VMPFC of HR subject 1.
This is done without loss of
generality since any other VMPFC can either be obtained by a rescaling of
the generated distances, or by modifying the frequencies in $\vec {\nu}$.
That is, we generate $n$ numbers in $\{0, 1, 2,{\ldots},11\}$ proportional
to the above frequencies, $\nu_i$, with replacement.
Then we generate as
many ${\cal U}(0,1)$ numbers for each $i\in \left\{ {0,1,2,\ldots,11} \right\}$
as $i$ occurs in the generated sample of 10000 numbers, and add these uniform
numbers to $i$.
More specifically, we independently generate $n$ numbers from
$\left\{ {0,1,2,,11} \right\}$ with the discrete probability mass
function
$P_N \left( {N_j=i} \right)={\nu_i} \mathord{\left/ {\vphantom
{{\nu_i} {11659}}} \right. \kern-\nulldelimiterspace} {11659}$ for
$i= 0,1,2,{\ldots},11$ and $j= 1,{\ldots},n$.
Let $n_i$ be the frequency of $i$ among
the $n$ generated numbers from $\left\{ {0,1,2,\ldots,11} \right\}$ with
distribution $P_N $, for $i= 0, 1, 2,{\ldots},11$.
Then we generate $U_{ik} \sim U\left( {0,1} \right)$ for $k=1,\ldots,n_i $ for each $i$, and the desired
distance values are $d_{ik}=\left( {i+U_{ik}} \right)$.
Hence the set of simulated distances is
\[
D_{mc}=\left\{ {d_{ik}=\left( {i+U_{ik}} \right)\big / 2:U_{ik} \sim U\left(
{0,1} \right)\;\text{for   } k=1,,N_i \;\text{  and  } \;N_i \,\sim P_N \;\;\text{for   }
\;i=0,1,2,\ldots,11} \right\}.
\]

We repeat these sample generations $N_{mc}=10000$ times. We count the
number of times the null hypothesis is rejected at $\alpha=0.05$ level for
B-F test of HOV, K-W test of distributional equality, and ANOVA $F$-tests (with
and without HOV) of equality of mean distances, thus obtain the estimated
significance levels under $H_{o}$.
The estimated significance levels for
various values of $n_x$, $n_y$, and $n_z$ are provided in Table \ref{tab:emp-sign-levels-3-sample},
where $\widehat {\alpha}_{BF}$ is the empirical size estimate for B-F test,
 $\widehat {\alpha}_{KW}$ is for K-W test, $\widehat {\alpha}_{F_1}$ is for ANOVA
$F$-test with HOV, and $\widehat {\alpha}_{F_2}$ is for ANOVA $F$-test without HOV;
furthermore, $\widehat {\alpha}_{KW,F_1}$ is the proportion of agreement
between K-W and ANOVA $F$-test with HOV, i.e., the number of times out of 10000
Monte Carlo replicates both K-W and ANOVA $F$-test with HOV reject the null hypothesis.
Similarly, $\widehat {\alpha}_{KW,F_2}$is the proportion of
agreement between K-W and ANOVA $F$-test without HOV, and $\widehat {\alpha}_{F_1 ,F_2}$
is the proportion of agreement between ANOVA $F$-test with HOV and
ANOVA $F$-test without HOV.
Using the asymptotic normality of the proportions,
we test the equality of the empirical size estimates with 0.05, and compare
the empirical sizes pairwise.
We observe that K-W and B-F tests are both at
the desired significance level, while ANOVA $F$-tests with and without HOV are
at the desired level or slightly conservative.
Notice also that under $H_{o}$,
the tests tend to be more conservative as the sample sizes increase.
Hence, if the distances are not that different; i.e., the frequency of
distances for each bin and the distances for each bin are identically
distributed for each group, the inherent spatial correlation does not seem
to influence the significance levels.
Moreover, we observe that for LCDM
distances K-W and $F_{1}$ tests have significantly different rejection (hence
acceptance) regions because, the proportion of agreement for these tests,
$\widehat {\alpha}_{KW,F_1}$ is significantly smaller than the minimum of
$\widehat {\alpha}_{KW}$ and $\widehat {\alpha}_{F_1}$.
Similarly, K-W and $F_{2}$ tests have significantly different rejection (hence acceptance)
regions because, the proportion of agreement for these tests, $\widehat {\alpha
}_{KW,F_2}$ is significantly smaller than the minimum of $\widehat {\alpha
}_{KW}$ and $\widehat {\alpha}_{F_2}$.
However, $F_{1}$ and $F_{2}$ tests have
about the same rejection (hence acceptance) regions because, the proportion
of agreement for these tests, $\widehat {\alpha}_{F_1 ,F_2}$ is not
significantly different from the minimum of $\widehat {\alpha}_{F_1}$ and
$\widehat {\alpha}_{F_2}$. This mainly results from the fact that K-W and
$F_{1}$ tests test different hypotheses, and so do the K-W and $F_{2}$ tests.
But, $F_{1}$ and $F_{2}$ tests basically test the same hypotheses.

\subsubsection{Empirical Power Estimates for Multi-Sample Case}
\label{sec:emp-power-multi-sample}
For the alternative hypotheses, we generate sample ${\cal X}$ as follows. We
generate as many ${\cal U}(0,1)$ numbers for each $i\in \left\{ {0,1,2,\ldots,11} \right\}$
as $i$ occurs in the generated sample of $n_{x}$ numbers,
and add these uniform numbers to $i$ for sample ${\cal X}$.
For sample ${\cal Y}$, we generate numbers in {\{}0,1,2,{\ldots},12{\}} with replacement
proportional to the frequencies $\vec {\nu}_y=\left( {\nu_0^y ,\nu_1^y
,\ldots ,\nu_{12}^y} \right)$ where $\nu_i^y $ is the $i^{th}$ value after
the entries $\vert \nu_i -\eta_y \vert $ are sorted in descending order
for $i=0,1,2,\ldots,11$ and
$\nu_{12}^y=11659-\sum\nolimits_{i=0}^{11} {\vert\nu_i -\eta_y \vert}$.
Then we generate as many ${\cal U}(0,r_y )$
numbers for each $i\in \left\{ {0,1,2,\ldots,12} \right\}$ as $i$ occurs in the
generated sample of $n_{y}$ numbers, and add these uniform numbers to $i$. For
sample ${\cal Z}$, we generate numbers in {\{}0,1,2,{\ldots},12{\}} with
replacement proportional to the frequencies $\vec {\nu}_z=\left( {\nu_0^z
,\nu_1^z ,\ldots ,\nu_{12}^z} \right)$ where $\nu_i^z $is the $i^{th}$
value after the entries $\vert \nu_i -\eta_z \vert $ are sorted in
descending order for $i=0,1,2,\ldots,11$ and
$\nu_{12}^z=11659-\sum\nolimits_{i=0}^{11} {\vert \nu_i -\eta_z \vert}$. Then we
generate as many ${\cal U}(0,r_z )$ numbers for each
$i\in \left\{{0,1,2,\ldots,12} \right\}$ as $i$ occurs in the generated sample of $n_{z}$
numbers, and add these uniform numbers to $i$.

\begin{figure}[htbp]
\centering
\rotatebox{-90}{ \resizebox{3. in}{!}{ \includegraphics{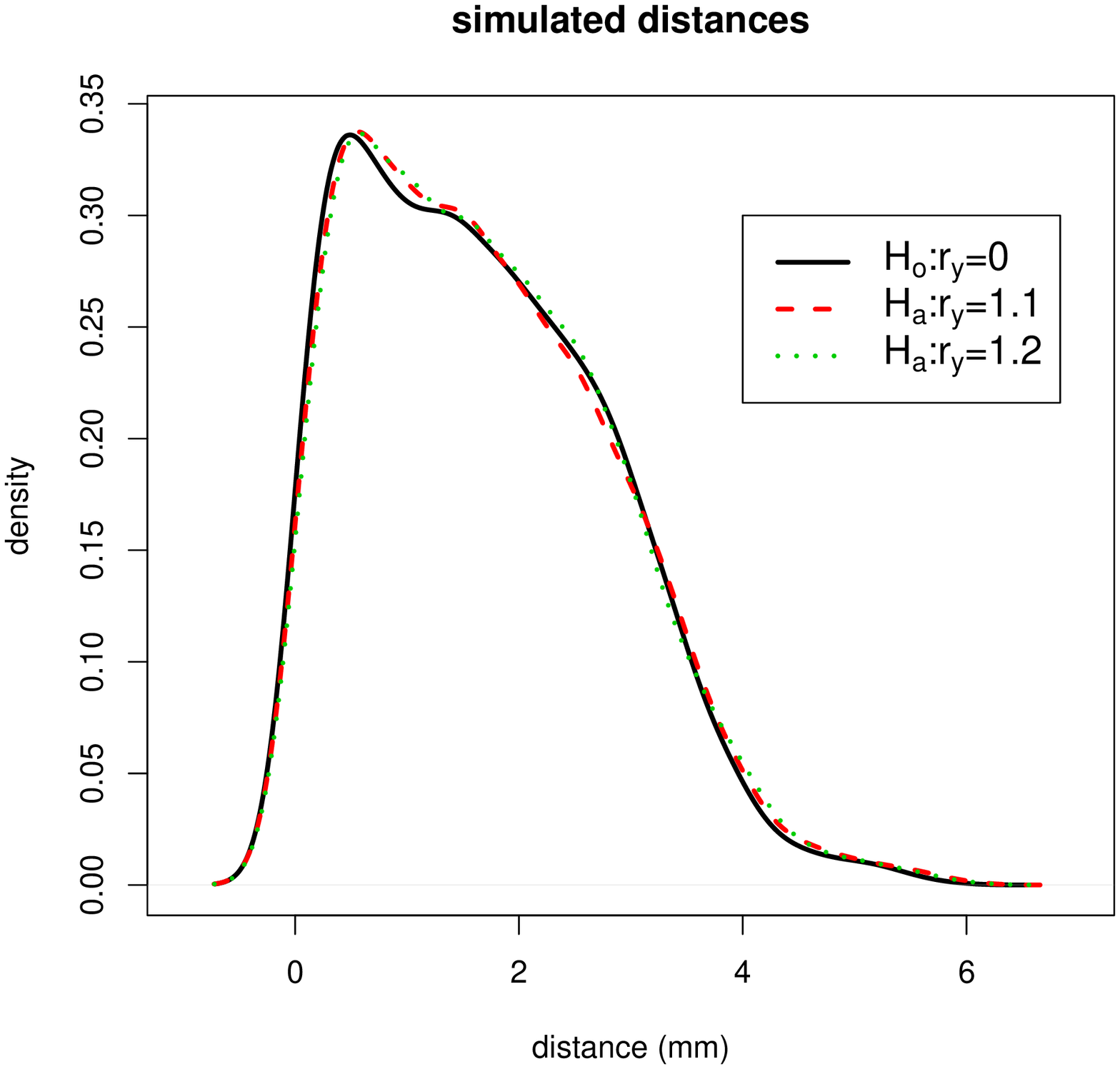} } }
\rotatebox{-90}{ \resizebox{3. in}{!}{ \includegraphics{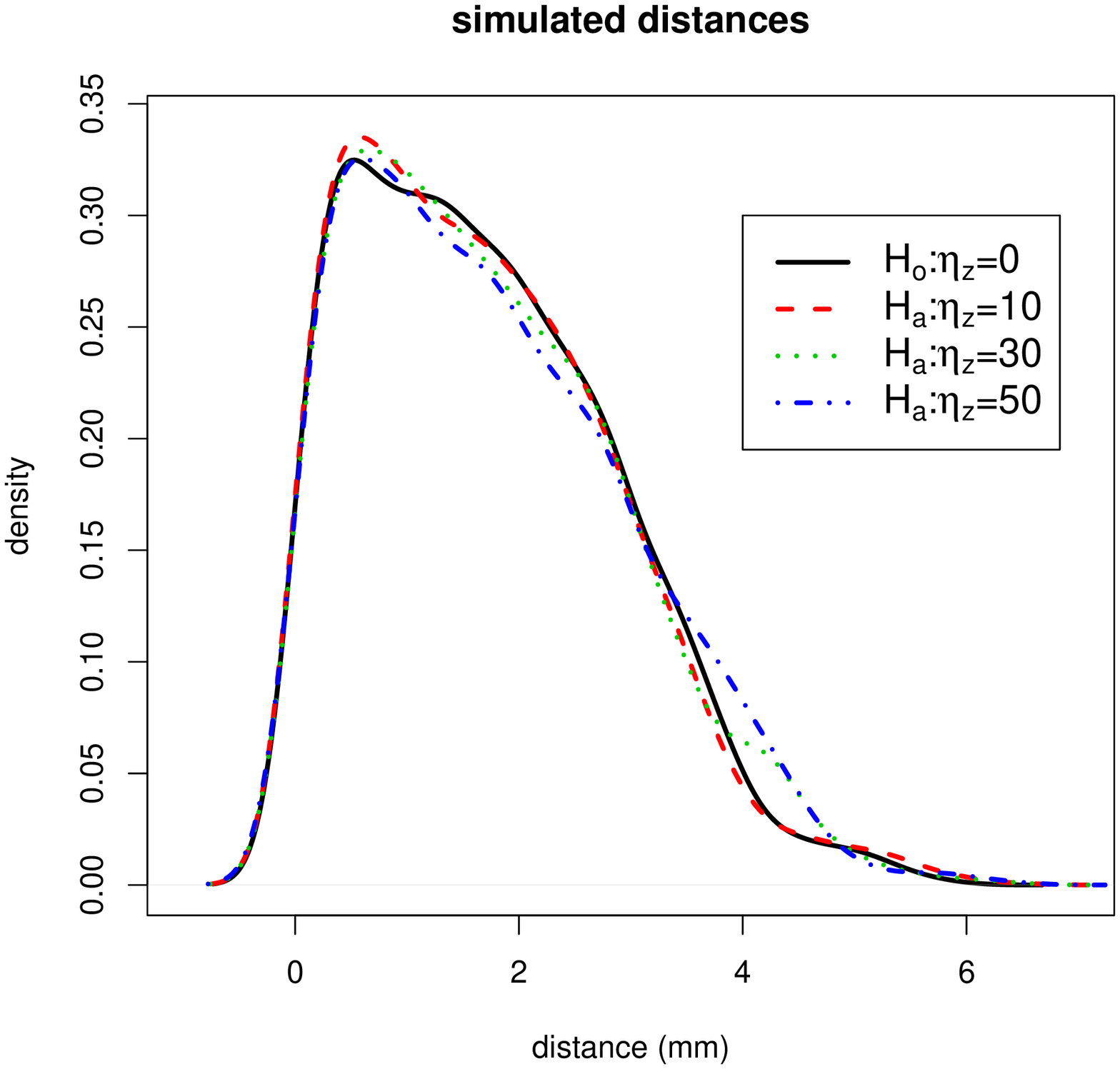} } }
\caption{
\label{fig:MC-alt-dist}
Depicted are the plots of the kernel density estimates of the Monte Carlo simulated LCDM distances under the null case
and alternatives with $\eta_z=0$ and $r_y \in \{1.1,1.2\}$ (left);
null case and alternatives with $r_y=1.0$ and $\eta_z \in \{10,30,50\}$.
For the parameters $r_y$, $r_z$, $\eta_y$, and $\eta_z$, see Section \ref{sec:emp-power-multi-sample}.
}
\end{figure}

More formally, the samples are generated as
\[
\begin{array}{l}
 N_{\cal X}=\left\{ {J\sim P_X ,J=1,\ldots,n_x} \right\}, \\
 \\
 N_{\cal Y}=\left\{ {J\sim P_Y ,J=1,\ldots,n_y} \right\}, \\
 \\
 N_{\cal Z}=\left\{ {J\sim P_Z ,J=1,\ldots,n_z} \right\}, \\
 \end{array}
\]
where $P_X \left( {X_j=i} \right)=\nu_i^x /11659$ with $\nu_i^x $ is the
$i^{th}$ entry in $\vec {v}$; $P_Y \left( {J=i} \right)={\nu_i^y}
\mathord{\left/ {\vphantom {{\nu_i^y} {\sum\nolimits_{i=0}^{12} {\nu_i^y
}}}} \right. \kern-\nulldelimiterspace} {\sum\nolimits_{i=0}^{12} {\nu_i^y}}$
with $\nu_i^y $ is the $i^{th}$entry in $\vec {\nu}_y$; and
$P_Z \left( {J=i} \right)={\nu_i^z} \mathord{\left/ {\vphantom {{\nu_i^z}
{\sum\nolimits_{i=0}^{12} {\nu_i^z}}}} \right. \kern-\nulldelimiterspace}
{\sum\nolimits_{i=0}^{12} {\nu_i^z}}$
with $\nu_i^z $ is the $i^{th}$ entry in $\vec {\nu}_z$.
Let $n_i^x $ be the frequency of $i$ among the
$n_{x}$ generated numbers from $P_{X}$, $n_i^y $ be the frequency of $i$ among
the $n_{y}$ generated numbers from $P_{Y}$ , and $n_i^z $ be the frequency of
$i$ among the $n_{z}$ generated numbers from $P_{Z}$.
Then we generate
$U_{ik} \sim \mbox{{\cal U}}\left( {0,1} \right)$ for $k $= 1,{\ldots},$n_i^x $
for each $i$.
Hence the set of simulated distances for set ${\cal X}$ is
\[
\begin{array}{l}
 \quad \quad \quad \quad \quad \quad {\cal X}=\left\{ {  \left(i+U_{ik} \right) \big / 2 :U_{ik} \sim
U\left( {0,1} \right)\mbox{ for }i=0,1,,11\mbox{ and }k=1,\ldots,N_{\cal X}
} \right\}, \\
 \mbox{similarly,} \\
 \quad \quad \quad \quad \quad \quad {\cal Y}=\left\{ { \left(i+U_{ik} \right) \big / 2 :U_{ik} \sim
U\left( {0,r_y} \right)\mbox{ for }i=0,1,,12\mbox{ and }k=1,\ldots,N_{\cal
Y}} \right\}, \\
 \mbox{and} \\
 \quad \quad \quad \quad \quad \quad {\cal Z}=\left\{ {\left(i+U_{ik} \right) \big / 2 :U_{ik} \sim
U\left( {0,r_z} \right)\mbox{ for }i=0,1,\ldots,12\mbox{ and }k=1,\ldots,N_{\cal
Z}} \right\}. \\
 \end{array}
\]
Note that when $r_y=r_z=1$ and $\eta_y=\eta_z=0$, we obtain the null
case of distributional equality between ${\cal X},\;{\cal Y}$ and ${\cal Z}$.
The alternative cases we consider are
$$\left( {r_y ,r_z ,\eta_y ,\eta_z} \right)\in \left\{ {\left(
{1.1,1.0,0,0} \right),\left( {1.1,1.2,0,0} \right),\left( {1.0,1.0,10,0}
\right),\left( {1.0,1.0,10,10} \right),\left( {1.0,1.0,10,30} \right)} \right\}.$$
See Figure \ref{fig:MC-alt-dist} for the kernel density estimates of sample distances
under the null case and various alternatives.

We repeat these sample generations $N_{mc}=10000$ times.
We count the number of times the null hypothesis is rejected for B-F test of HOV, K-W
test of distributional equality, and ANOVA $F$-tests (with and without HOV) of
equality of mean distances, thus obtain the empirical power estimates under
$H_{a}$ which are provided in Table \ref{tab:emp-power-3-sample},
where $\widehat {\beta}_{BF}$ is the
empirical power estimate for B-F test, $\widehat {\beta}_{KW}$ is for K-W
test, $\widehat {\beta}_{F_1}$ is for ANOVA $F$-test with HOV, and
$\widehat {\beta}_{F_2}$ is for ANOVA $F$-test without HOV.
Using the asymptotic normality of
the empirical power estimates, we observe that under $H_{a}$ with
$\left({r_y ,r_z ,\eta_y ,\eta_z} \right)\in \left\{ {\left( {1.1,1.0,0,0}
\right),\left( {1.1,1.2,0,0} \right)} \right\}$
the variances of the
distances are not that different, so we still have power estimates for B-F
test around .05 (see Figure \ref{fig:MC-alt-dist} (left) and Table \ref{tab:emp-power-3-sample}).
But the distributions are different,
so the larger the $r_y$ and $r_z$ from 1.0, the higher the power estimates
for K-W and ANOVA $F$-tests.
Furthermore, as the sample size $n$ increases, the
power estimates for K-W and ANOVA $F$-tests also increase.
Notice that under
these alternatives, K-W test tends to be more powerful than ANOVA $F$-tests,
since such alternatives influence the ranking (hence the distribution) of
the distances, more than the mean of the distances.
Furthermore, under these
alternatives, it is not the size or scale that is really different; it is
the shape that is more emphasized.
This size component is distance with
respect to the GM/WM surface; i.e., if the GM voxels from the GM/WM surface
are at about the same distance, K-W test is more sensitive to the differences
in LCDM distances.
We also note that ANOVA $F$-tests have about the same power
estimates.

Under $H_{a}$ with
$\left( {r_y ,r_z ,\eta_y ,\eta_z} \right)\in \left\{
{\left( {1.0,1.0,10,0} \right),\left( {1.0,1.0,10,10} \right),\left(
{1.0,1.0,10,30} \right)} \right\}$
the variances of the distances start to
differ (see Figure \ref{fig:MC-alt-dist} (right) and Table \ref{tab:emp-power-3-sample});
as $\eta_y ,\eta_z$ deviate more from 0,
the power estimates for B-F test increase, and so do the power estimates of
K-W and ANOVA $F$-tests.
Note that as $n$ increases, the power estimates also
increase under these alternative cases.
Under these second type of
alternatives, ANOVA $F$-tests tend to be more powerful, since the right skewness
(tail) of distances are more emphasized, which in turn implies that the
differences in the mean distances are emphasized more.
Under these alternatives, both the size or scale and shape are different. If the GM
voxels from the GM/WM surface are at different distances, ANOVA $F$-tests are
more sensitive to the differences in LCDM distances.
We also note that both
ANOVA $F$-tests have about the same power estimates.

Therefore, based on our Monte Carlo analysis, the spatial correlation
between distances has a mild influence on our results.
That is, the results
based on B-F, K-W, and ANOVA $F$-tests on multiple samples are still reliable,
although the assumption of within sample independence is violated.
Since normality is also violated, K-W test has fewer assumption violations than
the ANOVA $F$-tests.
However, our Monte Carlo analysis suggests that K-W test is
more sensitive against the shape differences for GM of VMPFCs with similar
distances to the GM/WM boundary; on the other hand, ANOVA $F$-tests are more
sensitive against the shape differences for GM of VMPFCs with different
distances to the boundary.

\subsubsection{Empirical Size Estimates for the Two-Sample Case}
\label{sec:emp-size-two-sample}
For the null hypothesis for the two-sample case, we generate two samples
${\cal X}$ and ${\cal Y}$each of size $n_x$ and $n_y$, respectively.
Each sample is generated as described above.
We repeat the sample generation $N_{mc}=10000$ times.

We count the number of times the null hypothesis is rejected at
$\alpha=0.05$ for Lilliefor's test of normality, B-F test of HOV, Wilcoxon rank sum
test of distributional equality, Welch's $t$-test of equality of mean distances, and
K-S test of equality of cdfs, thereby obtain the estimated significance levels.
Unlike the multi-sample case, for the two-sample case, except for
Lilliefor's test there are three types of alternative hypotheses possible:
two-sided, left, and right-sided alternatives.
The estimated significance
levels are provided in Table \ref{tab:emp-size-2-sample},
where $\widehat {\alpha}_{BF}$ is the
empirical size estimate for B-F test, $\widehat {\alpha}_W $ is for Wilcoxon
rank sum test, $\widehat {\alpha}_t $ is for Welch's $t$-test, $\widehat {\alpha}_{KS}$ is for K-S test.
Furthermore, $\widehat {\alpha}_{W,t}$ is the proportion of
agreement between Wilcoxon rank sum and Welch's $t$-tests, $\widehat {\alpha}_{W,KS}$ is
the proportion of agreement between Wilcoxon rank sum and K-S tests, and
$\widehat {\alpha}_{t,KS}$ is the proportion of agreement between Welch's $t$-test and K-S test.
We omit the Lilliefor's test, since by construction, our samples
are severely non-normal, so normality is rejected for almost all samples generated.
Observe that under $H_{o}$, the empirical significance levels are
about the desired level for all three types of alternatives, although B-F
and Wilcoxon tests are slightly liberal, while K-S test is slightly conservative.
Hence, if the distances are not that different; i.e., the
frequency of distances for each bin and the distances for each bin are
identically distributed for each group, the inherent spatial correlation
does not influence the significance levels.
However, Wilcoxon rank sum,
Welch's $t$-test, and K-S tests test different hypotheses, so their acceptance and
rejection regions are significantly different for LCDM distances, since the
proportion of agreement for each pair is significantly smaller than the
minimum of the empirical size estimates for each pair of tests.

\subsubsection{Empirical Power Estimates for the Two-Sample Case}
\label{sec:emp-power-two-sample}
For the alternative hypotheses, we generate samples ${\cal X}$ and ${\cal Y}$
as in Section \ref{sec:emp-size-multi-sample} also.
Note that when $r_y=1$ and $\eta_y=0$, we
obtain the null case.
The alternative cases we consider are
$\left( {r_y,\eta_y} \right) \in \{ (1.1,0),(1.2,0),(1.0,10),\\
(1.0,30),(1.0,50) \}$.
We count the number of times the null hypothesis is
rejected for Lilliefor's test of normality, B-F test of HOV, Wilcoxon rank
sum test of distributional equality, Welch's $t$-test of equality of mean distances,
and K-S test of equality of cdfs, thereby obtain the estimated significance levels.
The power estimates are provided in Table \ref{tab:emp-power-2-sample},
where $\widehat {\beta}_{BF}$ is the power estimate for B-F test, $\widehat {\beta}_W $ is for
Wilcoxon rank sum test, $\widehat {\beta}_t $ is for Welch's $t$-test, $\widehat {\beta}_{KS}
$ is for K-S test.

Under $H_a$ with $(r_y,\eta_y) \in \{(1.1,0),(1.2,0)\}$,
the variances of the distances are not that different (see Figure \ref{fig:MC-alt-dist} (left)
and Table \ref{tab:emp-power-2-sample}),
so we still have power estimates for B-F test around .05.
But the distributions start to differ;
so as $r_y$ deviates further away from 1.0, then the power estimates for
Wilcoxon rank sum, Welch's $t$-test, and K-S tests increase.
Furthermore, as the sample
size $n$ increases, the power estimates for Wilcoxon, Welch's $t$-test, and K-S tests also increase.
Observe that, as in the multi-sample case, under these
alternatives, Wilcoxon test is more powerful than Welch's $t$-test, since the ranking
of the distances are affected more than the mean distances under these alternatives.
But K-S test has the highest power estimates for sample sizes
larger than 1000.
Thus, for differences in shape rather than the distance
from the GM/WM surface, K-S test and Wilcoxon rank sum test are more
sensitive than Welch's $t$-test.
Furthermore, as the sample sizes increase, the
left-sided tests become more powerful than their two-sided counterparts.
Notice that we omit the power estimates for the right-sided alternatives.
By construction, ${\cal X}$ values tend to be smaller than ${\cal Y}$ values for
these alternatives; hence the tests virtually have no power for the
right-sided alternatives.

Under $H_a$ with $(r_y,\eta_y) \in \{(1.0,10),(1.0,30),(1.0,50)\}$,
the variances of the distances start to differ (see Figure \ref{fig:MC-alt-dist} (right)
and Table \ref{tab:emp-power-2-sample});
as $\eta_1$ deviates more from 0,
the power estimates for B-F test increase,
and so do the power estimates of Wilcoxon, Welch's $t$-test, and K-S tests.
Note that as $n$ increases, the power estimates also increase under each
alternative case.
Under these alternatives, $t$-test is more powerful than
Wilcoxon test, since mean distances are more affected than the rankings
under such alternatives.
However, K-S test has higher power estimates for
larger deviations from the null case.
These alternatives imply that the
distances of the GM voxels are at different scales, Welch's $t$-test has the best
performance for small differences, while for large differences, K-S has the
best performance.
Furthermore, as the sample sizes increase, the left-sided
tests become more powerful than their two-sided counterparts.
Again, we omit
the power estimates for the right-sided alternatives because by
construction, ${\cal X}$ values tend to be smaller than ${\cal Y}$ values for
these alternatives.

We do not report the power estimates for Lilliefor's test of normality,
since by construction our data is severely non-normal, and we get power
estimates of 1.000 under both null and alternative cases.

Therefore, based on our Monte Carlo analysis, the spatial correlation
between distances has a mild influence, if any, on our results.
That is, the
results based on Wilcoxon rank sum test, Welch's $t$-test, K-S, and B-F tests for two
samples are still reliable, although the assumption of within sample
independence is violated.
However, Wilcoxon rank sum test is more sensitive
against the shape differences of GM of VMPFCs with similar distance from the
GM/WM boundary; while the Welch's $t$-test is more sensitive against the differences of
GM tissue with different distances from the boundary.

\section{Discussion and Conclusions}
\label{sec:disc}
In this article, we investigate various uses of the LCDM distances to detect
differences in morphometry in brain tissues due to various factors such as a
disease or disorder.
As an illustrative example, we use GM tissue in Ventral
Medial Prefrontal Cortices (VMPFCs) for three groups of subjects; namely,
subjects with major depressive disorder (MDD), subjects at high risk for
depression (HR), and control subjects (Ctrl).
Our study comprises of (MDD,HR) and (Ctrl,Ctrl) co-twin pairs.
Since we focus on the use of LCDM distances, rather than the clinical implications
of the genetic influence (due to twinness),
we ignore the twin influence for most of the analysis in this article
(except in the comparison of MDD and HR volumes and descriptive measures).

LCDM distance data set comprises of a large set of distances, which
depends on the voxel size which is used to partition the tissue (GM of VMPFCs).
First, as a preliminary step, we use simple morphometric
measures based on LCDM distances.
These simple measures include,
volume (a multiple of the number of LCDM distances), and descriptive
measures such as median, mode, range, and variance of LCDM distances.
The location of the distribution (e.g., median) of LCDM distances provide information on the (average)
cortical mantle thickness,
on the other hand, the scale of the distribution (e.g., standard
deviation or variance) of the distribution of LCDM
distances provide information on variation in morphometry (shape and/or size).
More variation in the distances can be resulting from
the higher cortical mantle thickness or more
variation in the surface structure.
In the analysis of these descriptive summary measures,
we can both use nonparametric or parametric tests,
since most of the time the assumptions were met for both types of tests.
For example, for multi-group comparisons we
could apply Kruskal-Wallis (K-W) test or ANOVA $F$-test.
Each of these measures
conveys information on some aspect of the morphometry of VMPFCs.
The analysis of these measures might provide a preliminary assessment of
differences in morphometry, although by summarizing
most of the information LCDM distances convey is lost.
For example, in our data
set volumes and descriptive measures do not indicate much separation
between groups due to depressive disorders.

Since the descriptive measures of LCDM distances are summary statistics,
they tend to oversimplify the data, and hence we lose most of the information conveyed
by the LCDM distances.
To avoid the loss of information when using the descriptive summary
statistics from LCDM distance as performed [33], we pool LCDM
distances of subjects from the same group
assuming distances in the same group have the similar distributions.
As a precautionary step,
we find the extreme (outlier) subjects; i.e.,
the subjects whose VMPFCs have much different distributions than the rest.
Note that the kernel density estimates (or normalized histograms) can be
used as an exploratory tool to detect outliers.
The pooled LCDM distances can be
used to detect group differences in morphometry, left-right morphometric
asymmetry, and stochastic ordering of the distances. Since LCDM is designed
to measure cortical mantle thickness (with respect to the GM/WM surface), it
naturally provides size differences in the normal direction from the surface.
However, the ``width'' (i.e., the thickness of VMPFC parallel to the
surface) is less relevant, so LCDM tends to ignore the size differences in
the parallel (to the surface) direction.

We apply the parametric tests (e.g., ANOVA $F$-test and Welch's $t$-test)
and nonparametric tests (K-W test and Wilcoxon rank sum tests) for multi-group
and two-group comparisons of LCDM distances.
We use Brown-Forsythe (B-F) test for homogeneity of the variances (HOV)
and Kolmogorov-Smirnov (K-S) test for cdf comparisons.
But the parametric tests require normality and all of these tests require within
sample independence.
The pooled LCDM distances are extremely non-normal due to heavy right skewness,
and within sample independence is violated due to the spatial correlation
between LCDM distances of neighboring voxels.
However, our extensive Monte Carlo study reveals
that the influence of these violations is very mild if not negligible.
Applying K-S test on LCDM distances may provide
the stochastic ordering of LCDM distances, if present.
Although, Wilcoxon rank sum test and K-S test have the same null hypothesis,
the alternatives and information they provide are different.
If K-S test fails to reject the null, it means no significant distributional differences
over the whole range of the variable, while if Wilcoxon test
fails to reject the null hypothesis,
it means that the location parameter or more precisely
the ranking of the variables is not significantly different.
If K-S test rejects the null hypothesis and there is no stochastic ordering,
then it means that the direction of distributional differences
vary at different values of the variable (LCDM distance).

Left-right morphometric asymmetry can also be detected by the use of
LCDM distances (with Wilcoxon rank sum test and $t$-test).
Such asymmetry might be due to both asymmetry in
shape and/or size.
In terms of size asymmetry, LCDM emphasizes
mantle thickness asymmetry, rather than the mantle width asymmetry.

We demonstrate that pooled LCDM distances are a powerful tool to detect
various types of morphometric differences.
For the illustrative example we
used in the article, the analysis on LCDM distances indicate that VMPFC left
and right distances tend to decrease due to the depressive disorders or
being at high risk for depression, possibly due to a thinning in left and right VMPFCs;
the morphometric variation reduces in left and right VMPFCs
due to suffering from or being at high risk for depressive disorders compared
to Ctrl subjects and is smallest for the HR subjects for both left and right VMPFCs;
there is significant left-right asymmetry in LCDM distances in the
sense that, the cortical mantle in left VMPFC is thinner for MDD and Ctrl
subjects and thicker for HR subjects compared to their right counterparts.
Moreover, the analysis of LCDM distances yield a stochastic ordering as
HR $<^{ST}$ MDD $<^{ST}$ Ctrl for both left and right VMPFCs; i.e.,
it is more likely for HR-left distances to be smaller compared to MDD-left and
Ctrl-left distances, and more likely for MDD-left distances to be smaller
than Ctrl-left distances.
That is, it is more likely for left VMPFC of HR
subjects to be thinner compared to those of MDD subjects, and more likely
for VMPFC of MDD subjects to be thinner compared to those of Ctrl subjects.
The same holds for right VMPFCs.
The corresponding clinical findings, together
with the interpretations, have been described elsewhere [44].
Note that, pooled distances are only suggestive of
morphometric differences, but do not provide information on the location of these differences.
This aspect of LCDM analysis is a topic of ongoing research.

Observe also that LCDM distances provide information on morphometry
(both shape and size (especially in the normal direction from the interface, i.e., thickness).
One can adjust the distances for size (e.g., volume),
then LCDM distances will only provide shape information.
The size (or scale) adjustment for LCDM distances is also a topic of ongoing research.

Finally, we emphasize that the methodology used in this article for VMPFC
shape differences can be used for other tissues or organs of humans and
animals, as well as distances similar to the LCDM distances.

\section*{Acknowledgments}
Research supported by R01-MH62626-01, P41-RR15241, R01-MH57180.

\section*{References}
1. Fjell, A.M., et al., \textit{Selective increase of cortical thickness in high-performing elderly--structural indices of optimal cognitive aging.} Neuroimage, 2006. \textbf{29}(3): p. 984-94.

2. Goghari, V.M., et al., \textit{Regionally Specific Cortical Thinning and Gray Matter Abnormalities in the Healthy Relatives of Schizophrenia Patients.} Cereb Cortex, 2006.

3. Lyoo, I.K., et al., \textit{Regional cerebral cortical thinning in bipolar disorder.} Bipolar Disord, 2006. \textbf{8}(1): p. 65-74.

4. Lu, L.H., et al., \textit{Normal Developmental Changes in Inferior Frontal Gray Matter Are Associated with Improvement in Phonological Processing: A Longitudinal MRI Analysis.} Cereb Cortex, 2006.

5. Hardan, A.Y., et al., \textit{An MRI study of increased cortical thickness in autism.} Am J Psychiatry, 2006. \textbf{163}(7): p. 1290-2.

6. Haidar, H. and J.S. Soul, \textit{Measurement of cortical thickness in 3D brain MRI data: validation of the Laplacian method.} J Neuroimaging, 2006. \textbf{16}(2): p.
146-53.

7. Sigalovsky, I.S., B. Fischl, and J.R. Melcher, \textit{Mapping an intrinsic MR property of gray matter in auditory cortex of living humans: A possible marker for primary cortex and hemispheric differences.} Neuroimage, 2006.

8. Shaw, P., et al., \textit{Longitudinal mapping of cortical thickness and clinical outcome in children and adolescents with attention-deficit/hyperactivity disorder.} Arch Gen Psychiatry, 2006. \textbf{63}(5): p. 540-9.

9. Colliot, O., et al., \textit{In vivo profiling of focal cortical dysplasia on high-resolution MRI with computational models.} Epilepsia, 2006. \textbf{47}(1): p. 134-42.

10. Hutsler, J.J., T. Love, and H. Zhang, \textit{Histological and Magnetic Resonance Imaging Assessment of Cortical Layering and Thickness in Autism Spectrum Disorders.} Biol Psychiatry, 2006.

11. Luders, E., et al., \textit{Hemispheric asymmetries in cortical thickness.} Cereb Cortex, 2006. \textbf{16}(8): p. 1232-8.

12. Luders, E., et al., \textit{Gender effects on cortical thickness and the influence of scaling.} Hum Brain Mapp, 2006. \textbf{27}(4): p. 314-24.

13. Makris, N., et al., \textit{Cortical Thinning of the Attention and Executive Function Networks in Adults with Attention-Deficit/Hyperactivity Disorder.} Cereb Cortex, 2006.

14. Park, H.J., et al., \textit{Cortical surface-based analysis of 18F-FDG PET: measured metabolic abnormalities in schizophrenia are affected by cortical structural abnormalities.} Neuroimage, 2006. \textbf{31}(4): p. 1434-44.

15. Hadjikhani, N., et al., \textit{Anatomical differences in the mirror neuron system and social cognition network in autism.} Cereb Cortex, 2006. \textbf{16}(9): p. 1276-82.

16. Thompson, P.M., et al., \textit{Thinning of the cerebral cortex visualized in HIV/AIDS reflects CD4+ T lymphocyte decline.} Proc Natl Acad Sci U S A, 2005.
\textbf{102}(43): p. 15647-52.

17. Milad, M.R., et al., \textit{Thickness of ventromedial prefrontal cortex in humans is correlated with extinction memory.} Proc Natl Acad Sci U S A, 2005. \textbf{102}(30):
p. 10706-11.

18. Barta, P., M.I. Miller, and A. Qiu, \textit{A stochastic model for studying the laminar structure of cortex from MRI.} IEEE Trans Med Imaging, 2005.
\textbf{24}(6): p. 728-42.

19. Scott, M.L. and N.A. Thacker, \textit{Robust tissue boundary detection for cerebral cortical thickness estimation.} Med Image Comput Comput Assist Interv Int
Conf Med Image Comput Comput Assist Interv, 2005. \textbf{8}(Pt 2): p.
878-85.

20. Zivadinov, R., et al., \textit{Reproducibility and accuracy of quantitative magnetic resonance imaging techniques of whole-brain atrophy measurement in multiple sclerosis.} J Neuroimaging, 2005. \textbf{15}(1): p. 27-36.

21. Preul, C., et al., \textit{Morphometry demonstrates loss of cortical thickness in cerebral microangiopathy.} J Neurol, 2005. \textbf{252}(4): p. 441-7.

22. Miller, M.I., et al., \textit{Bayesian construction of geometrically based cortical thickness metrics.} Neuroimage, 2000. \textbf{12}(6): p. 676-87.

23. Miller, M.I., et al., \textit{Labeled cortical mantle distance maps of the cingulate quantify differences between dementia of the Alzheimer type and healthy aging.} Proc Natl Acad Sci U S A, 2003. \textbf{100}(25):
p. 15172-7.

24. Wang, L., et al., \textit{Abnormalities of cingulate gyrus neuroanatomy in schizophrenia.} Schizophrenia Research, 2007. \textbf{93}(1-3): p.
66-78.

25. Elkis, H., et al., \textit{Increased prefrontal sulcal prominence in relatively young patients with unipolar major depression.} Psychiatry Res, 1996. \textbf{67}(2): p. 123-34.

26. Drevets, W.C., et al., \textit{Subgenual prefontal cortex abnormalities in mood disorders.} Nature, 1997. \textbf{386}: p. 824-827.

27. Botteron, K.N., \textit{Genetic Analysis of brain imaging abnormalities.} Psychiatric Clinics of North America, 2001.
\textbf{10}(2): p. 241-258.

28. Botteron, K.N., et al., \textit{Volumetric reduction in the left subgenual prefrontal cortex in early onset depression.} Biological Psychiatry, 2002. \textbf{51}(4): p.
342-344.

29. Botteron, K.N., et al., \textit{Volumetric reduction in the left subgenual prefrontal cortex in early onset depression.} Biol. Psych., 2002. \textbf{51}(4): p. 342-344.

30. Wiegand, L.C., et al., \textit{Prefrontal cortical thickness in first-episode psychosis: a magnetic resonance imaging study.} Biol Psychiatry, 2004. \textbf{55}(2): p.
131-40.

31. O'Donnell, S., et al., \textit{Cortical thickness of the frontopolar area in typically developing children and adolescents.} Neuroimage, 2005. \textbf{24}(4): p. 948-54.

32. Ratnanather, J.T., et al., \textit{Validating cortical surface analysis of medial prefrontal cortex.} Neuroimage, 2001. \textbf{14}(5): p.
1058-69.

33. Ceyhan, E., et al. \textit{Statistical Analysis of Morphometric Measures Based on Labeled Cortical Distance Maps}. in \textit{Fifth International Symposium on Image and Signal Processing and Analysis (ISPA 2007)}. 2007. Istanbul, Turkey.

34. Joshi, M., et al., \textit{Brain segmentation and the generation of cortical surfaces.} NeuroImage, 1999. \textbf{9}(5): p. 461-76.

35. Gueziec, A. and R. Hummel, \textit{Exploiting triangulated surface extraction using tetrahedral decomposition.} IEEE Transactions on Visualization and
Computer Graphics, 1995. \textbf{1}(4): p. 328-342.

36. Miller, M.I., et al., \textit{Labeled cortical mantle distance maps of the cingulate quantify differences between dementia of the Alzheimer type and healthy aging.} Proceedings of the National Academy of Sciences
of the United States of America, 2003. \textbf{100}(25): p. 15172-15177.

37. Miller, M.I., et al., \textit{Bayesian construction of geometrically based cortical thickness metrics.} Neuroimage, 2000. \textbf{12}(6): p. 676-687.

38. Ratnanather, J.T., et al., \textit{Validating cortical surface analysis of medial prefrontal cortex.} NeuroImage, 2001. \textbf{14}(5): p.
1058-1069.

39. Holm, S., \textit{A simple sequentially rejective multiple test procedure.} Scandinavian Journal of Statistics, 1979. \textbf{6}: p. 65--70.

40. Conover, W., \textit{Practical Nonparametric Statistics}. 3rd ed. 1999: Wiley {\&} Sons.

41. Manly, B., \textit{Multivariate Statistical Methods: A Primer}. 2nd ed. 1994: Chapman {\&} Hall.

42. Makris, N., et al., \textit{Cortical Thinning of the Attention and Executive Function Networks in Adults with Attention-Deficit/Hyperactivity Disorder.} Cerebral Cortex, 2006.

43. Thode Jr., H., \textit{Testing for Normality}. 2002, New York: Marcel Dekker.

44. Botteron, K.N., et al., \textit{Ventral Medial Prefrontal Cortex Metrics in Early Onset Major Depressive Disorder: A Twin MRI Study.} Submitted for Publication, 2007.

45. Dalgaard, P., \textit{Introductory Statistics with R}. 2002, New York: Springer Verlag.

46. Olejnik, S. F. and Algina, J., \textit{Type i error rates and power estimates of selected parametric
and non-parametric tests of scale.} Journal of Educational Statistics, 1987. \textbf{12}: p. 45-61.

47. Glass, G. V. and Hopkins, K. D. \textit{Statistical methods in psychology and education (3rd ed.)}. 1996, Needham Heights, MA: Allyn \& Bacon.

\vspace{2 cm}
\section*{Tables}

\begin{table}[htbp]
\centering
\begin{tabular}{|c|c|c|c||c|c|}
\hline
\multicolumn{2}{|c|}{}& \multicolumn{4}{|c|}{volume ($mm^3$)} \\
\hline
\multicolumn{2}{|c|}{}&  \multicolumn{2}{c||}{left}  & \multicolumn{2}{|c|}{right} \\
\hline
group    & n & mean & std dev & mean &  std dev   \\
\hline
MDD      & 20  & 1680.7 & 248.2 & 1607.8 & 220.0  \\
\hline
HR       & 20  & 1601.6 & 235.6 & 1589.8 & 239.5  \\
\hline
Ctrl     & 28  & 1700.6 & 295.3 & 1676.3 & 297.5 \\
\hline
overall  & 68  & 1665.6 & 264.9 & 1630.7 & 259.2  \\
\hline
\end{tabular}
\caption{
\label{tab:describe-vol}
The sample sizes ($n$), means, and standard deviations (std dev) of the volumes
for left and right VMPFCs overall and for each group.}
\end{table}

\begin{table}[htbp]
\centering
\begin{tabular}{|c|c|c||c|c|}
\multicolumn{5}{c}{$p$-values for pairwise volume ($mm^3$) comparisons}\\
\hline
 & \multicolumn{2}{|c||}{left} & \multicolumn{2}{|c|}{right} \\
\hline
pair & $p_W$ & $p_t$ & $p_W$ & $p_t$ \\
\hline
MDD,HR   & .2145 $(g)$ & .2400 $(g)$ & .3990 $(g)$  & .4068 $(g)$  \\
\hline
MDD,Ctrl & .4794 $(g)$ & .4006 $(g)$ & .3990 $(\ell)$ & .4068 $(\ell)$ \\
\hline
HR,Ctrl  & .2145 $(\ell)$ & .2400 $(\ell)$ &.3990 $(\ell)$ & .4068 $(\ell)$  \\
\hline
\end{tabular}
\caption{
\label{tab:pairwise-pval-vol}
The $p$-values for the pairwise comparisons of the mean volumes with
pairwise Wilcoxon tests and Welch's $t$-tests.
$p_W$: $p$-value based on Wilcoxon rank sum test and
$p_t$: $p$-value based on Welch's $t$-test;
$g$ ($\ell$) stands for the greater (less) than alternative.}
\end{table}

\begin{table}[htbp]
\centering
\begin{tabular}{|c|c||c|c|c|}
\multicolumn{5}{c}{$p$-values for left-right volume ($mm^3$) asymmetry}\\
\hline
 &overall & MDD   & HR  & Ctrl \\
\hline
$p_W$ &.0064* $(g)$  & .0360* $(g)$ & .2545 $(g)$ & .2545 $(g)$ \\
\hline
$p_t$ &.0087* $(g)$  & .0233* $(g)$ & .3376 $(g)$ & .2436 $(g)$ \\
\hline
\end{tabular}
\caption{
\label{tab:paired-pval-vol}
The $p$-values for the tests of left-right volume asymmetry
by Wilcoxon signed rank test.
$p_W$: $p$-value based on Wilcoxon signed rank test and
$p_t$: $p$-value based on Welch's $t$-test;
$g$ ($\ell$) stands for the greater (less) than alternative.
Significant $p$-values at $\al=0.05$ are marked with an $^*$.}
\end{table}

\begin{table}[htp]
\centering
\begin{tabular}{|c|c||c|c|c||c|c|}
\multicolumn{7}{c}{correlation coefficients} \\
\hline
&overall (L,R) &   MDD (L,R)    &  HR (L,R)     & Ctrl (L,R) & MDDL,HRL & MDDR,HRR\\
\hline
$\rho_S$   & .8882 & .8120 & .8556 & .9425 & .4120 &.3158 \\
\hline
$p$   & $<.0001$*  & $<.0001$* & $<.0001$* & $<.0001$* & .0359* & .0868  \\
\hline
\end{tabular}
\caption{
\label{tab:correlation-vol}
The Spearman correlation coefficients (denoted by $\rho_S$) between left and right VMPFC volumes
and the associated $p$-values for the alternative that correlation coefficient is non-zero.
Significant $p$-values at $\al=0.05$ are marked with an $^*$.
}
\end{table}

\begin{table}[htbp]
\centering
\begin{tabular}{|c|c||c|}
\multicolumn{3}{c}{$p$-values for cdf comparisons}\\
\hline
& \multicolumn{2}{|c|}{volume ($mm^3$)} \\
\hline
pair & left & right \\
\hline
MDD,Ctrl & .2735 $(\ell)$ & .1489 $(g)$ \\
\hline
HR,Ctrl  & .1792 $(g)$ & .1489 $(g)$\\
\hline
\end{tabular}
\caption{
\label{tab:cdf-pval-vol}
The $p$-values based on K-S test for the cdf comparisons (overall and by group)
of the volumes.
$g$ ($\ell$) stands for the greater (less) than alternative.}
\end{table}

\begin{table}[htbp]
\centering
\begin{tabular}{|c|c|c|c|c||c|c|c|c|}
\hline
&\multicolumn{4}{|c||}{Left VMPFC}  &\multicolumn{4}{|c|}{Right VMPFC}  \\
\hline Group & $n$ & mean & median & std dev & $n$ & mean & median &
std dev \\
\hline
MDD & 238937 & 1.62 & 1.46 & 1.13 & 170534 & 1.63 & 1.49 & 1.10 \\
\hline
HR & 228224 & 1.61 & 1.46 & 1.11 & 216978 & 1.59 & 1.46 & 1.08 \\
\hline
Ctrl & 308498 & 1.66 & 1.50 & 1.14 & 293479 & 1.66 & 1.53 & 1.12 \\
\hline
Overall & 775659 & 1.63 & 1.48 & 1.13 & 680991 & 1.63 & 1.50 & 1.10 \\
\hline
\end{tabular}
\caption{
\label{tab:stat-pooled-LCDM}
The sample sizes ($n$), means, medians, and standard deviations (std dev) of the pooled LCDM distances
(in $mm$) for left and right VMPFCs overall and for each group
(after extreme subjects are removed).}
\end{table}

\begin{table}[htbp]
\centering
\begin{tabular}{|c|c|c||c|c|}
\hline
\multicolumn{5}{|c|}{With $t$-test}  \\
\hline
 & \multicolumn{2}{|c||}{all subjects included}  & \multicolumn{2}{|c|}{extreme subjects removed}  \\
\hline
Pair  & Left & Right & Left & Right \\
\hline
MDD, HR & .0383* ($g)$ & .0041* ($g)$ & $<.0001^\ast$ $(g)$ & $<.0001^\ast$ $(g)$ \\
\hline
MDD, Ctrl & $<.0001^\ast$ ($\ell )$ & $<.0001^\ast$ ($\ell )$ & $<.0001^\ast$ ($\ell )$ & $<.0001^\ast$ ($\ell )$ \\
\hline
HR, Ctrl & $<.0001^\ast$ ($\ell )$ & $<.0001^\ast$ ($\ell )$ & $<.0001^\ast$ ($\ell )$ & $<.0001^\ast$ ($\ell )$ \\
\hline
\multicolumn{5}{|c|}{With Wilcoxon rank sum test}  \\
\hline
 &\multicolumn{2}{|c||}{all subjects included}  &\multicolumn{2}{|c|}{extreme subjects removed}  \\
\hline
Pair  & Left & Right & Left & Right \\
\hline
MDD, HR & .3022 ($\ell )$ & .0776 ($g)$ & .0084* ($g)$ & $<.0001^\ast$ $(g)$ \\
\hline MDD, Ctrl & $<.0001^\ast$ ($\ell )$ & $<.0001^\ast$ ($\ell )$
& $<.0001^\ast$ ($\ell )$ &
$<.0001^\ast$ ($\ell )$ \\
\hline
HR, Ctrl & $<.0001^\ast$ ($\ell )$ & $<.0001^\ast$ ($\ell )$ & $<.0001^\ast$ ($\ell )$ & $<.0001^\ast$ ($\ell )$ \\
\hline
\end{tabular}
\caption{
\label{tab:pairwise-pooled-LCDM}
The $p$-values for the simultaneous pairwise comparisons of the pooled distances by
Welch's $t$ and Wilcoxon rank sum tests.
The $p$-values are adjusted by Holm's correction method.
$g$ ($\ell$) stands for the greater (less) than alternative.
Significant $p$-values at $\al=0.05$ are marked with an $^*$.}
\end{table}

\begin{table}[htbp]
\centering
\begin{tabular}{|c|c|c||c|c|}
\hline
 &
\multicolumn{2}{|c||}{with all subjects included}  & \multicolumn{2}{|c|}{extreme subjects removed}  \\
\hline
Pair  & Left & Right & Left & Right \\
\hline
MDD, HR & $<.0001^\ast$ $(g)$ & $<.0001^\ast$ $(g)$ & $<.0001^\ast$ $(g)$ & $<.0001^\ast$ $(g)$ \\
\hline
MDD, Ctrl & $<.0001^\ast$ ($\ell )$ & $<.0001^\ast$ ($\ell )$ & $<.0001^\ast$ ($\ell )$ & $<.0001^\ast$ ($\ell )$ \\
\hline
HR, Ctrl & $<.0001^\ast$ ($\ell )$ & $<.0001^\ast$ ($\ell )$ & $<.0001^\ast$ ($\ell )$ & $<.0001^\ast$ ($\ell )$ \\
\hline
\end{tabular}
\caption{
\label{tab:pairwise-HOV-LCDM}
The $p$-values for the simultaneous pairwise comparisons of the variances of distances by
B-F HOV test.
The $p$-values are adjusted by Holm's correction method.
$g$ ($\ell$) stands for the greater (less) than alternative.
Significant $p$-values at $\al=0.05$ are marked with an $^*$.}
\end{table}

\begin{table}[htbp]
\centering
\begin{tabular}{|c|c||c|c|c|}
\hline \multicolumn{5}{|c|}{With $t$-test}  \\ \hline  & overall &
MDD & HR &
Ctrl \\
\hline
all subjects & $<.0001^\ast$ $(g)$ & $<.0001^\ast$ $(g)$ & $<.0001^\ast$ $(g)$ & $<.0001^\ast$ $(g)$ \\
\hline
outliers removed & .1439 ($g)$ & .0227* ($\ell )$ & $<.0001^\ast$ $(g)$ & .0681* ($\ell )$ \\
\hline
\multicolumn{5}{|c|}{With Wilcoxon test}  \\
\hline
all subjects & $<.0001^\ast$ $(g)$ & $<.0001^\ast$ $(g)$ & $<.0001^\ast$ $(g)$ & $<.0001^\ast$ $(g)$ \\
\hline
outliers removed & $<.0001^\ast$ ($\ell )$ & $<.0001^\ast$($\ell )$ & .0015* $(g)$ & $<.0001^\ast$ ($\ell )$ \\
\hline
\end{tabular}
\caption{
\label{tab:pairwise-LR-asymmetry-LCDM}
The $p$-values for the tests of left-right distance asymmetry by Welch's $t$ and Wilcoxon rank sum tests.
The $p$-values for groups are adjusted by Holm's correction method.
$g$ ($\ell$) stands for the greater (less) than alternative.
Significant $p$-values at $\al=0.05$ are marked with an $^*$.}
\end{table}

\begin{table}[htbp]
\centering
\begin{tabular}{|c|c|c|c||c|c|c|}
\hline
 &
\multicolumn{6}{|c|}{$p-$values for cdf comparisons when all subjects included}  \\
\hline
 &
\multicolumn{3}{|c||}{Left}  & \multicolumn{3}{|c|}{Right}  \\
\hline
Pair & 2-sided & $1^{st}<2^{nd}$  & $1^{st}>2^{nd}$ & 2-sided & $1^{st}<2^{nd}$  & $1^{st}>2^{nd}$ \\
\hline
MDD, HR & $<.0001^\ast$ & $<.0001^\ast$ & .0073* & .0316* & .0158* & .6017 \\
\hline
MDD, Ctrl & $<.0001^\ast$ & .8340 & $<.0001^\ast$ & $<.0001^\ast$ & .0138* & $<.0001^\ast$ \\
\hline
HR, Ctrl & $<.0001^\ast$ & .8340 & $<.0001^\ast$ & $<.0001^\ast$ & .0129* & $<.0001^\ast$ \\
\hline
 &
\multicolumn{6}{|c|}{$p-$values for cdf comparisons when extreme subjects removed}  \\
\hline
 &
\multicolumn{3}{|c||}{Left}  & \multicolumn{3}{|c|}{Right}  \\
\hline
Pair & 2-sided & $1^{st}<2^{nd}$  & $1^{st}>2^{nd}$ & 2-sided & $1^{st}<2^{nd}$  & $1^{st}>2^{nd}$ \\
\hline
MDD, HR & $<.0001^\ast$ & $<.0001^\ast$ & .7519 & $<.0001^\ast$ & $<.0001^\ast$ & .9585 \\
\hline
MDD, Ctrl & $<.0001^\ast$ & .9544 & $<.0001^\ast$ & $<.0001^\ast$ & 1.000 & $<.0001^\ast$ \\
\hline
HR, Ctrl & $<.0001^\ast$ & .9544 & $<.0001^\ast$ & $<.0001^\ast$ & 1.000 & $<.0001^\ast$ \\
\hline
\end{tabular}
\caption{
\label{tab:pairwise-cdf-LCDM}
The $p$-values based on K-S test for the cdf comparisons (overall and by group) of the pooled LCDM distances.
The $p$-values for each type of alternative are adjusted by Holm's
correction method.
Significant $p$-values at $\al=0.05$ are marked with an $^*$.}
\end{table}

\begin{table}[htbp]
\centering
\begin{tabular}{|c|c|c|c|c||c|c|c|}
\hline
& \multicolumn{4}{|c||}{Empirical size} & \multicolumn{3}{|c|}{Prop. of agreement}  \\
\hline $(n_x ,n_y ,n_z )$& $\widehat {\alpha}_{BF}$& $\widehat
{\alpha}_{KW}$& $\widehat {\alpha}_{F_1}$& $\widehat{\alpha}_{F_2}$&
$\widehat {\alpha}_{KW,F_1}$& $\widehat {\alpha}_{KW,F_2}$&$\widehat {\alpha}_{F_1 ,F_2}$ \\
\hline
(1000,1000,1000)& .0508$^{a}$& .0511$^{a}$& .0508$^{a}$&
.0506$^{a}$& .0417$^{ l}$& .0419$^{l}$&.0499$^{\approx}$ \\
\hline
(5000,5000,10000)& .0516$^{a}$& .0495$^{a}$& .0498$^{a}$&
.0497$^{a}$& .0386$^{l}$ & .0386$^{l}$& .0491$^{\approx}$ \\
\hline
(5000,7500,10000)& .0499$^{a}$& .0480$^{a}$& .0451$^{a,<}$&
.0449$^{a,<}$& .0368$^{l}$& .0369$^{l}$&.0446$^{\approx}$ \\
\hline
(10000, 10000, 10000)& .0485$^{a}$& .0483$^{a}$& .0483$^{a}$&
.0480$^{a}$& .0392$^{l}$& .0392$^{l}$ & .0477$^{\approx}$ \\
\hline
\end{tabular}
\caption{
\label{tab:emp-sign-levels-3-sample}
Estimated significance levels and proportion of agreement between the tests
based on Monte Carlo simulation of distances with three groups,
${\cal X}$, ${\cal Y}$, and ${\cal Z}$ each with size $n_x$, $n_y$,
and $n_z$, respectively, with $N_{mc}=10000$ Monte Carlo replicates.
$\widehat {\alpha}_{BF}$ is for empirical size estimate for B-F test,
$\widehat {\alpha}_{KW}$ is for K-W test, $\widehat {\alpha}_{F_1}$
and $\widehat {\alpha}_{F_2}$ are for ANOVA $F$-tests with and without HOV, respectively;
$\widehat {\alpha}_{KW,F_1}$ is the proportion of agreement
between K-W and ANOVA $F$-test with HOV,
$\widehat {\alpha}_{KW,F_2}$is the proportion of
agreement between K-W and ANOVA $F$-test without HOV,
and $\widehat {\alpha}_{F_1 ,F_2}$
is the proportion of agreement between ANOVA $F$-tests with and without HOV.
The empirical sizes with the same superscript are not significantly
different from each other.
$^{>}$:Empirical size is significantly larger than 0.05; i.e. method is liberal.
$^{<}$:Empirical size is significantly smaller than 0.05; i.e. method is conservative.
$^{l}$:The proportion of agreement significantly less than the minimum of the empirical sizes.
$^{\approx}$:The proportion of agreement not significantly less than the minimum of the empirical
sizes. }
\end{table}

\begin{table}[htbp]
\centering
\begin{tabular}{|c|c|c|c|c|}
\hline
\multicolumn{5}{|c|}{$(r_y ,r_z )=$(1.1,1.0); $(\eta_y ,\eta_z )$=(0,0)}  \\
\hline
$(n_x ,n_y ,n_z )$ & $\widehat {\beta}_{BF}$ & $\widehat {\beta}_{KW}$ & $\widehat {\beta}_{F_1}$ & $\widehat {\beta}_{F_2}$ \\
\hline
(1000,1000,1000) & .0511 & .0778 & .0770 & .0768 \\
\hline
(5000,5000,10000) & .0511 & .2281 & .2137 & .2114 \\
\hline
(5000,10000,5000) & .0512 & .2936 & .2731 & .2745 \\
\hline
(5000,10000,7500) & .0508 & .3244 & .2939 & .2947 \\
\hline
(10000,10000,10000) & .0482 & .3900 & .3564  & .3559 \\
\hline
\multicolumn{5}{|c|}{$(r_y ,r_z )=$(1.1,1.2); $(\eta_y ,\eta_z )$=(0,0)}  \\
\hline
(1000,1000,1000) & .0516 & .1396 & .1316 & .1313 \\
\hline
(5000,5000,10000) & .0519 & .6725 & .6315 & .6317 \\
\hline
(10000,5000,5000) & .0503 & .6651 & .6262 & .6253 \\
\hline
(5000,10000,5000) & .0516 & .5296 & .4828 & .4828 \\
\hline
(10000,10000,10000) & .0490 & .8410 & .8050 & .8050 \\
\hline
\multicolumn{5}{|c|}{$(r_y ,r_z )=$(1.0,1.0); $(\eta_y ,\eta_z )$=(10,0)}  \\
\hline
(1000,1000,1000) & .0899 & .0574 & .0728 & .0721 \\
\hline
(5000,5000,10000) & .3408 & .0767 & .1930 & .1854 \\
\hline
(5000,10000,5000) & .4275 & .0884 & .2341 & .2381 \\
\hline
(5000,7500,10000) & .4378 & .0832 & .2415 & .2360 \\
\hline
(5000,10000,7500) & .4713 & .0878 & .2571 & .2584 \\
\hline
(10000,10000,10000) & .5564 & .1006 & .3127 & .3061 \\
\hline
\multicolumn{5}{|c|}{$(r_y ,r_z )=$(1.0,1.0); $(\eta_y ,\eta_z )$=(10,30)}  \\
\hline
(1000,1000,1000) & .2236 & .0963 & .1519 & .1512 \\
\hline
(5000,5000,10000) & .9255 & .3986 & .7436 & .7537 \\
\hline
(10000,5000,5000) & .9186 & .3556 & .7175 & .7071 \\
\hline
(5000,10000,5000) & .8083 & .2908 & .5826 & .5831 \\
\hline
(5000,7500,10000) & .9214 & .4191 & .7578 & .7627 \\
\hline (10000,7500,5000) & .9144 & .3652 & .7229 &
.7147 \\
\hline
(10000,5000,7500) & .9643 & .4554 & .8260 & .8226 \\
\hline
(7500,5000,10000) & .9644 & .4739 & .8331 & .8363 \\
\hline
(7500,10000,5000) & .8765 & .3421 & .6743 & .6702 \\
\hline
(5000,10000,7500) & .8811 & .3752 & .6938 & .6983 \\
\hline
(10000,10000,10000) & .9851 & .5352 & .8842 & .8835 \\
\hline
\end{tabular}
\caption{
\label{tab:emp-power-3-sample}
The power estimates based on Monte Carlo simulation of distances with three groups,
${\cal X}$, ${\cal Y}$, and ${\cal Z}$ each with size $n_x$, $n_y$, and
$n_z$, respectively, with $N_{mc}=10000$ Monte Carlo replicates.
For the parameters $r_y$, $r_z$, $\eta_y$, and $\eta_z$, see Section \ref{sec:emp-power-multi-sample}.
$\widehat {\beta}_{BF}$ is the empirical power estimate for B-F test,
$\widehat {\beta}_{KW}$ is for K-W test,
$\widehat {\beta}_{F_1}$ and $\widehat {\beta}_{F_2}$ are for ANOVA $F$-tests
with and without HOV, respectively.
}
\end{table}

\begin{table}[htbp]
\centering
\begin{tabular}{|c|c|c|c|c||c|c|c|}
\hline
\multicolumn{8}{|c|}{Two-Sided Tests}  \\
\hline
&\multicolumn{4}{|c||}{Empirical size}  & \multicolumn{3}{|c|}{Prop. of agreement}  \\
\hline $(n_x ,n_y )$ & $\widehat {\alpha}_{BF}$ & $\widehat
{\alpha}_W $ & $\widehat {\alpha}_t $ & $\widehat {\alpha}_{KS}$ &
$\widehat
{\alpha}_{W,t}$ & $\widehat {\alpha}_{W,KS}$ & $\widehat {\alpha}_{t,KS}$ \\
\hline
(1000,1000) & .0514 & .0517 & .0505 & .0486 & .0403$^{l}$ & .0305$^{l}$  & .0273$^{l}$ \\
\hline
(5000,10000) & .0533 & .0457$^{<}$ & .0463$^{<}$ & .0465 & .0356$^{l}$  & .0273$^{l}$  & .0244$^{l}$  \\
\hline
(7500,10000) & .0486 & .0493 & .0463$^{<}$ & .0464 & .0385$^{l}$  & .0282$^{l}$ & .0246$^{l}$  \\
\hline
(10000, 10000) & .0525 & .0518 & .0525 & .0501 & .0421$^{l}$ & .0320$^{l}$ & .0281$^{l}$  \\
\hline
\multicolumn{8}{|c|}{Left-Sided Tests (i.e., ${\cal X}$ values tend to be smaller than ${\cal Y}$ values)}  \\
\hline
 &\multicolumn{4}{|c||}{Empirical size}  &\multicolumn{3}{|c|}{Prop. of agreement}  \\
\hline
(1000,1000) & .0503 & .0517 & .0527 & .0486 & .0440$^{l}$ & .0329$^{l}$  & .0305$^{l}$ \\
\hline
(5000,10000) & .0512 & .0470 & .0489 & .0492 & .0382$^{l}$  & .0311$^{l}$  & .0282$^{l}$ \\
\hline
(7500,10000) & .0521 & .0490 & .0493 & .0478 & .0399$^{l}$  & .0322$^{l}$ & .0284$^{l}$ \\
\hline
(10000, 10000) & .0489 & .0517 & .0514 & .0494 & .0426$^{l}$ & .0330$^{l}$ & .0301$^{l}$ \\
\hline
\multicolumn{8}{|c|}{Right-Sided Tests (i.e., ${\cal X}$ values tend to be larger than ${\cal Y}$ values)}  \\
\hline
 &\multicolumn{4}{|c||}{Empirical size}  &\multicolumn{3}{|c|}{Prop. of agreement}  \\
\hline
(1000,1000) & .0514 & .0521 & .0502 & .0491 & .0409$^{l}$ & .0337$^{l}$ & .0294$^{l}$ \\
\hline
(5000,10000) & .0485 & .0486 & .0502 & .0478 & .0405$^{l}$ & .0308$^{l}$ & .0285$^{l}$  \\
\hline
(7500,10000) & .0493 & .0479 & .0469 & .0495 & .0391$^{l}$ & .0325$^{l}$ & .0287$^{l}$ \\
\hline
(10000, 10000) & .0549$^{>}$ & .0532 & .0517 & .0469 & .0435$^{l}$ & .0354$^{l}$ & .0311$^{l}$ \\
\hline
\end{tabular}
\caption{
\label{tab:emp-size-2-sample}
Estimated significance levels based on Monte Carlo simulation of distances with two groups
${\cal X}$ and ${\cal Y}$ each with size $n_x$ and $n_z$,
respectively, with $N_{mc}=10000$ Monte Carlo replicates.
$\widehat {\alpha}_{BF}$ is the empirical size estimate for B-F test,
$\widehat {\alpha}_W $ is for Wilcoxon rank sum test,
$\widehat {\alpha}_t $ is for Welch's $t$-test,
$\widehat {\alpha}_{KS}$ is for K-S test;
$\widehat {\alpha}_{W,t}$ is the proportion of agreement between Wilcoxon rank sum and Welch's $t$-tests,
$\widehat {\alpha}_{W,KS}$ is the proportion of agreement between Wilcoxon rank sum and K-S tests, and
$\widehat {\alpha}_{t,KS}$ is the proportion of agreement between Welch's $t$-test and K-S test.
$^{>}$:Empirical size is significantly larger than 0.05; i.e. method is liberal.
$^{<}$:Empirical size is significantly smaller than 0.05; i.e. method is conservative.
$^{l}$:The proportion of agreement significantly less than the minimum of the empirical sizes.
$^{\approx}$:The proportion of agreement not significantly less than
the minimum of the empirical sizes.}
\end{table}

\begin{table}[htbp]
\centering
\begin{tabular}{|c|c|c|c|c||c|c|c|c|}
\hline
\multicolumn{9}{|c|}{$r_y=1.1; \; \eta_y=0$}  \\
\hline
 & \multicolumn{4}{|c||}{Two-Sided}  & \multicolumn{4}{|c|}{Left-Sided}  \\
\hline $(n_x ,n_y )$ & $\widehat {\beta}_{BF}$ & $\widehat {\beta}_W
$ & $\widehat {\beta}_t $ & $\widehat {\beta}_{KS}$ & $\widehat
{\beta}_{BF}$ & $\widehat {\beta}_W $ & $\widehat {\beta}_t $ & $\widehat {\beta}_{KS}$ \\
\hline
(1000,1000) & .0517 & .1317 & .1264 & .0788 & .0524 & .0742 & .0712 & .0750 \\
\hline
(5000,10000) & .0529 & .2723 & .2520 & .3734 & .0568 & .3816 & .3600 & .5122 \\
\hline
(10000,5000) & .0471 & .2720 & .2507 & .3753 & .0522 & .3838 & .3572 & .5157 \\
\hline
(7500,10000) & .0491 & .3242 & .3046 & .4731 & .0570 & .4425 & .4178 & .6139 \\
\hline
(10000,7500) & .0515 & .3305 & .3100 & .4850 & .0551 & .4455 & .4204 & .6253 \\
\hline
(10000,10000) & .0498 & .3662 & .3362 & .5504 & .0521 & .4924 & .4588 & .6861 \\
\hline
\multicolumn{9}{|c|}{$r_y=1.2; \; \eta_y=0$}  \\
\hline
(1000,1000) & .0511 & .2635 & .2533 & .1838 & .0527 & .1695 & .1630 & .1813 \\
\hline
(5000,10000) & .0547 & .7606 & .7331 & .9401 & .0663 & .8463 & .8250 & .9755 \\
\hline
(10000,5000) & .0521 & .7588 & .7269 & .9421 & .0627 & .8437 & .8178 & .9765 \\
\hline
(7500,10000) & .0526 & .8514 & .8282 & .9839 & .0666 & .9121 & .8950 & .9945 \\
\hline
(10000,7500) & .0545 & .8561 & .8300 & .9845 & .0672 & .9133 & .8969 & .8882 \\
\hline
(10000,10000) & .0512 & .8976 & .8750 & .9935 & .0624 & .9468 & .9312 & .9982 \\
\hline
\multicolumn{9}{|c|}{$r_y=1.0; \quad \eta_y=10$}  \\
\hline
(1000,1000) & .0965 & .0772 & .1173 & .0514 & .1009 & .0506 & .0677 & .0477 \\
\hline
(5000,10000) & .3885 & .0871 & .2222 & .0673 & .5111 & .1361 & .3297 & .1089 \\
\hline
(10000,5000) & .3687 & .0841 & .2186 & .0670 & .4941 & .1390 & .3232 & .1076 \\
\hline
(7500,10000) & .4767 & .0951 & .2638 & .0737 & .5964 & .1497 & .3786 & .1159 \\
\hline
(10000,7500) & .4647 & .0995 & .2630 & .0748 & .5927 & .1560 & .3725 & .1161 \\
\hline
(10000,10000) & .5253 & .1018 & .2978 & .0743 & .6505 & .1628 & .4132 & .1200 \\
\hline
\multicolumn{9}{|c|}{$r_y=1.0; \; \eta_y=30$}  \\
\hline
(1000,1000) & .2858 & .1760 & .2887 & .0878 & .2905 & .1028 & .1885 & .0793 \\
\hline
(5000,10000) & .9528 & .4677 & .8254 & .7080 & .9769 & .5927 & .8881 & .8911 \\
\hline
(10000,5000) & .9452 & .4668 & .8094 & .6901 & .9707 & .5918 & .8807 & .8659 \\
\hline
(7500,10000) & .9859 & .5578 & .8987 & .9078 & .9938 & .6773 & .9435 & .9792 \\
\hline
(10000,7500) & .9837 & .5509 & .8976 & .8983 & .9919 & .6750 & .9438 & .9713 \\
\hline
(10000,10000) & .9932 & .6188 & .9369 & .9691 & .9971 & .7339 & .9679 & .9942 \\
\hline
\multicolumn{9}{|c|}{$r_y=1.0; \; \eta_y=50$}  \\
\hline
(1000,1000) & .4724 & .3361 & .4865 & .2041 & .4789 & .2266 & .3521 & .2048 \\
\hline
(5000,10000) & .9989 & .8876 & .9842 & .9980 & .9998 & .9363 & .9936 & .9998 \\
\hline
(10000,5000) & .9981 & .8830 & .9844 & .9980 & .9996 & .9325 & .9931 & 1.000 \\
\hline
(7500,10000) & .9999 & .9478 & .9964 & 1.000 & .9732 & .9932 & .9986 & 1.000 \\
\hline
(10000,7500) & .9998 & .9473 & .9961 & 1.000 & 1.000 & .9741 & .9984 & 1.000 \\
\hline
(10000,10000) & .9999 & .9716 & .9984 & 1.000 & 1.000 & .9847 & .9995 & 1.000 \\
\hline
\end{tabular}
\caption{
\label{tab:emp-power-2-sample}
The power estimates based
on Monte Carlo simulation of distances with two groups, ${\cal X}$,
and ${\cal Y}$, each with size $n_x $ and $n_y$, respectively, with
$N_{mc}=10000$ Monte Carlo replicates.
For the parameters $r_y$ and $\eta_y$, see Section \ref{sec:emp-power-two-sample}.
$\widehat {\beta}_{BF}$ is the power estimate for B-F test,
$\widehat {\beta}_W $ is for Wilcoxon rank sum test,
$\widehat {\beta}_t $ is for Welch's $t$-test,
and $\widehat {\beta}_{KS}$ is for K-S test.
}
\end{table}

\newpage
\section*{Figures}

\end{document}